\documentclass[preprint,11pt]{article}
\usepackage{amssymb,amsfonts,amsmath,amsthm,amscd}
\usepackage{epic,graphicx}

\newenvironment{pseudocode}[1]{
\begin{enumerate}\tt\small\linewidth=11cm
\item[]\underline{{#1}}}
{\end{enumerate}}

\def\T{{\sf T}}
\def\poisson{{\sf Poisson}}
\def\ome{\overline{\nu}}
\def\cX{{\cal X}}
\def\hu{\underline{h}}
\def\uu{\underline{u}}
\def\ox{\overline{x}}

\def\Fbe{F_{\rm B}}

\def\ub{\underline{b}}

\def\hf{{\sf f}}
\def\pd{{\sf p}}
\def\dens{{\sf a}}
\def\hdens{\widehat{\sf a}}
\def\atanh{{\rm atanh}}
\def\normeq{\propto}
\def\di{\partial i}
\def\da{\partial a}
\def\me{\nu}
\def\mh{\widehat{\nu}}
\def\M{{\sf M}}

\def\cA{{\cal A}}
\def\0t{{\tt 0}}
\def\1t{{\tt 1}}
\def\ux{\underline{x}}
\def\uy{\underline{y}}
\def\uz{\underline{z}}
\def\enc{{\sf F}}
\def\dec{\widehat{\sf F}}
\def\Code{{\mathfrak C}}
\def\ed{\stackrel{{\rm d}}{=}}
\def\uX{\underline{X}}
\def\uY{\underline{Y}}
\def\uh{\widehat{u}}
\def\ind{{\mathbb I}}
\def\E{{\mathbb E}}
\def\capac{{\tt C}}
\def\entro{{\mathfrak h}}
\def\cN{{\cal N}}
\def\ve{\varepsilon}
\def\H{{\mathbb H}}
\def\ux{\underline{x}}
\def\u0t{\underline{\tt 0}}
\def\ecN{\overline{\cal N}}
\def\coeff{{\sf coeff}}
\def\de{{\rm d}}
\def\neigh{{\sf B}}

\def\sB{\mbox{\tiny{B}}}
\def\sb{\mbox{\tiny{b}}}

\def\BMS{{\rm BMS}}
\def\BSC{{\rm BSC}}
\def\BEC{{\rm BEC}}
\def\sBP{\mbox{\tiny BP}}

\def\prob{{\mathbb P}}

\def\uxh{\widehat{\underline{x}}}
\def\xh{\widehat{x}}

\def\ball{{\sf B}}
\def\block{{\rm P}_{{\rm B}}}
\def\bit{{\rm P}_{{\rm b}}}

\def\P{{\mathfrak P}}
\def\Ph{\widehat{\mathfrak P}}
\def\cF{{\cal F}}

\newcommand{\graphtextsize}{\small}
\newcommand{\graphtextsizesmall}{\tiny}
\newcommand{\iteration}{t}
\DeclareMathOperator{\expectation}{\ensuremath{\mathbb{E}}} 

\newenvironment{example}[1]{\begin{itemize}{\setlength{\rightmargin}{\leftmargin}}\item[]\small\underline{Example {#1}:}}{\end{itemize}}

\newenvironment{exercise}[1]{\begin{itemize}{\setlength{\rightmargin}{\leftmargin}}\item[]\small\underline{Exercise {#1}:}
}{\end{itemize}}

\newtheorem{propo}{Proposition}[section]

\newtheorem{thm}[propo]{Theorem}

\newcommand{\reals}{{\mathbb R}}

\newcommand{\naturals}{{\mathbb N}}

\newcommand{\eqnsection}{\renewcommand{\theequation}{\thesection.\arabic{equation}}
      \makeatletter \csname @addtoreset\endcsname{equation}{section}\makeatother}
\begin{document}
\eqnsection

\title{Modern Coding Theory: \\
The Statistical Mechanics and Computer Science Point of View}

\author{Andrea Montanari$^1$\, and R\"udiger Urbanke$^2$ \\[1mm]
        {\small\mbox{}$^1$Stanford University, montanari@stanford.edu,\;\;\;\;
        \mbox{}$^2$EPFL, ruediger.urbanke@epfl.ch}
\thanks{The work of A.~Montanari was partially supported by the European Union under the project EVERGROW. The work of R.~Urbanke was partially supported by
the NCCR-MICS, a center supported by the Swiss National Science Foundation under grant number 5005-67322.}}

\date{February 12, 2007}

\maketitle

\abstract{These are the notes for a set of lectures delivered by the two 
authors at the Les Houches Summer School on `Complex Systems' in 
July 2006. They provide an introduction to the basic concepts in
modern (probabilistic) coding theory, highlighting connections 
with statistical mechanics. 
We also stress common concepts with other disciplines dealing 
with similar problems that can be generically referred to as
`large graphical models'.

While most of the lectures are devoted to the classical channel coding
problem over simple memoryless channels, we present a discussion of
more complex channel models. We conclude with an overview of the main open 
challenges in the field.}       

%
%
\section{Introduction and Outline}

The last few years have witnessed an impressive convergence 
of interests between disciplines which are {\it a priori} well separated:
coding and information theory, statistical inference, statistical mechanics
(in particular, mean field disordered systems), as well as theoretical 
computer science. The underlying reason for this convergence is the importance of
probabilistic models and/or probabilistic techniques in each of 
these domains. This has long been obvious in information theory \cite{Sha48},
statistical mechanics \cite{Boltzmann}, and statistical inference 
\cite{Per86}. 
In the last few years it has also become
apparent in coding theory and theoretical computer science.
In the first case, the invention of Turbo codes \cite{BGT93} and the 
re-invention
of Low-Density Parity-Check (LDPC) codes \cite{MaN95,Mac97} has motivated 
the use of random constructions for coding information 
in robust/compact ways \cite{RiU07}. 
In the second case (theoretical computer science) the relevance 
of randomized algorithms has steadily increased (see for instance 
\cite{MoR95}), thus motivating
deep theoretical developments. A particularly important 
example is provided by the Monte Carlo Markov Chain method for counting 
and sampling random structures.

Given this common probabilistic background, some analogies
between these disciplines is not very surprising nor is it particularly
interesting. The key new ingredient which lifts the connections beyond some superficial commonalities
is that one can name specific problems, 
questions, and results which lie at the intersection of 
these fields while being of central interest for each of them.
The set of problems and techniques thus defined can be somewhat loosely
named  ``theory of large graphical models.'' 
The typical setting is the following: a large set of 
random variables taking values in a finite (typically quite small) alphabet with a ``local'' dependency
structure; this local dependency structure is conveniently described by an appropriate graph.

In this lecture we shall use ``modern'' coding theory 
as an entry point to the domain. There are several motivations for this:
$(i)$ theoretical work on this topic is strongly motivated 
by concrete and well-defined practical applications; 
$(ii)$ the probabilistic approach mentioned above has been
quite successful and has substantially changed the field (whence the 
reference to \emph{modern} coding theory);
$(iii)$ a sufficiently detailed picture exists illustrating the interplay
among different view points.

We start in Section~\ref{sec:background} with a brief 
outline of the (channel coding) problem. This allows us to introduce the
standard definitions and terminology used in this field.
In Section~\ref{sec:sparsegraphcode} we introduce ensembles of codes
defined by sparse random graphs and discuss their most basic property -- the weight distribution.
In Section~\ref{sec:decodingproblem} we phrase the decoding problem as an inference problem on a graph and
consider the performance of the efficient (albeit in general suboptimal) message-passing
decoder. We show how the performance of such a combination (sparse graph code and message-passing decoding)
can be analyzed and we discuss the relationship of the performance under message-passing decoding
to the performance of the optimal decoder.
In Section~\ref{sec:beyondcoding} we briefly touch on some problems 
beyond coding, in order to show as similar concept emerge there.
In particular, we discuss how message passing techniques can be successfully 
used in some families of counting/inference problems.
In Section~\ref{sec:beyondbms} we show that several of the simplifying assumptions (binary case,
symmetry of channel, memoryless channels) are convenient in that they allow for a simple theory but
are not really necessary. In particular, we discuss a simple channel with memory and we see
how to proceed in the asymmetric case. Finally, we conclude in Section~\ref{sec:openproblems}
with a few fundamental open problems. 

To readers who would like to find current contributions on this topic we recommend the
{\em IEEE Transactions on Information Theory}. A considerably more in-depth discussion 
can be found in the two upcoming books
{\em Information, Physics and Computation} \cite{MeM07}
and {\em Modern Coding Theory} \cite{RiU07}.
Standard references on coding theory are \cite{Ber84,Bla84,LiC04}
and very readable introductions to information theory can be found in \cite{CoT91,Gal68}.
Other useful reference sources are the book by Nishimori \cite{NishimoriBook} as well as the book
by MacKay \cite{MacKay}.

%
%
\section{Background: The Channel Coding Problem}
\label{sec:background}
The central problem of communications is how to 
transmit information reliably through a noisy 
(and thus unreliable) communication channel. Coding theory 
aims at accomplishing this task by adding a properly designed redundancy to 
the transmitted message. This redundancy is then used at the receiver
to reconstruct the original message despite the noise introduced by the 
channel.
%
%
\subsection{The Problem}

In order to model the situation described above we shall assume that the 
noise is random with some known 
distribution.\footnote{It is worth mentioning that an alternative approach
would be to consider the noise as `adversarial' (or worst case)
under some constraint on its intensity.}
To keep things simple we shall assume that the communication
channel admits as input binary symbols $x\in\{\0t,\1t\}$,
while the output belongs to some finite alphabet $\cA$.
We denote the probability of observing the output $y\in\cA$ given that the input
was $x\in\{\0t,\1t\}$ by $Q(y|x)$.
The channel model is defined by the transition probability matrix
\begin{eqnarray}
Q = \{Q(y|x):\, x\in\{\0t,\1t\},\; y\in\cA\}\, .
\end{eqnarray}
Of course, the entries of this matrix must be non-negative and normalized
in such a way that $\sum_y Q(y|x)=1$.
\begin{figure}[t]
\begin{tabular}{ccc}
\phantom{a}\hspace{1.5cm}\includegraphics[angle=0,width=0.15\columnwidth]{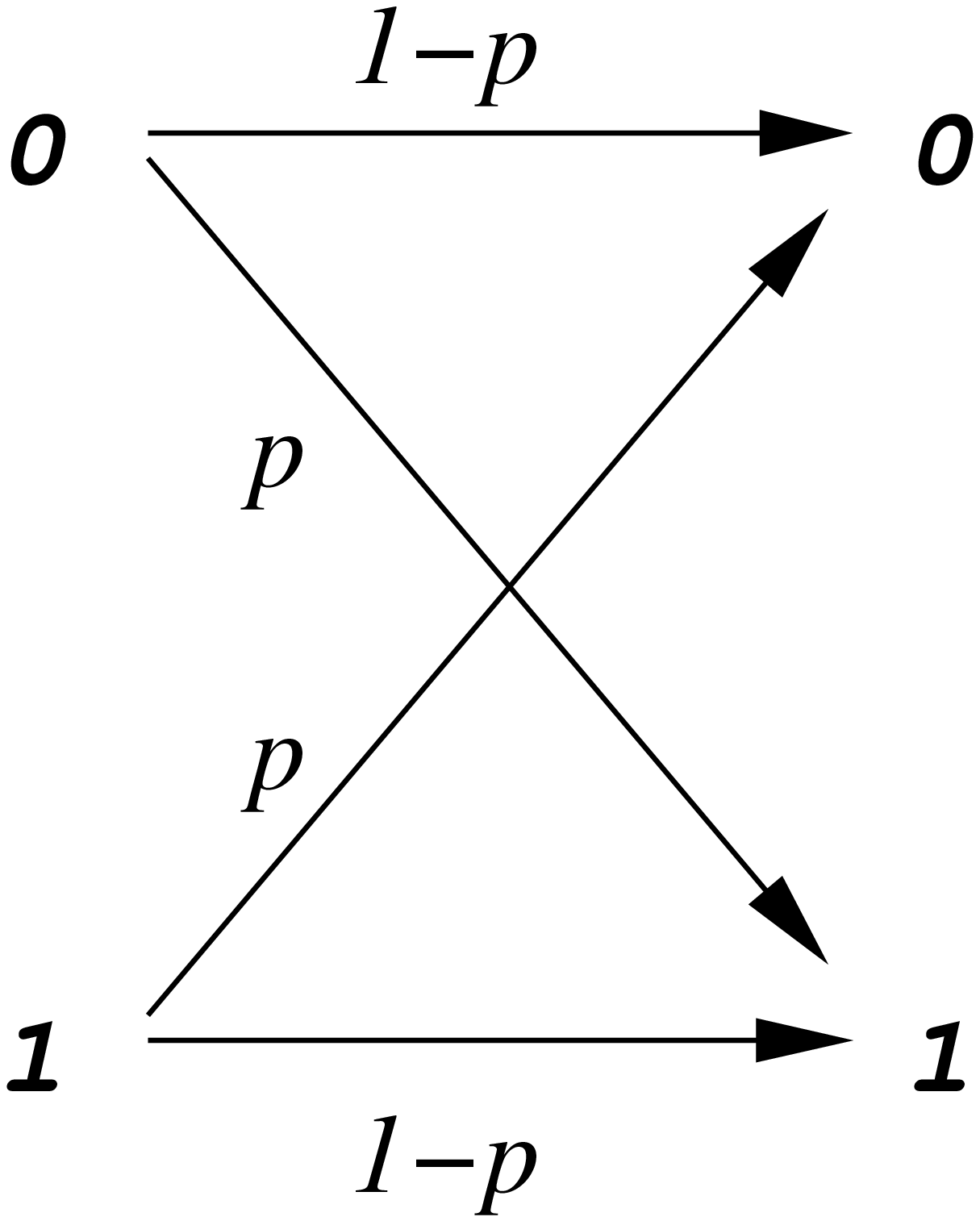}&
\hspace{2.3cm}\includegraphics[angle=0,width=0.155\columnwidth]{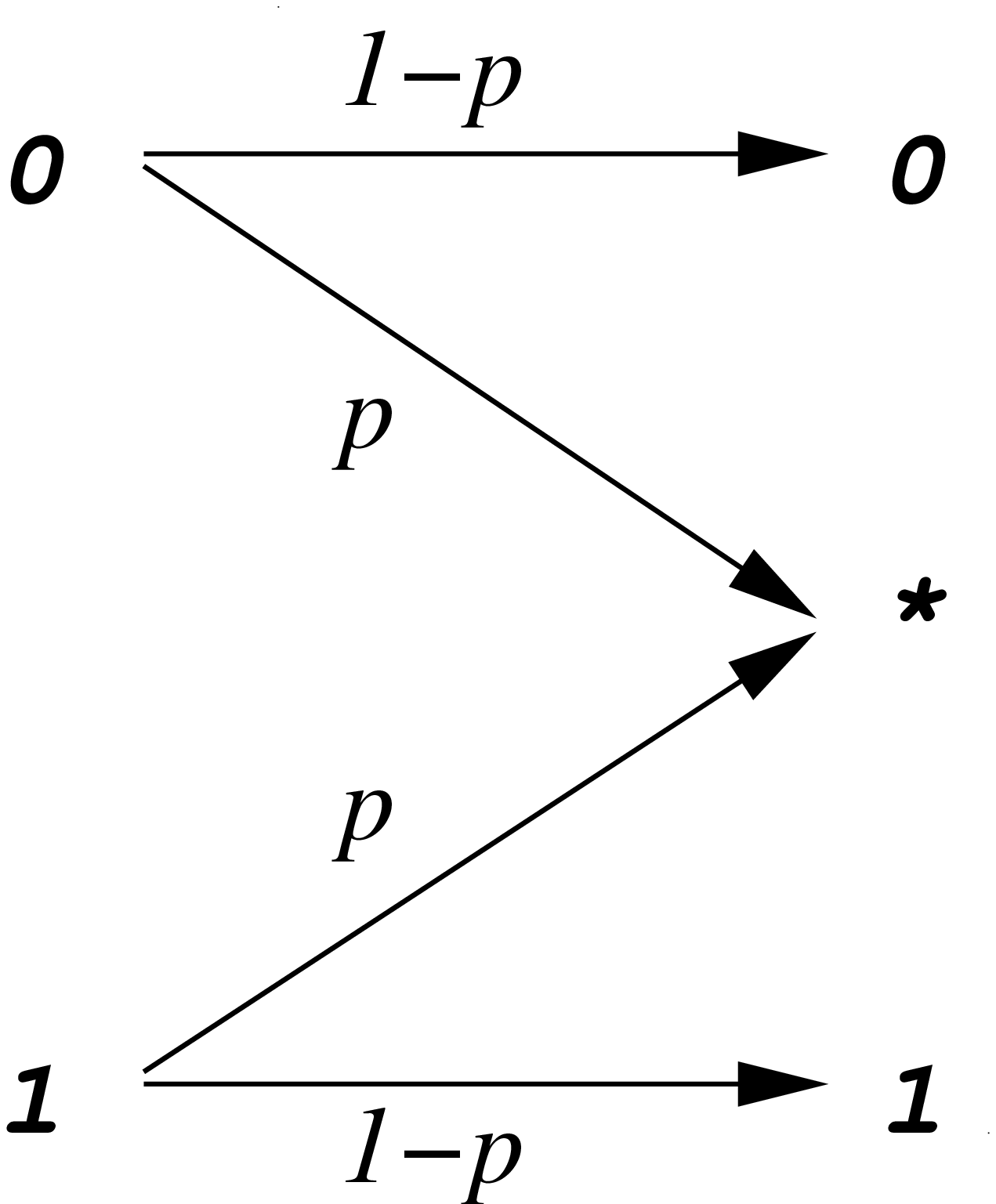}&
\hspace{1.8cm}\includegraphics[angle=0,width=0.15\columnwidth]{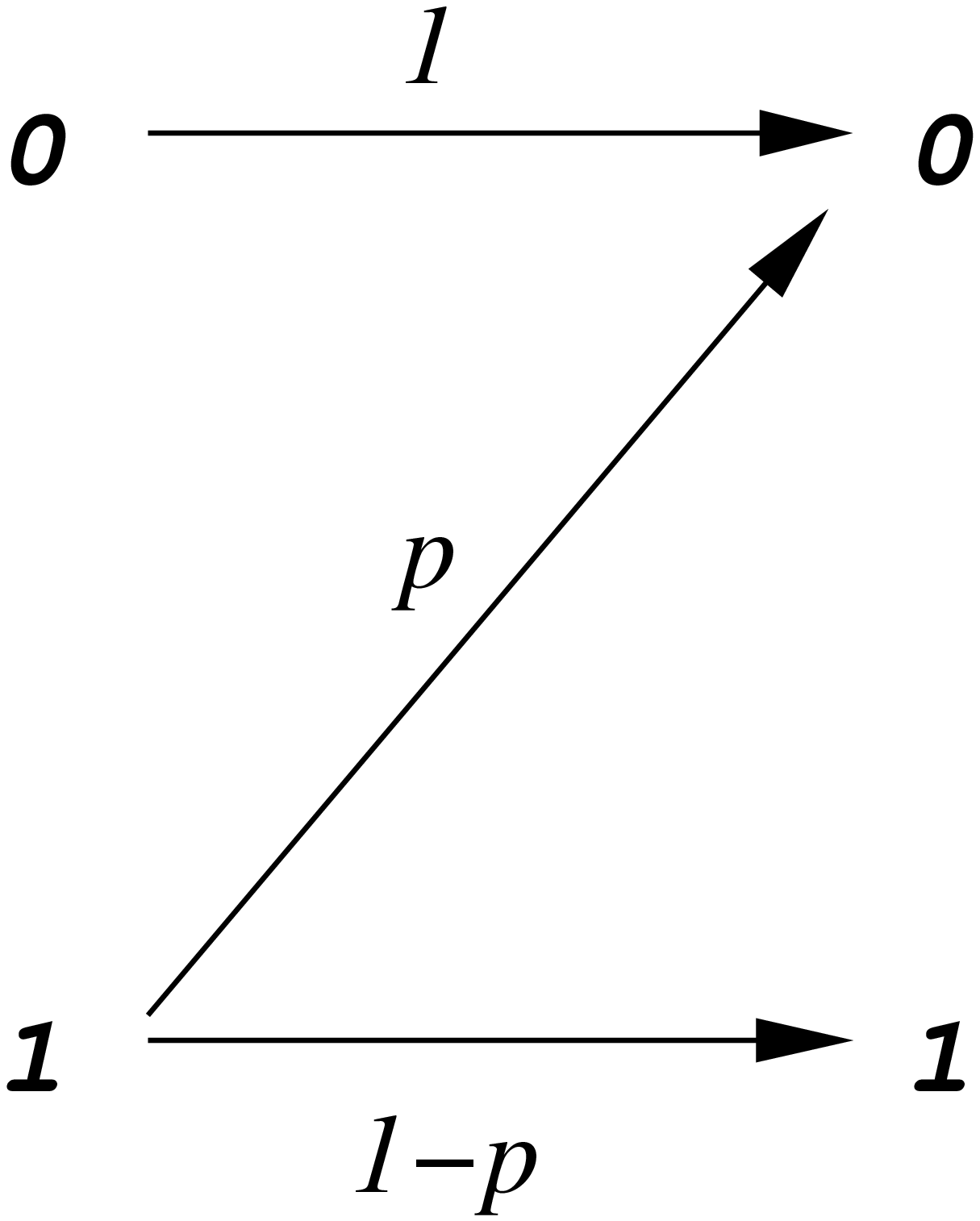}
\end{tabular}
\caption{Schematic description of three simple binary memoryless channels.
From left to right: binary symmetric channel $\BSC(p)$,
binary erasure channel $\BEC(\epsilon)$, and Z channel ZC$(p)$.}
\label{fig:Channels}
\end{figure}
It is convenient to have a few simple examples in mind.
We refer to Fig.~\ref{fig:Channels} for an illustration of the
channel models which we introduce in the following three examples.
\begin{example}{1}
The \emph{binary symmetric channel} $\BSC(p)$ is defined by 
letting $\cA = \{\0t,\1t\}$ and $Q(\0t|\0t)=Q(\1t|\1t) = 1-p$; the normalization
then enforces $Q(\1t|\0t)=Q(\0t|\1t) = p$. In words, 
the channel ``flips'' the input bit with probability $p\in[0,1]$.
Since flips are introduced for each bit independently we say that the channel is
{\em memoryless}. Except for an example in Section~\ref{sec:channelwithmemory} 
all channels which we consider are memoryless.
\end{example}
\begin{example}{2}
The \emph{binary erasure channel} $\BEC(\epsilon)$ is defined by 
$\cA = \{\0t,\1t,\ast\}$ and $Q(\0t|\0t)=Q(\1t|\1t) = 1-\epsilon$
while $Q(\ast|\0t)=Q(\ast|\1t) = \epsilon$. 
In words, the channel input is erased with probability $\epsilon$ and it is transmitted correctly otherwise.
\end{example}
\begin{example}{3}
The \emph{Z-channel} ZC$(p)$ has an output alphabet $\cA = \{\0t,\1t\}$ 
but acts differently on input $\0t$
(that is transmitted correctly) and $\1t$ (that is flipped with probability 
$p$). We invite the reader to write the transition probability 
matrix.
\end{example}
Since in each case the input is binary we speak of a {\em binary-input} channel.
Since further in all models each input symbol is distorted independently from all other ones
we say that the channels are {\em memoryless}. It is convenient to further restrict
our attention to {\em symmetric} channels:
this means that there is an involution on $\cA$
(i.e. a mapping $\iota:\cA\to\cA$ such that $\iota\circ\iota=1$)
so that $Q(y|\0t)=Q(\iota(y)|\1t)$. (E.g., if $\cA=\reals$ then we could require
that $Q(y|\0t)=Q(-y|\1t)$.)
This condition is satisfied by the first
two examples above but not by the third one. 
To summarize these three properties one refers to such models as BMS 
channels.

In order to complete the problem description we need to formalize 
the information which is to be transmitted. We shall model this 
probabilistically as well and assume that the transmitter has an information 
source that provides an infinite stream of i.i.d. fair coins:
$\{z_i;i=0,1,2,\dots\}$, with $z_i\in\{\0t,\1t\}$ uniformly at random.
The goal is to reproduce this stream faithfully after communicating it over 
the noisy channel. 

Let us stress that, despite its simplification, the present setting contains
most of the crucial and challenges of the channel coding problem. 
Some of the many generalizations are described in Section~\ref{sec:beyondbms}. 
%
%
\subsection{Block Coding}

The (general) coding strategy we shall consider here is 
\emph{block coding}. It works as follows:
\begin{itemize} 
\item The source stream
$\{z_i\}$ is chopped into blocks of length $L$. 
Denote one such block by $\uz$, $\uz = (z_1,\dots,z_L)\in\{\0t,\1t\}^L$.
\item Each block is fed into an \emph{encoder}. This is a map
$\enc:\{\0t,\1t\}^L\to\{\0t,\1t\}^N$, for some fixed $N>L$ 
(the \emph{blocklength}). In words, the encoder introduces redundancy in the source message.
Without loss of generality we can assume $\enc$ to be injective. 
It this was not the case, even 
in the absence of noise, we could not uniquely recover the transmitted information
from the observed codeword.
\item The image of $\{\0t,\1t\}^L$ under the map $\enc$ is called the \emph{codebook}, or sometimes the 
\emph{code}, and it will be denoted by $\Code$. 
The code contains $|\Code| = 2^L$ strings of length $N$ called \emph{codewords}.
These are the possible channel inputs. 
The codeword $\ux = \enc(\uz)$ is sent through the channel, bit by bit.
\item Let $\uy = (y_1,\cdots,y_N)\in\cA^N$ be the channel output.
Conditioned on $\ux$ the $y_i$, $i=1, \cdots, L$, are independent random variables with
distribution $y_i\ed Q(\cdot\,|x_i)$ (here and below $\ed$ denotes
identity in distribution and $x\ed P(\,\cdot\,)$ means that $x$
is a random variable with distribution $P(\,\cdot\,)$).
\item The channel output is fed into a \emph{decoder}, which is a map
$\dec:\cA^N\to\{\0t,\1t\}^L$. It is the objective of the decoder to reconstruct the source $\uz$
from the noisy channel output $\uy$.
\end{itemize}
\begin{figure}[t]
\setlength{\unitlength}{0.9bp}%
\center{\includegraphics[angle=0,width=0.81\columnwidth]{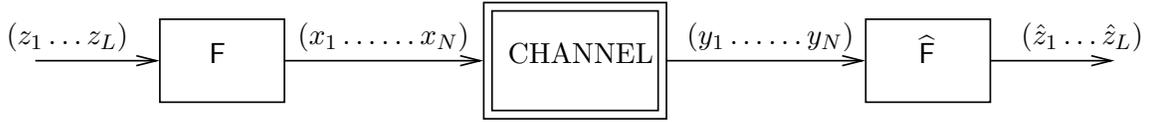}}
\put(-380,24){$\enc$}
\put(-255,24){CHANNEL}
\put(-83,24){$\dec$}
\put(-465,32){$(z_1\dots z_L)$}
\put(-343,32){$(x_1\dots\dots x_N)$}
\put(-180,32){$(y_1\dots\dots y_N)$}
\put(-40,32){$(\hat{z}_1\dots\hat{z}_L)$}
\caption{Flow chart of a block coding scheme.}
\label{fig:BlockCoding}
\end{figure}
The flow chart describing this coding scheme is shown in Fig.~\ref{fig:BlockCoding}.
It is convenient to slightly modify the above scheme.
Notice that, under the hypothesis that the encoder is injective,
the codebook is in one-to-one correspondence with the source sequences.
Since these are equiprobable, the transmitted codewords are 
equiprobable as well. We can therefore equivalently assume that
the transmitter picks a codeword uniformly at random and transmits it. Every reference
to the source stream can be eliminated if we redefine the decoder to
be a map $\dec:\cA^N\to\{\0t,\1t\}^N$, i.e., the decoder aims to reconstruct the
transmitted codeword. 
If $\dec(\uy)\not\in\Code$ we declare an 
error.\footnote{More precisely, if we are interested only in
the block probability of error, i.e., the frequency at which the whole block
of data is decoded correctly, then indeed any one-to-one mapping between information word
and codeword performs identical. If, on the
other hand, we are interested in the fraction of {\em bits} that we decode correctly
then the exact mapping from information word to codeword does come into play.
We shall ignore this somewhat subtle point in the sequel.}
In the following we shall also use the notation
$\dec(\uy) = \uxh(\uy) = (\xh_1(\uy),\dots,\xh_N(\uy))$.

One crucial parameter of a code is its \emph{rate}: it quantifies how many 
bits of information are transmitted per channel use,
\begin{eqnarray}
R \equiv \frac{L}{N} = \frac{1}{N}\,\log_2 |\Code|\, .
\end{eqnarray}
Two fundamental performance parameters are the 
\emph{bit} (or `symbol') and \emph{block} (or `word') \emph{error rates}.
The block error rate is the probability that the input codeword 
is not recovered correctly at the end of the process,
\begin{eqnarray}
\block \equiv\prob\left\{\uxh(\uy)\neq\ux\right\}\, .
\end{eqnarray}
The bit error rate is the expected fraction of bits
that are not recovered correctly,
\begin{eqnarray}
\bit \equiv\frac{1}{N}\sum_{i=1}^N\prob\left\{\xh_i(\uy)\neq x_i\right\}\, .
\end{eqnarray}

It should not be too surprising that one can trade-off rate and error probability.
We want to achieve a high rate and achieve a low probability of error.
However, increasing
the rate decreases the redundancy built into the codeword, thus inducing a higher
error probability. The aim of coding theory is to choose the code
$\Code$ and the decoding function $\uxh(\,\cdot\,)$ in a way 
to optimize this trade-off.
%
%
\subsection{Decoding}

Given the code there is a simple 
(although in general not computationally efficient) prescription
for the decoder. If we want to minimize the block error rate, we 
must chose the most likely codeword,
\begin{eqnarray}
\uxh^{\sB}(\uy)\equiv \arg\max_{\ux}\prob\{\uX=\ux|\uY=\uy\}\, .
\end{eqnarray}
To minimize the bit error rate we must instead return the sequence of 
most likely bits,
\begin{eqnarray}
\xh_i^{\sb}(\uy)\equiv \arg\max_{x_i}\prob\{X_i=x_i|\uY=\uy\}\, .
\end{eqnarray}
The reason of these prescriptions is the object of the next exercise.
\begin{exercise}{1}
Let $(U,V)$ be a pair of discrete random variables. Think of $U$ as 
a `hidden' variable and imagine you observe $V=v$. We want to
understand what is the optimal estimate for $U$ given $V=v$.
Show that the function
$v\mapsto\uh(v)$ that minimizes the error probability 
${\sf P}(\uh) \equiv \prob\left\{U\neq\uh(V)\right\}$
is given by
\begin{eqnarray}
\uh(v) = \arg\max_u \, \prob\left\{U=u|V=v\right\}\, .
\end{eqnarray}
\end{exercise}

It is instructive to explicitly write down the conditional distribution
of the channel input given the output. We shall denote
it as $\mu_{\Code,y}(\ux) = \prob\{\uX=\ux|\uY=\uy\}$ (and sometimes 
drop the subscripts $\Code$ and $y$ if they are clear from the context).
Using Bayes rule we get
\begin{eqnarray}
\mu_{\Code,y} (\ux) = \frac{1}{Z(\Code,y)}\,
\prod_{i=1}^{N}Q(y_i|x_i)\, \ind_{\Code}(\ux)\, ,
\end{eqnarray}
where $\ind_{\Code}(\ux)$ denotes the code membership function
($\ind_{\Code}(\ux)= 1$ if $\ux\in\Code$ and $=0$ otherwise).

According to the above  discussion, decoding amounts to computing the 
marginals (for symbol MAP) or the mode\footnote{We recall that the mode of
a distribution with density $\mu(\,\cdot\,)$ is the value
of $x$ that maximizes $\mu(x)$.} (for word MAP)
of $\mu(\,\cdot\,)$. 
More generally, we would like to understand the
properties of $\mu(\,\cdot\,)$: is it concentrated on a single 
codeword or spread over many of them? In the latter case, are these close to
each other or very different? And what is their relationship with the 
transmitted codeword?

The connection to statistical mechanics emerges in the study of the 
decoding problem \cite{Sou89,Rujan}.
To make it completely transparent we rewrite the distribution 
$\mu(\,\cdot\,)$ in Boltzmann form
\begin{eqnarray}
\mu_{\Code,y} (\ux) & = &\frac{1}{Z(\Code,y)}\, e^{-E_{\Code,y}(\ux)}\, ,\\
E_{\Code,y}(\ux) & = &\left\{\begin{array}{ll}
-\sum_{i=1}^N\log Q(y_i|x_i), & \mbox{if $\ux\in\Code$}\, ,\\
+\infty, &\mbox{otherwise}\, .
\end{array}
\right.\label{eq:GeneralEnergy}
\end{eqnarray}
The word MAP and bit MAP rule can then be written as
\begin{eqnarray}
\uxh^{\sB}(\uy) & = &\arg\min_{\ux}E_{\Code,y}(\ux)\, ,\\
\xh_i^{\sb}(\uy) &= & \arg\max_{x_i}\sum_{x_j:j\neq i}
\mu_{\Code,y}(\ux)\, .
\end{eqnarray}
In words, word MAP amounts to computing the ground state of a certain
energy function, and bit MAP corresponds to computing the expectation with respect to the Boltzmann 
distribution.
Notice furthermore that $\mu(\,\cdot\,)$ is itself random because of the 
randomness in $y$ (and we shall introduce further randomness in the 
choice of the code). This is analogous to what happens in statistical physics 
of disordered systems, with $\uy$ playing the role of quenched random 
variables.
%
%
\subsection{Conditional Entropy and Free Energy}

As mentioned above, we are interested in understanding
the properties of the (random) distribution $\mu_{\Code,y}(\,\cdot\,)$.
One possible way of formalizing this idea is to consider
the entropy of this distribution.

Let us recall that the (Shannon) entropy of a discrete random variable
$X$ (or, equivalently, of its distribution) quantifies, in a very precise 
sense, the `uncertainty' associated with $X$.\footnote{For a very readable account
of information theory we recommend \cite{CoT91}.} It is given by
\begin{eqnarray}
H(X) = -\sum_{x}\prob(x)\log\prob(x)\, .
\end{eqnarray}
For two random variables $X$ and $Y$ one defines the conditional entropy of
$X$ given $Y$ as
\begin{eqnarray}
\label{equ:conditionalentropy}
H(X|Y) & = &-\sum_{x,y}\prob(x,y)\log \prob(x|y) =  
\E_y \left\{-\sum_{x}\prob(x|Y)\log \prob(x|Y)\right\}\, .
\end{eqnarray}
This quantifies the remaining uncertainty about $X$ when $Y$ is observed.

Considering now the coding problem. Denote by $\uX$ the
(uniformly random) transmitted codeword and by $\uY$ the channel output.
The right-most expression in Eq.~(\ref{equ:conditionalentropy}) states that
$H(\uX|\uY)$ is the expectation 
of the entropy of the conditional distribution $\mu_{\Code,y}(\,\cdot\,)$ with respect to $\uy$.

Let us denote by $\nu_{\Code}(x)$ the probability that a uniformly random
codeword in $\Code$ takes the value $x$ at the $i$-th position,
averaged over $i$. Then a straightforward 
calculation yields 
\begin{eqnarray}
H(\uX|\uY) & = & -\frac{1}{|\Code|}\sum_{\ux,\uy}\prod_{i=1}^N
Q(y_i|x_i)\, \log\left\{\frac{1}{Z(\Code,\uy)}\prod_{i=1}^NQ(y_i|x_i)\right\}\, ,\\
& = & -N\sum_{x,y} \nu_{\Code}(x) Q(y|x)\log Q(y|x)+ 
\E_{\uy}\log Z(\Code,\uy)\, .\label{eq:EntropyVsPartFun}
\end{eqnarray}
The `type' $\nu_{\Code}(x)$ is usually a fairly straightforward 
characteristic of the code. For most of the examples considered 
below we can take $\nu_{\Code}(\0t) =\nu_{\Code}(\1t)=1/2$.
As a consequence the first of the terms above is trivial to compute
(it requires summing over $2|\cA|$ terms). 

On the other hand the second term is highly non-trivial. The reader
will recognize the expectation of a free energy, with $\uy$ playing the
role of a quenched random variable.

The conditional entropy $H(\uX|\uY)$ provides an answer to the
question: how many codewords is $\mu_{\Code,y}(\,\cdot\,)$ spread over?
It turns out that about $e^{H(\uX|\uY)}$ of them carry most of the weight.
%
%
\subsection{Shannon Theorem and Random Coding}
\label{sec:shannontheorem}

As mentioned above, there exists an obvious tradeoff between high
rate and low error probability. 
In his celebrated 1948 paper \cite{Sha48}, Shannon derived the optimal 
error probability-vs-rate curve in the limit of large blocklengths.
In particular, he proved that if the rate is larger than a particular 
threshold, then the error probability can be made arbitrarily small.
The threshold depends on the channel and 
it is called the channel \emph{capacity}.
The capacity of a  BMS channel (measured in bits per channel use) 
is given by the following elementary expression,
\begin{align*}
\capac(Q) = &  H(X)-H(X|Y) \\
          = & 1+ \sum_{y}Q(y|\0t)\log_2 
\left\{\frac{Q(y|\0t)}{Q(y|\0t)+Q(y|\1t)}\right\} \, .
\end{align*}
For instance, the capacity of a $\BSC(p)$ is $\capac(p) = 1-\entro_2(p)$,
(where $\entro_2(p) = -p\log_2 p-(1-p)\log_2(1-p)$ is the entropy of 
a Bernoulli random variable of parameter $p$)
while the capacity of a $\BEC(\epsilon)$ is $\capac(\epsilon)=1-\epsilon$.
As an illustration, the capacity of a $\BSC(p)$ with flip probability
$p\approx 0.110028$ is $\capac(p) = 1/2$: such a channel can be used to transmit
reliably $1/2$ bit of information per channel use.

\begin{thm}[Channel Coding Theorem]
For any BMS channel with transition probability $Q$
and $R<\capac(Q)$ there exists a sequence of codes $\Code_N$
of increasing blocklength $N$ and rate $R_N\to R$ 
whose block error probability $\block^{(N)}\to 0$ as $N\to\infty$.

Vice versa, for any $R>\capac(Q)$ the block error probability of a 
code with rate at least $R$ is bounded away from $0$.
\end{thm}
The prove of the first part (`achievability') is one of the first examples of the so-called `probabilistic method'. In order to prove that there
exists an object with a certain property
(a code with small error probability), one constructs a probability 
distribution over all potential candidates (all codes of a certain 
blocklength and rate) and shows that a random element has the desired
property with non-vanishing probability. The power of this approach
is in the (meta-mathematical) observation that random constructions are often
much easier to produce than explicit, deterministic ones.

\begin{figure}[htp]
\center{\includegraphics[angle=0,width=0.4\columnwidth]{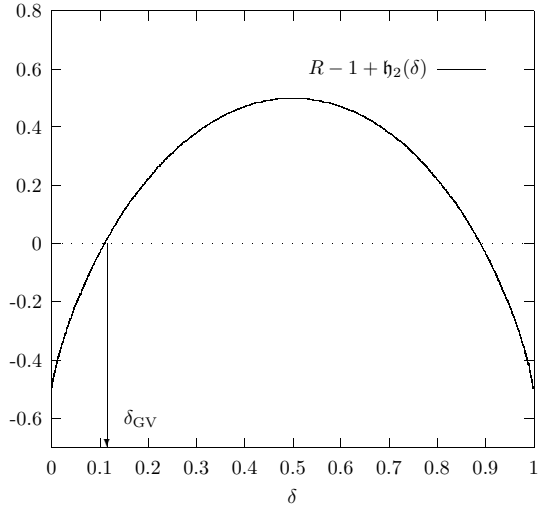}}
\caption{Exponential growth rate for the expected
distance enumerator $\E\,\cN_{\ux^{(0)}}(n\delta)$ within the random code
ensemble.}
\label{fig:RCEExponent}
\end{figure}
The distribution over codes proposed by Shannon is usually
referred to as the \emph{random code} (or, \emph{Shannon}) \emph{ensemble},
and is particularly simple. One picks a code $\Code$
uniformly at random among all codes of blocklength $N$ and rate $R$.
More explicitly, one picks $2^{NR}$ codewords 
as uniformly random points in the hypercube $\{\0t,\1t\}^N$.
This means that each codeword is a string of 
$N$ fair coins $\ux^{(\alpha)}=(x^{(\alpha)}_1,\dots,x^{(\alpha)}_N)$
for\footnote{The reader might notice two imprecisions
with this definition. First, $2^{NR}$ is not necessarily
an integer: one should rather use $\lceil 2^{NR}\rceil$
codewords, but the difference is obviously negligible. 
Second, in contradiction with our definition, two codewords
may coincide if they are independent. Again, only an exponentially 
small fraction of codewords will coincide and they can be 
neglected for all practical purposes.} $\alpha=1,\dots, 2^{NR}$.

Once the ensemble is defined, one can estimate its average block error
probability and show that it vanishes in the blocklength for $R<\capac(Q)$.
Here we will limit ourselves to providing some basic 
`geometric' intuition of why a 
random code from the Shannon ensemble performs well 
with high probability.\footnote{Here and in the rest of the lectures,
the expression \emph{with high probability} means
`with probability approaching one as $N\to\infty$'}

Let us consider a particular codeword, say $\ux^{(0)}$, and 
try to estimate the distance (from $\ux^{(0)}$) at which
other codewords in $\Code$ can be found. This information is conveyed by the 
\emph{distance enumerator}
\begin{eqnarray}
\cN_{\ux^{(0)}}(d) \equiv \#\left\{\; \ux\in\Code\backslash \ux^{(0)} 
\; \mbox{ such that }
\; d(\ux,\ux^{(0)})=d\; \right\}\, ,
\end{eqnarray}
where $d(\ux,\ux')$ is the Hamming distance between 
$\ux$ and $\ux'$ (i.e., the number of positions in which $\ux$ and $\ux'$ differ).
The expectation of this quantity is the number of
codewords different from $\ux^{(0)}$ (that is $(2^{NR}-1)$) 
times the probability that any given codeword has distance $d$ from
$\ux^{(0)}$. Since each entry is independent and different with probability
$1/2$, we get
\begin{eqnarray}
\E\,\cN_{\ux^{(0)}}(d) = (2^{NR}-1)\,\;\frac{1}{2^N}\binom{N}{d} \doteq 2^{N[R-1+\entro_2(\delta)]}\, , 
\end{eqnarray}
where $\delta= d/N$
and  $\doteq$ denotes equality to the leading exponential 
order.\footnote{Explicitly, we  write $f_N\doteq g_N$ if 
$\frac{1}{N}\log f_N/g_N\to 0$.}

The exponent $R-1+\entro_2(\delta)$ is plotted in Fig.~\ref{fig:RCEExponent}.
For $\delta$ sufficiently small (and $R<1$) this exponent is negative. Its first zero,
to be denoted as $\delta_{\rm GV}(R)$, is called the Gilbert-Varshamov distance.
For any $\delta<\delta_{\rm GV}(R)$ the expected number of codewords of distance
at most $N\delta$ from $\ux^{(0)}$ is exponentially small in $N$.
It follows that the probability to find \emph{any codeword}
at distance smaller than $N\delta$ is exponentially small in $N$.

Vice-versa, for $d = N\delta$, with $\delta>\delta_{\rm GV}(R)$,
$\E\,\cN_{\ux^{(0)}}(d) $ is exponentially large in $N$. 
Indeed, $\cN_{\ux^{(0)}}(d) $ is a binomial random variable,
because each of the $2^{NR}-1$ codewords is at distance $d$ 
independently and with the same probability. As a consequence, $\cN_{\ux^{(0)}}(d) $ 
is exponentially large as well with high probability.

\begin{figure}[htp]
\center{\includegraphics[angle=0,width=0.35\columnwidth]{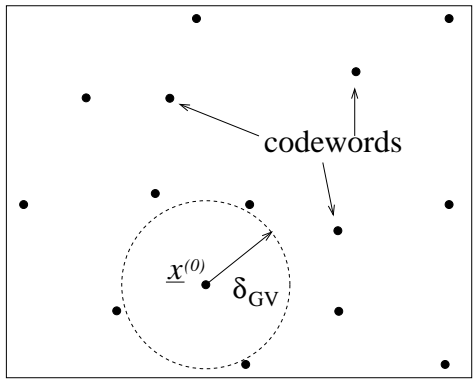}}
\caption{Pictorial description of a typical code from the random code ensemble.}
\label{fig:RCEHamming}
\end{figure}

The bottom line of this discussion is that, for any given codeword 
$\ux^{(0)}$ in $\Code$, the closest other codeword is, with high
probability, at distance $N(\delta_{\rm GV}(R)\pm\ve)$.
A sketch of this situation is provided in Fig.~\ref{fig:RCEHamming}.

Let us assume that the codeword $\ux^{(0)}$ is transmitted through a
$\BSC(p)$. Denote by $\uy\in\{\0t,\1t\}^N$ the channel output.
By the law of large numbers $d(\ux,\uy)\approx Np$ with high probability.
The receiver tries to reconstruct the transmitted codeword from $\uy$
using word MAP decoding. Using Eq.~(\ref{eq:GeneralEnergy}), 
we see that the `energy' 
of a codeword $\ux^{(\alpha)}$ (or, in more conventional terms, its 
log-likelihood) is given by
\begin{eqnarray}
E(\ux^{(\alpha)}) &=&-\sum_{i=1}^N\log Q(y_i|x_i) 
= 
-\sum_{i=1}^N\left\{\ind(y_i=x^{(\alpha)}_i)\log(1-p)+
\ind(y_i\neq x^{(\alpha)}_i)\log p\right\} \\
&=& NA(p)+ 2 B(p) d(\ux^{(\alpha)},\uy)\, ,
\end{eqnarray}
where $A(p) \equiv -\log p$ and $B(p) \equiv\frac{1}{2}\log(1-p)/p$.
For $p<1/2$, $B(p)>0$ and word MAP decoding amounts to finding the codeword
$\ux^{(\alpha)}$  which is closest in Hamming distance to the
channel output $\uy$. By the triangle inequality,
the distance between $\uy$ and any of the `incorrect' codewords
is $\gtrsim N(\delta_{\rm GV}(R)-p)$. For $p<\delta_{\rm GV}(R)/2$
this is with high probability larger than the distance from $\ux^{(0)}$.

The above argument implies that, for $p<\delta_{\rm GV}(R)/2$,
the expected block error rate of a random code from Shannon's ensemble 
vanishes as $N\to\infty$. Notice that the channel coding theorem promises 
instead vanishing error probability whenever $R<1-\entro_2(p)$,
that is (for $p<1/2$) $p<\delta_{\rm GV}(R)$.
The factor 2 of discrepancy can be recovered through a more careful
argument. 

Without entering into details, it is interesting to understand the
basic reason for the discrepancy between the Shannon Theorem and
the above argument. This is related to the geometry of high dimensional spaces.
Let us assume for simplicity that the minimum distance between 
\emph{any two} codewords in $\Code$ is at least 
$N(\delta_{\rm GV}(R)-\ve)$. In a given random code, this is the case
for most codeword pairs. We can then
eliminate the pairs that do not satisfy this constraint, thus modifying
the code rate in a negligible way (this procedure is called 
\emph{expurgation}).  The resulting code will have 
\emph{minimum distance} (the minimum distance among any two codewords 
in $\Code$)  $d(\Code) \approx N\delta_{\rm GV}(R)$.

Imagine that we use such a code to communicate through a BSC
and that exactly $n$ bits are flipped.
By the triangular inequality, as long as $n<d(\Code)/2$, the word MAP 
decoder will recover the transmitted 
message for \emph{all} error patterns.
If on the other hand $n>d(\Code)/2$, there are error patterns involving
$n$ bits such that the word-MAP decoder does not return the 
transmitted codeword. If for instance there exists a single
codeword $\ux^{(1)}$ at distance $d(\Code) = 2n-1$ from $\ux^{(0)}$,
any pattern involving $n$ out of the $2n-1$ such that
$x_i^{(0)}\neq x_i^{(1)}$, will induce a decoding error.
However, it might well be that \emph{most} error patterns with
the same number of errors can be corrected. 

Shannon's Theorem points out that this is indeed the case
until the number of bits flipped by the channel is roughly equal to the
minimum distance $d(\Code)$.
%
%
\section{Sparse Graph Codes}
\label{sec:sparsegraphcode}

Shannon's Theorem provides a randomized construction to find a code
with `essentially optimal' rate vs error probability tradeoff.
In practice, however, one cannot use random codes for communications.
Just storing the code $\Code$ requires a memory which grows exponentially in the blocklength.
In the same vein the optimal decoding procedure requires an exponentially increasing effort.
On the other hand, we can not use very short codes since their performance is not very good. To see this
assume that we transmit over the BSC with parameter $p$. 
If the blocklength is $N$ then the standard deviation
of the number of errors contained in a block is $\sqrt{N p (1-p)}$. Unless this quantity is very small
compared to $N p$ we have to either over-provision 
the error correcting capability of the code so as to 
deal with the occasionally large number of errors, waisting transmission rate most of the time, or we
dimension the code for the typical case, but then we will not be able to decode when the number of errors
is larger than the average. This means that short codes are either inefficient or unreliable (or both). 

The general strategy for tackling this problem is to
introduce more structure in the code definition, and to hope that such 
structure can be exploited for the encoding and the decoding.
In the next section we shall describe a way of introducing structure
that, while preserving Shannon's idea of random codes, opens the way to
efficient encoding/decoding.

There are two main ingredients that make {\em modern coding} work and
the two are tightly connected. 
The first important ingredient is to use codes which can be described by
{\em local} constraints only. 
The second ingredient is to use a local algorithm instead of
an high complexity global one (namely symbol MAP or word MAP
decoding). In this section we describe the first component. 
%
%
\subsection{Linear Codes}

One of the simplest forms of structure consists in requiring $\Code$
to be a linear subspace of $\{\0t,\1t\}^N$. One speaks then of a 
\emph{linear code}. For specifying such a code it is not necessary
to list all the codewords. In fact, any linear space can be seen
as the kernel of a matrix:
\begin{eqnarray}
\Code = \left\{\ux\in\{\0t,\1t\}^N:\; \H\ux = \u0t\right\}\, ,
\label{eq:NullSpace}
\end{eqnarray}
where the matrix vector multiplication is assumed to be performed modulo $2$.
The matrix $\H$ is called the \emph{parity-check matrix}. It has $N$ columns
and we let $M<N$ denote its number of rows. 
Without loss of generality we can assume $\H$ to have maximum rank $M$.
As a consequence, $\Code$ is a linear space of dimension $N-M$. The rate of $\Code$ 
is
\begin{eqnarray}
R=1-\frac{M}{N}\, .
\end{eqnarray}
The $a$-th line in $\H\ux = \u0t$ has the form
(here and below $\oplus$ denotes modulo $2$ addition)
\begin{eqnarray}
x_{i_1(a)}\oplus \cdots\oplus x_{i_k(a)} = \0t.
\end{eqnarray}
It is called a \emph{parity check}.
\begin{figure}[htp]
\center{\includegraphics[angle=0,width=0.3\columnwidth]{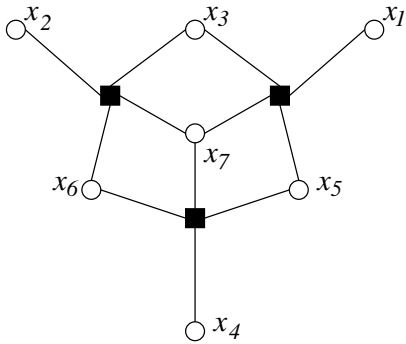}}
\caption{Factor graph for the parity-check matrix (\ref{eq:ParityCheck}).}
\label{fig:FactorHamming}
\end{figure}

The parity-check matrix is conveniently represented through a 
\emph{factor graph} (also called \emph{Tanner graph}).
This is a bipartite graph including two types of nodes: $M$ function nodes
(corresponding to the rows of $\H$, or the parity-check equations)
and $N$ variable nodes (for the columns of $\H$, or the variables).  
Edges are drawn whenever the corresponding entry in $\H$ is non-vanishing.
\begin{example}{4}
In Fig.~\ref{fig:FactorHamming} we draw the factor graph corresponding 
to the parity-check matrix (here $N=7$, $M = 3$)
\begin{eqnarray}
\H = \left[
\begin{array}{ccccccc}
\1t & \0t & \1t & \0t & \1t & \0t & \1t\\
\0t & \1t & \1t & \0t & \0t & \1t & \1t\\
\0t & \0t & \0t & \1t & \1t & \1t & \1t
\end{array}
\right]\,.\label{eq:ParityCheck}
\end{eqnarray}
\end{example}
In the following we shall use indices $i,j,\dots$ for variable nodes and 
$a,b,\dots$ for check nodes. We shall further denote
by $\di$ (respectively, $\da$) the set of nodes that are adjacent to 
variable node $i$ (to factor node $a$).

Remarkably, introducing the linear space structure does not 
deteriorate the performances of the resulting code. 
Let us introduce Shannon's \emph{parity-check ensemble}:
it is defined by letting the parity-check matrix $\H$ be a uniformly 
random matrix with the prescribed dimensions. Explicitly,
each of the $NM$ entries $H_{ai}$ is an independent Bernoulli
random variable of mean $1/2$.
Probabilistic arguments similar to the ones for the random code
ensemble can be developed for the random parity-check ensemble.
The conclusion is that random codes from this ensemble  allow
to communicate with arbitrarily small block error probability
at any rate $R<\capac(Q)$, where $\capac(Q)$ is the capacity of the given BMS channel.

Unfortunately, linearity is not sufficient to guarantee that a code admits a low-complexity
decoding algorithm. In particular, the algorithm which we discuss in the sequel works 
well only for codes that can be represented by a sparse parity-check 
matrix $\H$ (i.e. a parity check matrix with $O(N)$ non-vanishing entries). 
Notice that a given code $\Code$ has more than one representation 
of the form (\ref{eq:NullSpace}).
A priori one could hope that, given a uniformly random 
matrix $\H$, a new matrix $\H'$ could be built such that
$\H'$ is sparse and that its null space coincides with the one of $\H$.
This would provide a sparse representation of $\Code$.
Unfortunately, this is the case only for a vanishing fraction of 
matrices $\H$, as shown by the exercise below.
\begin{exercise}{2}
Consider a linear code $\Code$, with blocklength $N$, and dimension 
$N-M$ (as a linear space).
Prove the following sequence of arguments.
\begin{itemize}
\item[$(i)$] The total number of binary $N\times M$ 
parity-check matrices is $2^{NM}$.
\item[$(ii)$]  Each code $\Code$
has $2^{\binom{M}{2}} \prod_{i=1}^{M} \left(2^i-1 \right)$ distinct
$N \times M$ parity-check matrices $\H$. 
\item[$(iii)$] 
The number of such matrices with at most $a N$ non-zero entries is $\sum_{i=0}^{a N} \binom{NM}{i} \leq 2^{NM \entro_2(a/(N-M))}$.
\item[$(iv)$] Conclude from the above that, for any given $a$, the fraction of
parity-check matrices $\H$ that admit a sparse representation in terms 
of a matrix $\H'$ with at most $aN$ ones, is of order $e^{-N\gamma}$
for some $\gamma>0$. 
\end{itemize}
\end{exercise}
With an abuse of language in
the following we shall sometimes use the term `code' to denote
a pair code/parity-check matrix. 
 
%
%
\subsection{Low-Density Parity-Check Codes}

Further structure can be introduced by restricting the ensemble of
parity-check matrices.
Low-density parity-check (LDPC) codes are codes that have at least one sparse parity-check matrix.

Rather than considering the most general case let us limit ourselves to a particularly
simple family of LDPC ensembles, originally introduced by Robert
Gallager \cite{Gal62}. We call them `regular' ensembles. 
An element in this family is characterized by the blocklength
$N$ and two integer numbers $k$ and $l$, with $k>l$. 
We shall therefore refer to it as the $(k,l)$ regular ensemble). In order to 
construct a random Tanner graph from this ensemble, 
one proceeds as follows:
\begin{enumerate}
\item Draw $N$ variable nodes, each attached to $l$ half-edges and $M=Nl/k$
(we neglect here the possibility of $Nl/k$ not being an integer)
check nodes, each with $k$ half edges. 
\item Use an arbitrary convention to label the half edges form $1$ to $Nl$,
both on the variable node side as well as the check node side (note that this requires
that $Mk=Nl$).
\item Choose  a permutation $\pi$ uniformly at random among all permutations
over $Nl$ objects, and connect half edges accordingly.
\end{enumerate}
Notice that the above procedure may give rise to multiple edges. 
Typically there will be $O(1)$ multiple edges in a graph constructed as 
described. These can be eliminated easily without effecting the performance substantially. From the analytical
point of view, a simple choice consists in eliminating all the edges $(i,a)$
if $(i,a)$ occurs an even number of times, and replacing them by
a single occurrence $(i,a)$ if it occurs an odd number of times.

\begin{figure}[tp]
\centering
\setlength{\unitlength}{0.9bp}%
\begin{picture}(300,100)(0,-15)
\put(0,0){\includegraphics[scale=0.9]{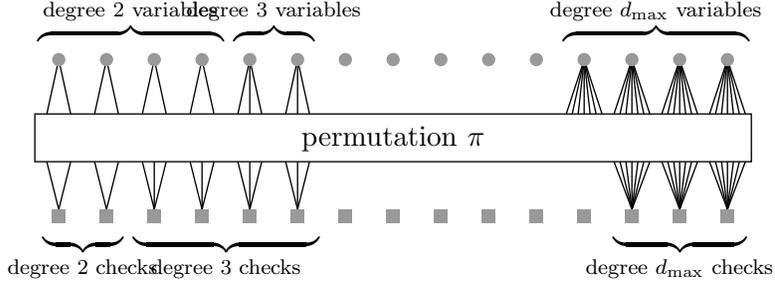}}
\put(20,0){\makebox(0,0)[t]{$\underbrace{\phantom{\hspace{25pts}}}_{\text{degree 2 checks}}$}}
\put(80,0){\makebox(0,0)[t]{$\underbrace{\phantom{\hspace{65pts}}}_{\text{degree 3 checks}}$}}
\put(270,0){\makebox(0,0)[t]{$\underbrace{\phantom{\hspace{45pts}}}_{\text{degree $d_{\text{max}}$ checks}}$}}
\put(40,70){\makebox(0,0)[b]{$\overbrace{\phantom{\hspace{65pts}}}^{\text{degree $2$ variables}}$}}
\put(100,70){\makebox(0,0)[b]{$\overbrace{\phantom{\hspace{25pts}}}^{\text{degree $3$ variables}}$}}
\put(260,70){\makebox(0,0)[b]{$\overbrace{\phantom{\hspace{65pts}}}^{\text{degree $d_{\text{max}}$ variables}}$}}
\put(150,35){\makebox(0,0)[c]{permutation $\pi$}}
\end{picture}
\caption{Factor graph of an irregular LDPC code. Variable nodes
and function nodes can have any degree between $2$ and $d_{\rm max}$.
Half edges on the two sides are joined through a uniformly random permutation.}
\label{fig:multiedgeldpc}
\end{figure} 
Neglecting multiple occurrences (and the way to resolve them), the
parity-check matrix corresponding to the graph constructed in this way
does include $k$ ones per row and $l$ ones per column. 
In the sequel we will keep $l$ and $k$ fixed and consider the behavior of
the ensemble as $N\to\infty$.
This implies that the matrix has only $O(N)$ non-vanishing entries. The matrix
is {\em sparse}.

For practical purposes it is important to maximize the rate at which 
such codes enable one to communicate with vanishing error probability.
To achieve this goal, several more complex ensembles have been introduced. 
As an example, one simple idea is to consider a generic row/column 
weight distribution (the weight being the number of non-zero elements),
cf. Fig.~\ref{fig:multiedgeldpc} for an illustration.
Such ensembles are usually referred to as `irregular', and were introduced 
in \cite{LMSSS97}.
%
%
\subsection{Weight Enumerator}
\label{sec:we}

As we saw in Section~\ref{sec:shannontheorem},
the reason of the good performance of Shannon ensemble
(having vanishing block error probability at rates arbitrarily close to the capacity),
can be traced back to its minimum distance properties. This is indeed only a partial explanation
(as we saw errors could be corrected well beyond half 
its minimum distance). It is nevertheless instructive and useful to understand
the geometrical structure (and in particular the minimum distance properties) 
of typical codes from the LDPC ensembles defined above.

Let us start by noticing that, for linear codes, the distance enumerator
does not depend upon the reference codeword. 
This is a straightforward consequence of the observation that, for any 
$\ux^{(0)}\in\Code$ the set $\ux^{(0)}\oplus\Code \equiv
\{\ux^{(0)}\oplus\ux\, :\, \ux\in\Code\}$ coincides with $\Code$.
We are therefore led to consider the distance enumerator with respect to
the all-zero codeword $\u0t$. This is also referred to as the 
\emph{weight enumerator},
\begin{eqnarray}
\cN(w) = \#\left\{\,\ux\in\Code\, :\;\; w(\ux) = w\,\right\}\, ,
\end{eqnarray}
where $w(\ux) = d(\ux,\u0t)$ is the number of non-zero entries in $\ux$.

Let us compute the expected weight enumerator $\ecN(w) \equiv
\E\cN(w)$. The final result is
\begin{eqnarray}
\ecN(w) = \frac{(lw)!(F-lw)!}{F!}\binom{N}{w}\, \coeff[q_k(z)^M,z^{lw}]\, .
\label{eq:ExpectedWeightEnumerator}
\end{eqnarray}
Here, $F=Nl=Mk$ denotes the number of edges in the Tanner graph,
$q_k(z) \equiv\frac{1}{2}[(1+z)^k+(1-z)^k]$, and, given a polynomial
$p(z)$ and an integer $n$, $\coeff[p(z),z^n]$ denotes the coefficient of 
$z^n$ in the polynomial $p(z)$.

We shall now prove Eq.~(\ref{eq:ExpectedWeightEnumerator}).
Let $\ux\in\{\0t,\1t\}^N$ be a binary word of length $N$ and weight $w$.
Notice that $\H\ux = 0$ if and only if the corresponding factor graph
has the following property. Consider all variable nodes $i$ such that
$x_i=\1t$, and color in red all edges incident on these nodes. 
Color in blue all
the other edges. Then all the check nodes must have an even number of incident
red edges. A little thought shows that $\ecN(w)$ is the number of `colored'
factor graphs having this property, divided by the total number of factor
graphs in the ensemble.

A valid colored graph must have $w l$ red edges. It can be
constructed as follows. First choose $w$ variable nodes. This can be done in
$ \binom{N}{w}$ ways. Assign to each node in this set $l$ red sockets, and to
each node outside the set $l$ blue sockets. Then, for each of the $M$
function nodes, color in red an even subset of its sockets in such a way that
the total number of red sockets is $E= w l$. 
The number of ways of doing this is\footnote{This is a standard generating function calculation,
and is explained in Appendix \ref{app:GenFun}.} $\coeff[q_k(z)^M,z^{lw}]$.
Finally we join the variable node and check node sockets in such a way that
colors are matched. There are $(l w)!(F-l w)!$ such matchings out of the
total number of $F!$ corresponding to different elements in the ensemble.

\begin{figure}
\begin{center}
\includegraphics[angle=0,width=0.5\columnwidth]{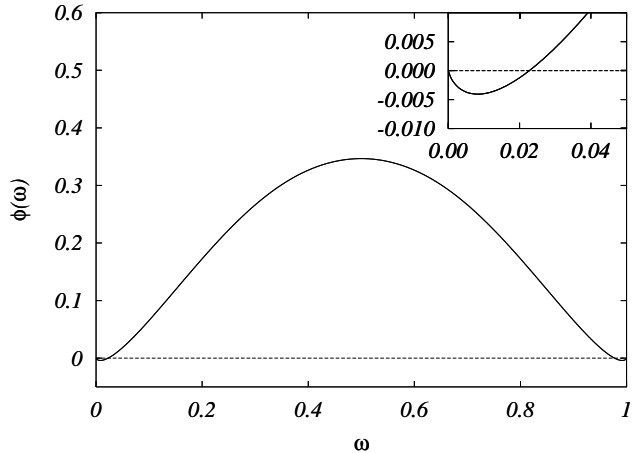}
\caption{Logarithm of the expected weight enumerator for the 
$(3,6)$ ensemble in the large blocklength limit. Inset: small weight region.
Notice that $\phi(\omega)<0$ for $\omega<\omega_* \approx 0.02$:
besides the `all-zero' word there is no codeword of weight smaller than $N\omega_*$
in the code with high probability.}
\label{fig:WE36}
\end{center}
\end{figure}
Let us compute 
the exponential growth rate $\phi(\omega)$ of $\ecN (w)$.
This is defined by
\begin{equation}
\ecN (w = N\omega) \doteq e^{N \phi(\omega)}\ .
\label{eq:weightphidef}
\end{equation}
In order to estimate the leading exponential behavior 
of Eq.~(\ref{eq:ExpectedWeightEnumerator}), we set
$w=N\omega$
and estimate the $\coeff[\dots,\dots]$ term 
using the Cauchy Theorem,
\begin{eqnarray}
\coeff\left[q_k(z)^{M},z^{wl}\right] =
\oint \frac{q_k(z)^{M}}{z^{lw+1}}
\, 
\frac{\de z}{2\pi i} = \oint 
\exp\left\{N[\left[\frac{l}{k}\log q_k(z)-l\omega\log z\right]\right\}\, 
\frac{\de z}{2\pi i} \, .
\label{eq:Cauchy}
\end{eqnarray}
Here the integral runs over any path encircling the origin in the complex $z$ 
plane.
Evaluating the integral using the saddle point method we finally get
$\ecN(w)\doteq e^{N\phi}$, where
\begin{eqnarray}
\phi(\omega) \equiv (1-l)\entro(\omega) + \frac{l}{k}\log q_k(z)-\omega
l\log z\, ,
\end{eqnarray}
and $z$ is a solution of the saddle point equation
\begin{eqnarray}
\omega = \frac{z}{k}\, \frac{q'_k(z)}{q_k(z)}\, .
%
\end{eqnarray}
The typical result of such a computation is shown in Fig.~\ref{fig:WE36}.
As can be seen, there exists $\omega_*>0$ such that
$\phi(\omega)<0$ for $\omega\in(0,\omega_*)$. This implies that a typical code
from this ensemble will not have any codeword of weight between $0$
and $N(\omega_*-\varepsilon)$.
By linearity the minimum distance of the code is at
least $\approx N\omega_*$. This implies in particular that such codes
can correct any error pattern over the binary symmetric channel of weight 
$\omega_*/2$ or less.

Notice that $\phi(\omega)$ is an `annealed average', in the terminology of
disordered systems. As such, it can be dominated by rare
instances in the ensemble. On the other hand, since
$\log\cN_N(N\omega) = \Theta(N)$ is  an `extensive' quantity,
we expect it to be 
\emph{self averaging} in the language of statistical physics. 
In mathematics terms one says that it should 
\emph{concentrate in probability}. Formally, this means that there
exists a function $\Phi_N(\omega)$ that is non-random (i.e.,
does not depend upon the code) and such that
\begin{eqnarray}
\lim_{N\to\infty}
\prob\left\{\left|\log\cN_N(N\omega)-\Phi_N(\omega)\right|\ge N\delta\right\}
=0\, .
\end{eqnarray}
Further we expect that $\Phi_N(\omega) = N\phi_{\rm q}(\omega)+o(N)$
as $N\to\infty$.
Despite being rather fundamental, both these statements are open conjectures.

The coefficient $\phi_{\rm q}(\omega)$ is the growth rate of the weight 
enumerator for typical codes in the ensembles. In statistical
mechanics terms, it is a `quenched' free energy (or rather, entropy).
By Jensen inequality $\phi_{\rm q}(\omega)\le\phi(\omega)$.
A statistical mechanics calculation reveals that the inequality is strict for
general (irregular) ensembles. On the other hand, for regular
ensembles as the ones considered here, 
$\phi_{\rm q}(\omega) = \phi(\omega)$: the annealed calculation
yields the correct exponential rate.
This claim has been supported rigorously by the results of 
\cite{Rat06,BaB05,Mac05}.

Let us finally comment on the relation between distance enumerator and 
the Franz-Parisi potential \cite{FraPar_pot}, 
introduced in the study of glassy systems.
In this context the potential is used to probe the structure of 
the Boltzmann measure. 
One considers a system with energy function $E(x)$,
a reference configuration $x_0$ and some notion of distance 
between configurations $d(x,x')$.  The constrained partition function
is then defined as
\begin{eqnarray}
Z(x_0,w) = \int  e^{-E(x)}\, \delta(d(x_0,x)-w)\; \de x\, .
\end{eqnarray}
One then defines the potential $\Phi_N(\omega)$ as the typical value of
$\log Z(x_0,w)$ when $x_0$ is a random configuration 
with the same Boltzmann distribution and $w=N\omega$.
Self averaging is expected to hold here too:
\begin{eqnarray}
\lim_{N\to\infty}\prob_{x_0}\left\{\,\left|\log Z(x_0,N\omega)
-\Phi_N(\omega)\right|
\ge N\delta\,\right\}=0 \, .
\end{eqnarray}
Here $N$ may denote the number of particles or the volume of the system 
and $\prob_{x_0}\left\{\,\cdots\,\right\}$ indicates probability 
with respect to $x_0$ distributed with the Boltzmann measure for
the energy function $E(x_0)$. 

It is clear that the two ideas are strictly related and 
can be generalized to any joint distribution of $N$
variables $(x_1,\dots,x_N)$. In both cases the structure 
of such a distribution is probed by picking a
reference configuration and restricting the measure to its neighborhood.

To be more specific, the weight enumerator can be seen as a special case
of the Franz-Parisi potential. 
It is sufficient to take as Boltzmann distribution the uniform measure over 
codewords of a linear code $\Code$. In other words,
let the configurations be binary strings of length $N$,
and set $E(\ux) = 0$ if $\ux\in\Code$, and $=\infty$ otherwise.
Then the restricted partition function is just the
distance enumerator with respect to the reference codeword,
which indeed does not depend on it.
%
%
\section{The Decoding Problem for Sparse Graph Codes}
\label{sec:decodingproblem}

As we have already seen, MAP decoding requires  computing 
either marginals or the mode of the conditional distribution of
$\ux$ being the channel input given output $\uy$.
In the case of LDPC codes the posterior probability distribution factorizes
according to underlying factor graph $G$:
\begin{eqnarray}
\mu_{\Code,y} (\ux) = \frac{1}{Z(\Code,y)}\,
\prod_{i=1}^{N}Q(y_i|x_i)\, \prod_{a=1}^M\ind(x_{i_1(a)}\oplus 
\cdots\oplus x_{i_k(a)}=\0t)\, .\label{eq:ProbLDPC}
\end{eqnarray}
Here $(i_1(a),\dots,i_k(a))$ denotes the set of variable indices involved in
the $a$-th parity check (i.e., the non-zero entries in the $a$-th row
of the parity-check matrix $\H$). 
In the language of spin models, the terms $Q(y_i|x_i)$ correspond
to an external random field. The factors 
$\ind(x_{i_1(a)}\oplus\cdots\oplus x_{i_k(a)}=\0t)$ can instead be regarded 
as hard core $k$-spins interactions. Under the mapping $\sigma_i 
= (-1)^{x_i}$, such interactions depend on the spins through
the product $\sigma_{i_1(a)}\cdots \sigma_{i_k(a)}$. The
model (\ref{eq:ProbLDPC}) maps therefore onto a $k$-spin model with 
random field.

For MAP decoding, minimum distance properties of the code play 
a crucial role in determining the performances. We investigated 
such properties in the previous section.
Unfortunately, there is no known way of implementing MAP decoding efficiently. 
In this section we discuss
two decoding algorithms that exploit the sparseness of the factor
graph to achieve efficient decoding. Although such  strategies are
sub-optimal with respect to word (or symbol) MAP decoding,
the graphical structure can itself be optimized, leading to
state-of-the-art performances. 

After briefly discussing bit-flipping decoding, most of this section will 
be devoted to message passing that is the approach most used in 
practice. Remarkably, both bit flipping as well as message passing 
are closely related to statistical mechanics.
%
%
\subsection{Bit Flipping}
\label{sec:BitFlipping}

For the sake of simplicity, let us assume that communication takes
place over a binary symmetric channel. We receive
the message $\uy\in\{\0t,\1t\}^N$ and try to find the transmitted codeword 
$\ux$ as follows:
\begin{pseudocode}{Bit-flipping decoder}
\item[0.] Set $\ux(0) =\uy$.
\item[1.] Find a bit belonging to more unsatisfied than satisfied parity 
checks.
\item[2.] If such a bit exists, flip  it: $x_i(t+1)=x_i(t)\oplus \1t$.
Keep the other bits:
$x_j(t+1) = x_j(t)$ for all $j\neq i$.\\
 If there is no such bit, 
return $\ux(t)$ and halt. 
\item[3.] Repeat steps 1 and 2.
\end{pseudocode}
The bit to be flipped is usually chosen uniformly at random
among the ones satisfying the condition at step {\tt 1}. However this is 
irrelevant for the analysis below.

In order to monitor the bit-flipping algorithm, it
is useful to introduce the function:
\begin{eqnarray}
U(t) \equiv\#\left\{
\mbox{ parity-check equations not satisfied by $\ux(t)$\, }
\right\} \, .\label{eq:BitFlipEn
ergy}
\end{eqnarray}
This is a non-negative integer, and if $U(t)=0$ the algorithm
is halted and it outputs $\ux(t)$. Furthermore, $U(t)$ cannot be larger
than the number of parity checks $M$ and decreases (by at least one)
at each cycle. Therefore, the  algorithm complexity is  
$O(N)$ (this is a commonly regarded as the ultimate goal for 
many communication problems).

\begin{figure}[tp]
\begin{tabular}{cc}
\hspace{0.5cm}
\includegraphics[angle=0,width=0.42\columnwidth]{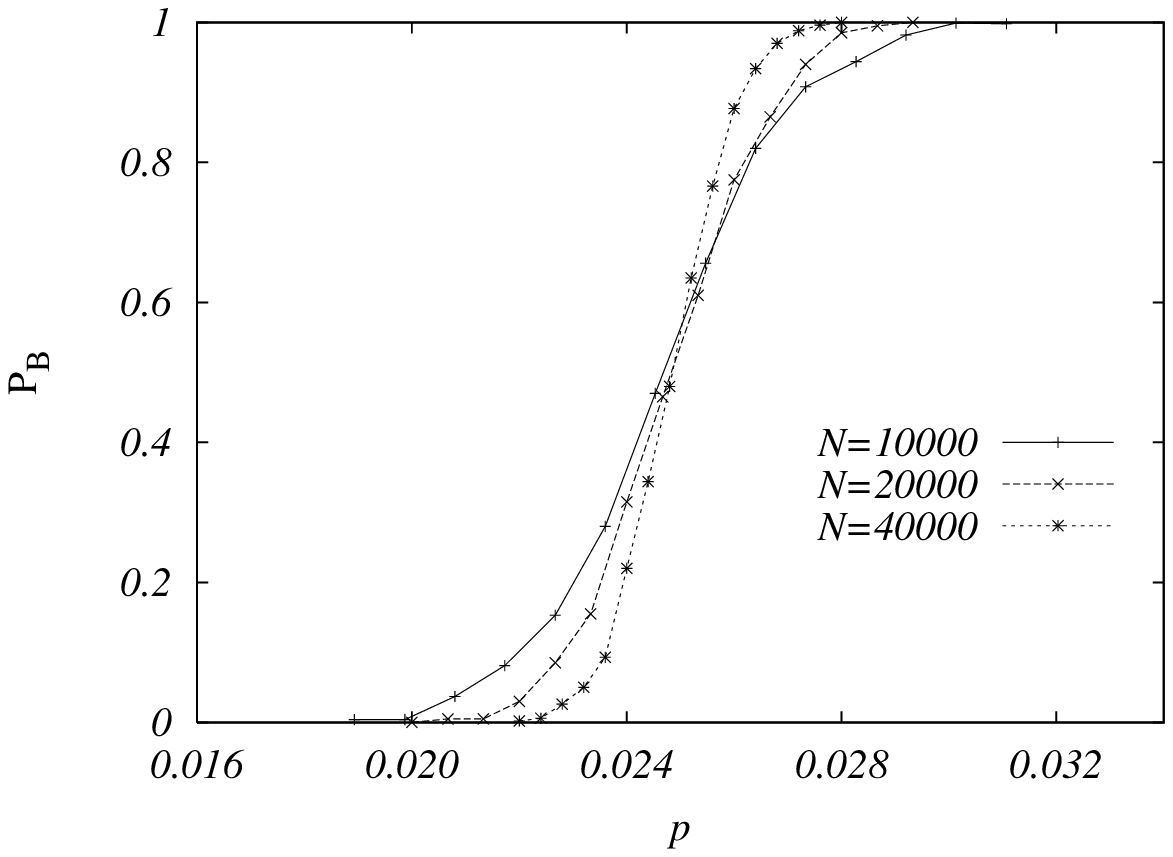}&
\hspace{0.5cm}
\includegraphics[angle=0,width=0.42\columnwidth]{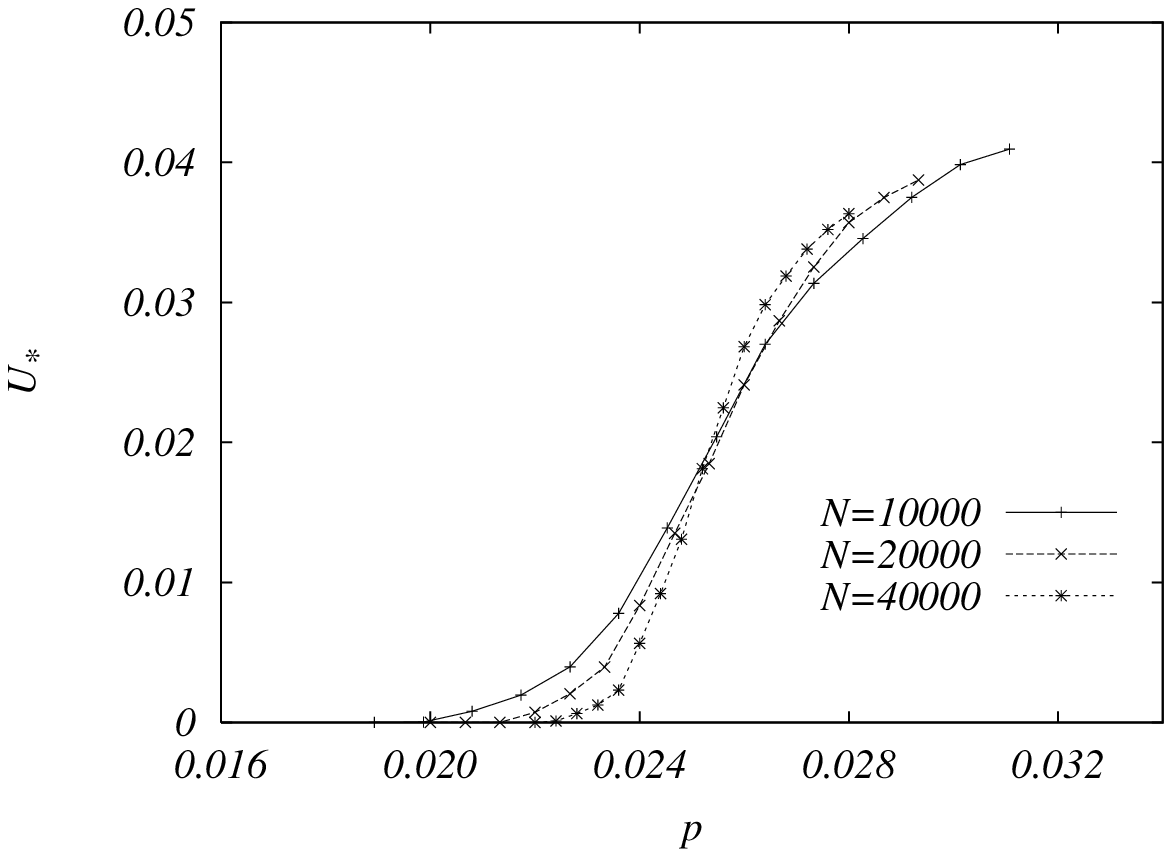}
\end{tabular}
\caption{Numerical simulations of bit-flipping decoding of 
random codes from the $(5,10)$ regular ensemble. On the left:
block error rate achieved by this scheme. On the right:
fraction of unsatisfied parity checks in the word found by the 
algorithm.}
\label{fig:Flip510}
\end{figure} 
It remains to be seen if the output of the bit-flipping algorithm 
is related to the
transmitted codeword. In Fig.~\ref{fig:Flip510} we present the results of a
numerical experiment. We considered the $(5,10)$ regular ensemble and
generated about $1000$ random code and channel realizations for each value of
the noise level $p$ in some mesh. 
Then, we applied the above algorithm and traced the
fraction of successfully decoded blocks, as well as the residual energy $U_* =
U(t_*)$, where $t_*$ is the total number of iterations of the algorithm. The
data suggests that bit-flipping is able to overcome a finite noise level: it
recovers the original message with high probability when less than about
$2.5\%$ of the bits are corrupted by the channel. Furthermore, the curves for
the block error probability $\block^{\rm bf}$ under bit-flipping decoding 
become steeper and steeper as
the system size is increased. It is natural to conjecture that
asymptotically, a phase transition takes place at a well defined noise level
$p_{\rm bf}$:  $\block^{\rm bf}\to 0$ for $p<p_{\rm bf}$ and $\block^{\rm
bf}\to 1$ for $p>p_{\rm bf}$. Numerically $p_{\rm bf} = 0.025\pm 0.005$.

This threshold can be compared with the one for 
word MAP decoding, that we will call $p_{\rm c}$: The bounds in \cite{WiS05} state that
$0.108188\le p_{\rm c}\le 0.109161$ for
the $(5,10)$ ensemble, while a statistical mechanics calculation
yields $p_{\rm c}\approx 0.1091$. Bit-flipping is significantly sub-optimal,
but it is still surprisingly good, given the extreme
simplicity of the  algorithm.

These numerical findings can be confirmed rigorously \cite{SiS96}.
\begin{thm}
Consider a regular $(l,k)$ LDPC ensemble and let $\Code$ be chosen uniformly at
random from the ensemble. If $l\ge 5$
then there exists $\ve>0$ such that, with high probability,
{\tt Bit-flipping} is able to correct any pattern of at most 
$N\ve$ errors produced by a binary symmetric channel.
\end{thm}
\begin{figure}[tp]
\vspace{-1.cm}

\center{\includegraphics[angle=0,width=0.65\columnwidth]{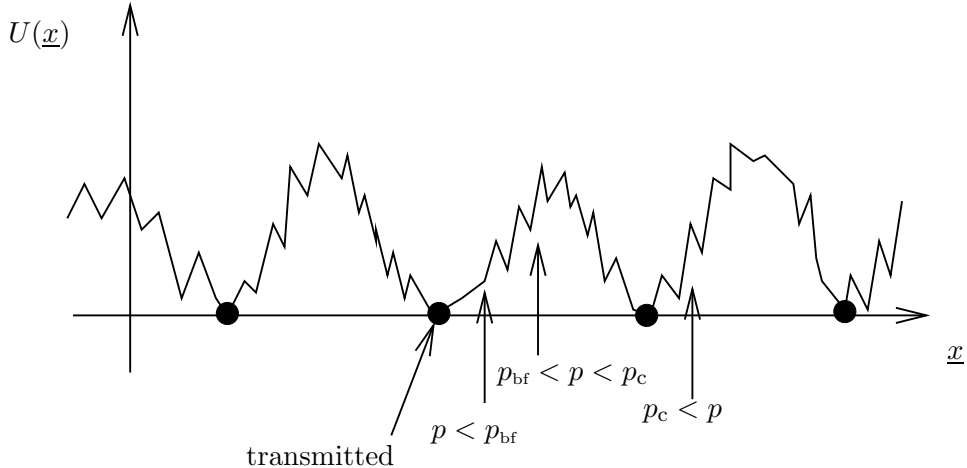}}
\put(-350,150){$U(\ux)$}
\put(5,30){$\ux$}
\put(-260,-10){transmitted}
\put(-190,0){$p<p_{\mbox{\tiny bf}}$}
\put(-165,22){$p_{\mbox{\tiny bf}}<p<p_{\rm c}$}
\put(-110,8){$p_{\rm c}<p$}
\caption{Sketch of the cost function $U(\ux)$ (number of unsatisfied 
parity checks) for a typical random LDPC code. Filled circles correspond to
codewords, and arrows to received messages in various possible regimes.}
\label{fig:Landscape}
\end{figure} 
Given a generic word $\ux$ (i.e., a length $N$ binary string that 
is not necessarily a codeword), let us denote, with a slight abuse of
notation, by $U(\ux)$ the number of parity-check equations that are not 
satisfied by $\ux$. 
The above result, together with the weight enumerator calculation
in the previous section, suggests the following picture of the function  
$U(\ux)$. If $\ux^{(0)}\in\Code$, than $U(\ux^{(0)})=0$. 
Moving away from $\ux^{(0)}$, $U(\ux)$ will become strictly positive.
However as long as $d(\ux^{(0)},\ux)$ is small enough, 
$U(\ux)$ does not have any local minimum distinct from
$\ux^{(0)}$. A greedy procedure with a starting point within such a
Hamming radius is able to reconstruct $\ux^{(0)}$. 
As we move further away, $U(\ux)$ stays positive (no other codewords
are encountered) but local minima start to appear. Bit flipping 
gets trapped in such minima.
Finally, for $d(\ux^{(0)},\ux)\ge N\omega_*$ new codewords, 
i.e., minima with $U(\ux)=0$, are encountered.

%
%
\subsection{Message Passing}

Message-passing algorithms are iterative and have low complexity.
Unlike the bit-flipping procedure in the previous section,
the basic variables are now associated to directed edges in the
factor graph. More precisely, for each edge $(i,a)$
(corresponding to a non-zero entry in the parity-check matrix at
row $a$ and column $i$), we introduce two
messages $\me_{i\to a}$ and $\mh_{a\to i}$.
Messages are elements of some set (the message alphabet)
that we shall denote by $\M$. Depending on the specific algorithm,
$\M$ can have finite cardinality, or be infinite, for instance
$\M = \reals$.
Since the algorithm is iterative, it is convenient to introduce 
a time index $t=0,1,2,\dots$ and label the messages with the time at which 
they are updated:  $\me^{(t)}_{i\to a}$ and $\mh^{(t)}_{a\to i}$
(but we will sometimes drop the label below).

\begin{figure}
\begin{center}
\includegraphics[width=9cm]{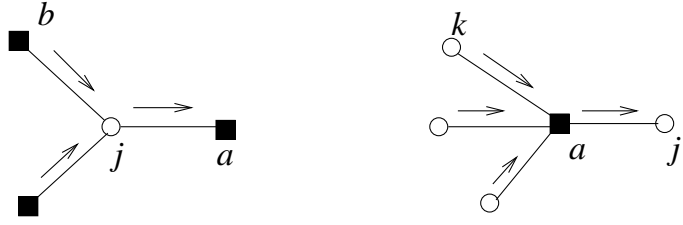}
\end{center}
\caption{Graphical representation of message passing updates.}
\label{fig:MPEquations}
\end{figure}

The defining property of message-passing algorithms is
that the message flowing from node $u$ to $v$ at a given time is a function
of messages entering $u$ from nodes $w$ distinct from $v$ at the previous 
time step. Formally, the algorithm is defined in terms of two sets
of functions $\Phi_{i\to a}(\,\cdot\,)$, $\Psi_{a\to i}(\,\cdot\,)$,
that define the update operations at variable and function nodes as follows
\begin{eqnarray}
\me_{i\to a}^{(t+1)} = 
\Phi_{i\to a}(\{\mh^{(t)}_{b\to i}; b\in\di\backslash a\};y_i)\, ,\;\;\;\;\;\;\;\;\;
\mh_{a\to i}^{(t)} = 
\Psi_{a\to i}(\{\me^{(t)}_{j\to a}; j\in\da\backslash i\})\, .\label{eq:BPUpdate}
\end{eqnarray}
Notice that messages are updated in parallel and that the time counter is
incremented only at variable nodes. Alternative scheduling schemes
can be considered but we will stick to this for the sake of simplicity.
After a pre-established number of iterations, the transmitted bits
are estimated using \emph{all} 
the messages incoming at the corresponding nodes. 
More precisely, the estimate at function $i$ is defined through a new 
function
\begin{eqnarray}
\xh^{(t)}_i(\uy) = \Phi_i(\{\mh^{(t)}_{b\to i}; b\in\di\};y_i)\, .
\end{eqnarray}
A graphical representation of message passing updates is provided in 
Fig.~\ref{fig:MPEquations}.

A specific message-passing algorithm requires the following 
features to be specified:
\begin{enumerate}
\item The message alphabet $\M$.
\item The initialization $\{\me^{(0)}_{i\to a}\}$, $\{\mh^{(0)}_{i\to a}\}$.
\item The update functions $\{\Phi_{i\to a}(\,\cdot\,)\}$,
 $\{\Psi_{a\to i}(\,\cdot\,)\}$.
\item The final estimate functions $\{\Phi_i(\,\cdot\,)\}$.
\end{enumerate}

The most prominent instance of a message-passing algorithm is the
Belief Propagation (BP) algorithm. In this case the messages
$\me_{i\to a}^{(t)}(x_i)$ and $\mh_{a\to i}^{(t)}(x_i)$ are distributions over 
the bit variables $x_i\in\{\0t,\1t\}$. The message $\mh_{a\to i}^{(t)}(x_i)$ is
usually interpreted as the \emph{a posteriori}
distributions of the bit $x_i$ given the information coming 
from edge $a\to i$.
Analogously, $\me_{i\to a}(x_i)$ is interpreted as the \emph{a posteriori}
distribution of $x_i$, given all the information collected through
edges distinct from $(a,i)$.
Since the messages normalization (explicitly 
$\me_{i\to a}(\0t)+\me_{i\to a}(\1t) = 1$) can be enforced at any time, we 
shall neglect overall factors in writing down the relation
between to messages (and correspondingly, we shall use the symbol $\propto$).

BP messages are updated according to the following rule,
whose justification we will discuss in the next section
\begin{eqnarray}
\me_{i\to a}^{(t+1)}(x_i) & \normeq & Q(y_i|x_i)\!\prod_{b\in\di\backslash a}
\mh^{(t)}_{b\to i}(x_i)\, ,\label{eq:BPdec1}\\
\mh_{a\to i}^{(t)}(x_i) & \normeq & \sum_{\{x_j\}}
\ind(x_i\oplus x_{j_1}\oplus \cdots\oplus x_{j_{k-1}}= \0t)
\prod_{j\in\da\backslash i}\me^{(t)}_{j\to a}(x_j)\, ,\label{eq:BPdec2}
\end{eqnarray}
where we used $(i,j_1,\dots,j_{k-1})$ 
to denote the neighborhood $\da$ of factor node $a$. 
After any number of iterations the single bit marginals
can be estimated as follows
\begin{eqnarray}
\me_{i}^{(t+1)}(x_i) \; \normeq \; Q(y_i|x_i)\!\prod_{b\in\di}
\mh^{(t)}_{b\to i}(x_i)\, .\label{eq:BPFinal}
\end{eqnarray}
The corresponding MAP decision for bit $i$ (sometimes called
`hard decision', while $\me_{i}(x_i)$ is the `soft decision') is
\begin{eqnarray}
\widehat{x}_i^{(t)} = \arg\max_{x_i}\;
\me_{i}^{(t)}(x_i) \, .\label{eq:HardDecision}
\end{eqnarray}
Notice that the above prescription is ill-defined when $\me_i(\0t) 
= \me_i(\1t)$. It turns out that it is not really important which rule to
use in this case. To preserve the $\0t-\1t$ symmetry, we shall 
assume that the decoder returns $\widehat{x}_i^{(t)}=\0t$ or
$=\1t$ with equal probability.

Finally, as initial condition one usually takes
$\mh_{a\to i}^{(-1)}(\,\cdot\,)$
to be the uniform distribution over $\{\0t,\1t\}$
(explicitly $\mh_{a\to i}^{(-1)}(\0t) = \mh_{a\to i}^{(-1)}(\1t)=1/2$). 

\begin{figure}
\center{
\includegraphics[angle=0,width=0.35\columnwidth]{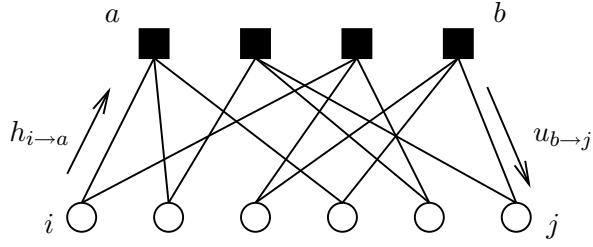}}
\put(5,0){$j$}
\put(-15,80){$b$}
\put(-185,0){$i$}
\put(-162,80){$a$}
\put(0,35){$u_{b\to j}$}
\put(-198,35){$h_{i\to a}$}
\caption{Factor graph of a regular LDPC code, and notation for
the belief propagation messages.}
\label{fig:FactorMess}
\end{figure}
Since for BP the messages are distributions over binary valued variables, they
can be described by a single real number, that is often chosen to be
the bit log-likelihood:\footnote{The conventional definition of log-likelihoods
does not include the factor $1/2$. We introduce this factor here for uniformity
with the statistical mechanics convention (the $h$'s and $u$'s being
analogous to effective magnetic fields).}
\begin{eqnarray}
h_{i\to a} = \frac{1}{2}\, \log\frac{\me_{i\to a}(\0t)}{\me_{i\to a}(\1t)}
\, ,\;\;\;\;\;\;\;\;
u_{a\to i} = \frac{1}{2}\, \log\frac{\mh_{a\to i}(\0t)}{\mh_{a\to i}(\1t)}
\, .\label{eq:LLR}
\end{eqnarray}
We refer to Fig.~\ref{fig:FactorMess} for a pictorial representation of
these notations.
 We further introduce the channel log-likelihoods
\begin{eqnarray}
B_i = \frac{1}{2}\,\log\frac{Q(y_i|\0t)}{Q(y_i|\1t)}\, .\label{eq:APrioriLLR}
\end{eqnarray}
The BP update equations (\ref{eq:BPdec1}), (\ref{eq:BPdec2}) read in this
notation
\begin{eqnarray}
h^{(t+1)}_{i\to a} =  B_i+\sum_{b\in\di\backslash a}u^{(t)}_{b\to i}\, ,
\;\;\;\;\;\;\;\;\;
u^{(t)}_{a\to i}  =  \atanh\Big\{\prod_{j\in\da\backslash i}
\tanh h^{(t)}_{j\to a}\Big\}\, .\label{eq:BPDec_LLR}
\end{eqnarray}                        
In this language the standard message initialization would be
$u^{(-1)}_{a\to i} =0$. Finally, the overall log-likelihood
at bit $i$ is obtained by combining \emph{all} the incoming messages 
in agreement with Eq.~(\ref{eq:BPFinal}).
One thus gets the decision rule
\begin{eqnarray}
\widehat{x}_i^{(t)} = \left\{
\begin{array}{cc}
\0t & \mbox{ if $B_i+ \sum_{b\in\di}u^{(t)}_{b\to i}>0$,}\\
\1t & \mbox{ if $B_i+ \sum_{b\in\di}u^{(t)}_{b\to i}<0$.}
\end{array}
\right.\label{eq:BPDecisionLLR}
\end{eqnarray}
Notice that we did not commit to any special decision if
$B_i+ \sum_{b\in\di}u^{(t)}_{b\to i}=0$. To keep complete symmetry
we'll establish that the decoder returns $\0t$ or $\1t$ with equal 
probability in this case.
%
%
\subsection{Correctness of Belief Propagation on Trees}

The justification for the BP update equations 
(\ref{eq:BPdec1}), (\ref{eq:BPdec2}) lies in the observation that, 
whenever the underlying factor graph is a tree, the estimated marginal 
$\me_i^{(t)}(x_i)$  converges after a finite number of iterations to the 
correct one $\mu_i(x_i)$.
In particular, under the tree assumption, and for any $t$ sufficiently large, 
$\xh^{(t)}_i(\uy)$ coincides with the symbol MAP decision.

In order to prove this statement,
consider a tree factor graph $G$. Given a couple of adjacent nodes
$u, v$, denote by $G(u\to v)$ the subtree rooted at the directed edge 
$u\to v$ (this contains all that can be reached from $v$ through a
non-reversing path whose first step is $v\to u$). 
If $i$ is a variable index and $a$ a parity-check index, let
$\mu_{i\to a}(\, \cdot\, )$ be the measure over $\ux = \{x_j:\,
j\in G(i\to a)\}$, that is obtained by retaining in 
Eq.~(\ref{eq:ProbLDPC}) only those terms that are related to nodes
in $G(i\to a)$:
\begin{eqnarray}
\mu_{i\to a}(\ux) = \frac{1}{Z(i\to a)}\, \prod_{j\in G(i\to a)}
Q(y_i|x_i)\;\prod_{b\in G(i\to a)} \ind(x_{i_1(b)}\oplus\cdots\oplus
x_{i_k(b)}=\0t)\, .
\end{eqnarray}
The measure $\widehat{\mu}_{a\to i}(\,\cdot\,)$ is defined analogously
for the subtree $G(a\to i)$. The marginals $\mu_{i\to a}(x_i)$
(respectively $\widehat{\mu}_{a\to i}(x_i)$) are easily seen to
satisfy the recursions 
\begin{eqnarray}
\mu_{i\to a}(x_i)  & \normeq&  Q(y_i|x_i)\!\prod_{b\in\di\backslash a}
\widehat{\mu}_{b\to i}(x_i)\, ,\label{eq:TreeRec1}\\
\widehat{\mu}_{a\to i}(x_i) & \normeq & \sum_{\{x_j\}}
\ind(x_i\oplus x_{j_1}\oplus \cdots\oplus x_{j_{k-1}}= \0t)
\prod_{j\in\da\backslash i}\mu_{j\to a}(x_j)\, ,\label{eq:TreeRec2}
\end{eqnarray}
which coincide, apart from the time index, with the BP recursion
(\ref{eq:BPdec1}), (\ref{eq:BPdec2}). That such recursions
converges to $\{\mu_{i\to a}(x_i),\widehat{\mu}_{a\to i}(x_i)\}$
follows by induction over the tree depth.

In statistical mechanics equations similar to
(\ref{eq:TreeRec1}), (\ref{eq:TreeRec2}) are 
often written as recursions on the constrained partition function.
They allow to solve exactly models on trees. 
However they have been often applied as mean-field approximation
to statistical models on non-tree graphs. 
This is often referred to as  the \emph{Bethe-Peierls approximation}
\cite{Bethe}.

The Bethe approximation presents several advantages with respect to 
`naive-mean field' \cite{YedidiaFreemanWeiss}
(that amounts to writing `self-consistency'
equations for expectations over single degrees of freedom).
It retains correlations among degrees of freedom that interact directly,
and is exact on some non-empty graph (trees). It is often asymptotically
(in the large size limit) exact on locally tree-like graphs.
Finally, it is quantitatively more accurate for non-tree like graphs and
offers a much richer modeling palette.

Within the theory of disordered systems
(especially, glass models on sparse random graphs), 
Eqs.~(\ref{eq:TreeRec1}) and (\ref{eq:TreeRec2}) are also referred
to as the \emph{cavity equations}. With respect to Bethe-Peierls, the cavity 
approach includes a hierarchy of (`replica symmetry breaking') refinements
of such equations that aim at capturing long range correlations
\cite{MezardParisiBethe}. 
This will briefly described in Section \ref{sec:beyondcoding}.

We should finally mention that several improvements over 
Bethe approximation have been developed within statistical physics.
Among them, Kikuchi's cluster variational method 
\cite{Kikuchi} is worth mentioning
since it motivated the development of a `generalized 
belief propagation' algorithm, which spurred a lot of
interest within the artificial intelligence community 
\cite{YedidiaFreemanWeiss}.
%
%
\subsection{Density Evolution}
Although BP  converges to the exact marginals on tree graphs,
this says little about its performances on practical codes
such as the LDPC ensembles introduced in Section~\ref{sec:sparsegraphcode}.
Fortunately, a rather precise picture on the performance of LDPC ensembles can be derived in the large 
blocklength limit $N\to\infty$. The basic reason for 
this is that the corresponding random factor graph is locally 
tree-like with high probability if we consider large blocklengths.

Before elaborating on this point, notice that the performance under BP decoding 
(e.g., the bit error rate) is independent on the transmitted codeword.
For the sake of analysis, we shall hereafter assume that the all-zero 
codeword $\u0t$ has been transmitted.

Consider a factor graph $G$ and let $(i, a)$ be one of its edges.
Consider the message $\me^{(t)}_{i\to a}$ sent by the BP decoder in iteration
$t$ along edge $(i, a)$. A considerable amount of information is contained in the 
distribution of $\me^{(t)}_{i\to a}$ with respect to the channel realization,
as well as in the analogous distribution for $\mh^{(t)}_{a\to i}$.
To see this, note that under the all-zero codeword 
assumption, the bit error rate after $t$ iterations is given by
\begin{eqnarray}
\bit^{(t)} = \frac{1}{n}\sum_{i=1}^n
\prob\left\{\Phi_i(\{\mh_{b\to i}^{(t)};b\in\di\};y_i)\neq \0t\right\}\, .
\end{eqnarray}
Therefore, if the messages $\mh_{b\to i}^{(t)}$ are independent, then the bit
error probability is determined by the distribution of $\mh^{(t)}_{a\to i}$.

Rather than considering one particular graph (code) and a specific
edge, it is much simpler to take the average over all edges and all graph realizations.
We thus consider the distribution $\dens_t^{(N)}(\,\cdot\,)$
of $\me^{(t)}_{i\to a}$ with respect to the channel, the edges, \emph{and} the graph
realization. While this is still a
quite difficult object to study rigorously, it is on the other
hand possible to characterize its large blocklength limit
$\dens_t(\,\cdot\,) = \lim_N\dens_t^{(N)}(\,\cdot\,)$.
This distribution satisfies a simple recursion.
\begin{figure}[htp]
\centering
\setlength{\unitlength}{0.8bp}%
\begin{picture}(322,490)(-22,-10)
\graphtextsizesmall
\put(0,400)
{
\multiputlist(-22,14.2)(0,14.2)[l]{$0.05$,$0.10$,$0.15$,$0.20$}
\multiputlist(0,-10)(23.65,0)[b]{$\text{-}10$,$0$,$10$,$20$,$30$,$40$}
\multiputlist(158,14.2)(0,14.2)[l]{$0.05$,$0.10$,$0.15$,$0.20$}
\multiputlist(180,-10)(23.65,0)[b]{$\text{-}10$,$0$,$10$,$20$,$30$,$40$}
\put(0,0){\includegraphics[scale=0.8]{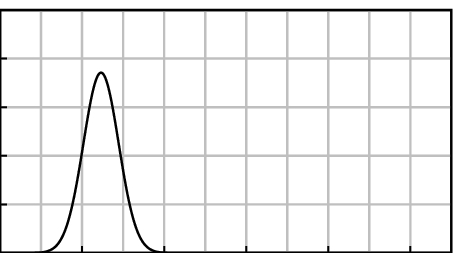}}
\put(180,0){\includegraphics[scale=0.8]{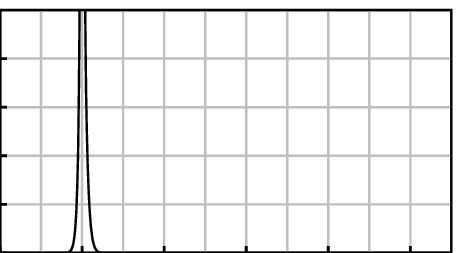}}
}
\put(0,300)
{
\multiputlist(-22,14.2)(0,14.2)[l]{$0.05$,$0.10$,$0.15$,$0.20$}
\multiputlist(0,-10)(23.65,0)[b]{$\text{-}10$,$0$,$10$,$20$,$30$,$40$}
\multiputlist(158,14.2)(0,14.2)[l]{$0.05$,$0.10$,$0.15$,$0.20$}
\multiputlist(180,-10)(23.65,0)[b]{$\text{-}10$,$0$,$10$,$20$,$30$,$40$}
\put(0,0){\includegraphics[scale=0.8]{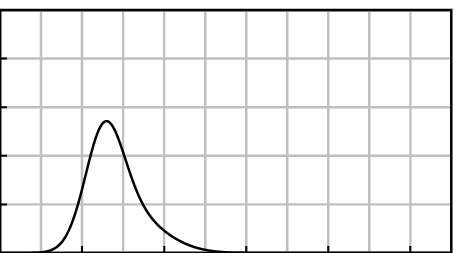}}
\put(180,0){\includegraphics[scale=0.8]{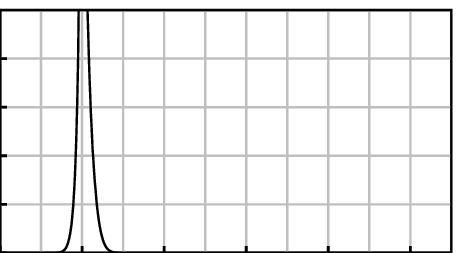}}
}
\put(0,200)
{
\multiputlist(-22,14.2)(0,14.2)[l]{$0.05$,$0.10$,$0.15$,$0.20$}
\multiputlist(0,-10)(23.65,0)[b]{$\text{-}10$,$0$,$10$,$20$,$30$,$40$}
\multiputlist(158,14.2)(0,14.2)[l]{$0.05$,$0.10$,$0.15$,$0.20$}
\multiputlist(180,-10)(23.65,0)[b]{$\text{-}10$,$0$,$10$,$20$,$30$,$40$}
\put(0,0){\includegraphics[scale=0.8]{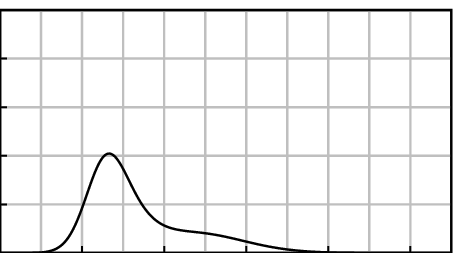}}
\put(180,0){\includegraphics[scale=0.8]{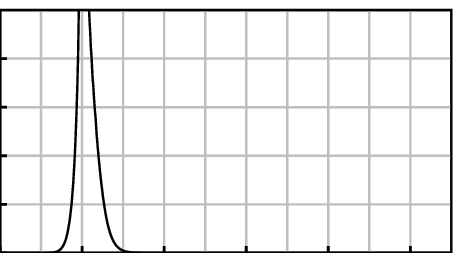}}
}
\put(0,100)
{
\multiputlist(-22,14.2)(0,14.2)[l]{$0.05$,$0.10$,$0.15$,$0.20$}
\multiputlist(0,-10)(23.65,0)[b]{$\text{-}10$,$0$,$10$,$20$,$30$,$40$}
\multiputlist(158,14.2)(0,14.2)[l]{$0.05$,$0.10$,$0.15$,$0.20$}
\multiputlist(180,-10)(23.65,0)[b]{$\text{-}10$,$0$,$10$,$20$,$30$,$40$}
\put(0,0){\includegraphics[scale=0.8]{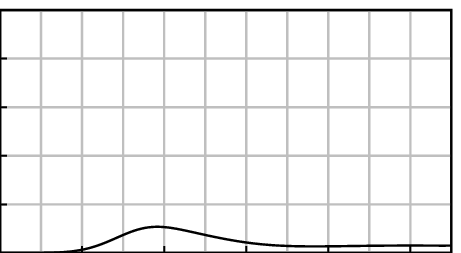}}
\put(180,0){\includegraphics[scale=0.8]{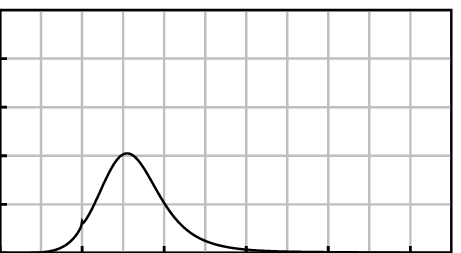}}
}
\put(0,0)
{
\multiputlist(-22,14.2)(0,14.2)[l]{$0.05$,$0.10$,$0.15$,$0.20$}
\multiputlist(0,-10)(23.65,0)[b]{$\text{-}10$,$0$,$10$,$20$,$30$,$40$}
\multiputlist(158,14.2)(0,14.2)[l]{$0.05$,$0.10$,$0.15$,$0.20$}
\multiputlist(180,-10)(23.65,0)[b]{$\text{-}10$,$0$,$10$,$20$,$30$,$40$}
\put(0,0){\includegraphics[scale=0.8]{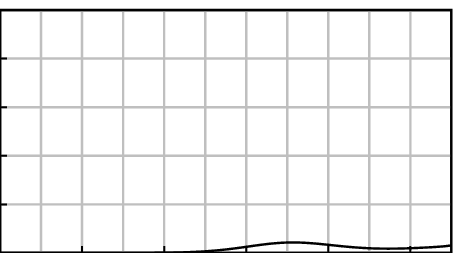}}
\put(180,0){\includegraphics[scale=0.8]{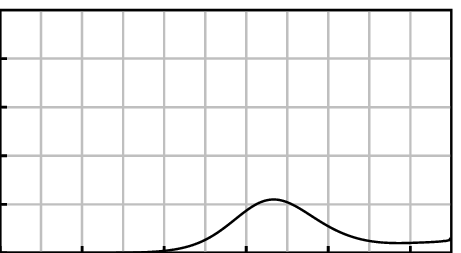}}
}

\end{picture}
\caption{\label{fig:densevolution}
Evolution of the probability density functions of $h^{(t)}$
an $u^{(t+1)}$ for an irregular LDPC code used over a gaussian channel.
From top to bottom $\iteration=0$, $5$, $10$, $50$, and $140$.}
\end{figure}

\begin{figure}
\center{
\includegraphics[angle=0,width=0.32\columnwidth]{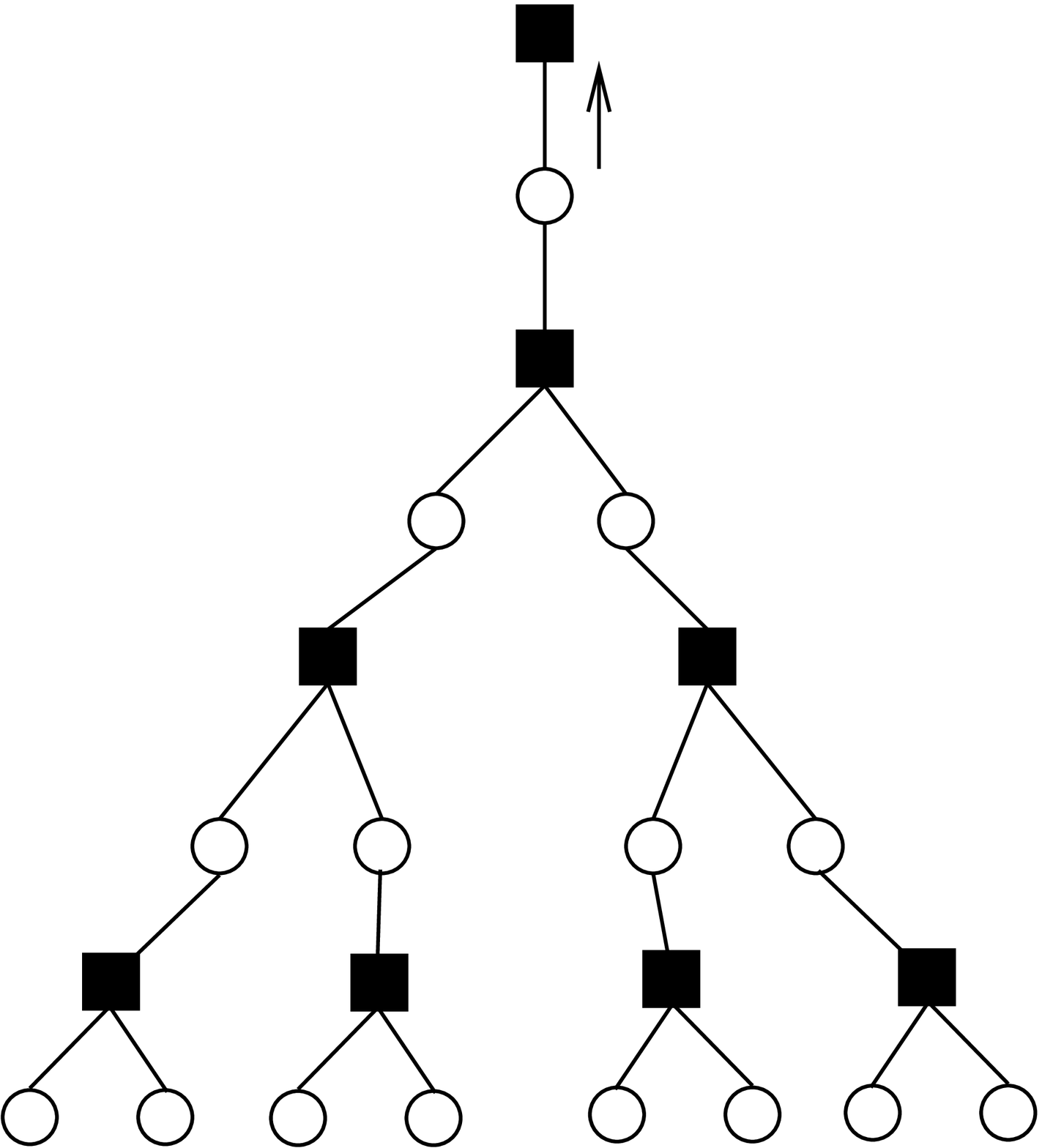}}
\put(-67,140){$i$}
\put(-67,175){$a$}
\caption{A radius $3$ directed neighborhood $\ball_{i\to a}(3;G)$.}
\label{fig:FactorTree}
\end{figure}

It is convenient to introduce the \emph{directed neighborhood} of radius $r$
of the directed edge $i\to a$ in $G$, call it $\ball_{i\to a}(r;G)$.
This is defined as the subgraph of $F$ that includes all
the variable nodes that can be reached from $i$ through a non-reversing path of 
length at most $r$, whose first  step {\em is not} the edge  $(i,a)$.
It includes as well all the function nodes connected {\em only} to the 
above specified variable nodes. In Fig.~\ref{fig:FactorTree} we 
reproduce an example
of a directed neighborhood of radius $r=3$ (for illustrative purposes
we also include the edge $(i, a)$) in a $(2,3)$ regular code.

If $F$ is the factor graph of a random code from the $(k,l)$ LDPC ensemble, 
then
$\neigh_{i\to a}(r; F)$ is with high probability a depth-$r$ 
regular tree with degree $l$ at variable nodes and degree $k$ at check nodes
(as in Fig.~\ref{fig:FactorTree} where $l =2 $ and $k=3$). 
The basic reason for this phenomenon is rather straightforward.
Imagine to explore the neighborhood progressively, moving away from the 
root, in a breadth first fashion. At any finite radius $r$,
 about $c^{r}/N$ vertices have been visited (here
$c= (k-1)(l-1)$). 
The vertices encountered at the next layer will be 
`more or less' uniformly random among all the ones not visited so far.
As a consequence they will be distinct with high probability,
and $\ball_{i\to a}(r+1;G)$ will be a tree as well. 
This argument breaks down when the probability that two of
the $\Theta(c^r)$ new vertices coincide, that is 
for
$c^{2r}=\Theta(N)$.\footnote{This is the famous birthday problem. The probability 
that two out of a party of $n$ peoples were born on the same day of the year,
scales like $n^2/N$ for $n^2\ll N$ ($N$ is here the number of days in a year).}
This is equivalent to $r\simeq\frac{1}{2}\log_c N$.

\begin{figure}
\center{
\includegraphics[angle=0,width=0.48\columnwidth]{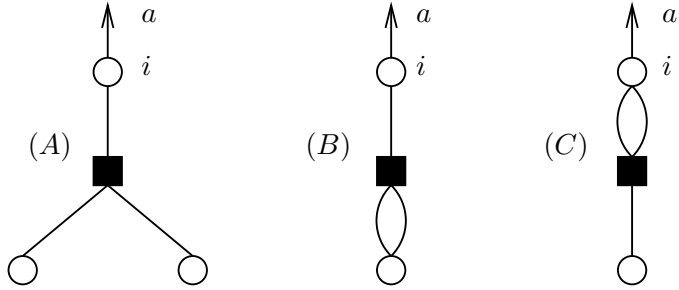}}
\put(-192,80){$i$}
\put(-192,100){$a$}
\put(-88,80){$i$}
\put(-88,100){$a$}
\put(5,80){$i$}
\put(5,100){$a$}
\put(-235,50){$(A)$}
\put(-130,50){$(B)$}
\put(-40,50){$(C)$}
\caption{The three possible radius--$1$ directed neighborhoods 
in a random factor graph from the regular $(2,3)$ graph ensemble.}
\label{fig:FactorTreeRadius1}
\end{figure}

The skeptical reader is invited to solve the following exercise.
\begin{exercise}{3}
In order to illustrate the above statement, consider the example
of a random code from the regular $(2,3)$ ensemble 
(each variable has degree 2 and each check has degree 3).
The three possible radius-$1$ neighborhoods appearing 
in the associated factor graph are depicted in 
Fig.~\ref{fig:FactorTreeRadius1}. 
\begin{enumerate}
\item[$(a)$] Show that the probability that a given edge 
$(i,a)$ has neighborhoods as in (B) or (C) is $O(1/N)$.
\item[$(b)$] What changes for a generic  radius $r$?
\end{enumerate}
\end{exercise}
For illustrative reasons, 
we shall occasionally add a `root edge' to $\ball_{i\to a}(r;G)$,
as for $i\to a$ in Fig.~\ref{fig:FactorTree}.

Now consider the message $\me^{(t)}_{i\to a}$. This is a 
function of the
factor graph $G$ and of the received message $\uy$. 
However, a moment's thought 
shows that it will depend on $G$ only through its directed 
neighborhood $\ball_{i\to a}(t+1;G)$, and only on
the received symbols $y_j$, $j\in \ball_{i\to a}(t;G)$. 

In view of the above discussion, let us consider
the case in which  $\ball_{i\to a}(t+1; G)$ is a  $(k,l)$-regular tree. We further assume that the 
received symbols $y_j$ are i.i.d. with distribution $Q(y|\0t)$,
and that the update rules (\ref{eq:BPUpdate}) do not depend on
the edge we are considering (i.e., $\Phi_{i\to a}(\,\cdot\,) = \Phi(\,\cdot\,)$
and $\Psi_{i\to a}(\,\cdot\,) = \Psi(\,\cdot\,)$
independent of $i$, $a$).
 
Let $\me^{(t)}$ be the 
message passed through the root edge of such a tree after $t$ 
BP iterations. 
Since the actual neighborhood  $\neigh_{i\to a}(t+1;G)$
is with high probability a tree,  
$\me^{(t)}_{i\to a}\stackrel{{\text d}}{\to} \me^{(t)}$ as $N\to\infty$.
The symbol $\stackrel{{\text d}}{\to}$ denotes convergence in 
distribution. In other words, for large blocklengths, the message distribution
after $t$ iterations is asymptotically the same that we would have obtained 
if the graph were a tree.

Consider now a $(k,l)$-regular tree,
and let $j\to b$ an edge directed towards the root, at distance $d$
from it. It is not hard to realize that the message passed through it 
after $r-d-1$ (or more) iterations is distributed as $\me^{(r-d-1)}$. Furthermore,
if $j_1\to b_1$ and $j_2\to b_2$ are both directed upwards
and none belongs to the subtree rooted at the other one, then
 the corresponding messages are independent.
Together with Eq.~(\ref{eq:BPUpdate}), these observation imply that
\begin{eqnarray}
\me^{(t+1)}
\ed  \Phi(\mh_1^{(t)},\dots, \mh_{l-1}^{(t)};y)
\, ,\;\;\;\;\;
\mh^{(t)}
\ed  \Psi(\me_1^{(t)},\dots, \me_{k-1}^{(t)})\, .\label{eq:GeneralDE}
\end{eqnarray}
Here $\mh_1^{(t)},\dots,\mh_{l-1}^{(t)}$ are i.i.d. copies
of $\mh^{(t)}$, and $\me_1^{(t)},\dots, \me_{k-1}^{(t)}$
i.i.d. copies of $\me^{(t)}$. Finally, $y$ is a received symbol
independent from the previous variables and distributed according to
$Q(y|\0t)$. 

Equations (\ref{eq:GeneralDE}), or the sequence of distributions
that they define, are usually referred to as \emph{density evolution}.
The name is motivated by the identification
of the random variables with their densities (even if these do not 
necessarily exist).
They should be parsed as follows (we are refer here to the first equation
in (\ref{eq:GeneralDE}); an analogous phrasing holds for the second): 
pick $l-1$ i.i.d. copies $\mh^{(t)}$
and $y$ with distribution $Q(y|\0t)$, compute
$\Phi(\mh_1^{(t)},\dots, \mh_{l-1}^{(t)};y)$. The resulting quantity will have
distribution $\me^{(t+1)}$. Because of this description, they are also called
`recursive distributional equations'.

Until this point we considered a generic message passing 
procedure. If we specialize to BP decoding,  we can use the parametrization of
messages in terms of log-likelihood ratios, cf. Eq.~(\ref{eq:LLR}),
and use the above arguments to characterize the limit
random variables $h^{(t)}$ and  $u^{(t)}$. 
The update rules (\ref{eq:BPDec_LLR}) then imply
\begin{eqnarray}
h^{(t+1)} \ed  B+ u^{(t)}_{1}+\cdots+u^{(t)}_{l-1}\, ,
\;\;\;\;\;\;\;\;\;
u^{(t)} \ed  \atanh\left\{
\tanh h^{(t)}_{1}\cdots \tanh h_{k-1}^{(t)}\right\}\,. 
\label{eq:DensityEvolution}
\end{eqnarray}
Here $u^{(t)}_1,\dots,u^{(t)}_{l-1}$ are i.i.d. copies of $u^{(t)}$,
$h^{(t)}_1,\dots,h^{(t)}_{k-1}$ are i.i.d. copies of $h^{(t)}$,
and $B = \frac{1}{2}\log \frac{Q(y|\0t)}{Q(y|\1t)}$, where $y$ is independently 
distributed according to $Q(y|\0t)$. It is understood that the 
recursion is initiated with $u^{(-1)} = 0$.

Physicists often write distributional recursions explicitly in terms of 
densities. For instance, the first of the equations above reads
\begin{eqnarray}
\dens_{t+1}(h) = \int\!\!\prod_{b=1}^{l-1} \de\hdens_{t}(u_b)\;
\de\pd (B) \;\;\;\delta\left(h-B-\sum_{b=1}^{l-1}u_b\right)\, ,
\end{eqnarray}
where $\hdens_{t}(\,\cdot\,)$ denotes the density of $u^{(t)}$,
and $\pd(\,\cdot\,)$ the density of $B$. We refer to 
Fig.~\ref{fig:densevolution}
for an illustration of how the densities $\dens_t(\,\cdot\,)$,
$\hdens_t(\,\cdot\,)$ evolve during the decoding process.

In order to stress the importance of density evolution
notice that, for any continuous function $f(x)$,
\begin{eqnarray}
\lim_{N\to\infty}\E\Big\{\frac{1}{N}\sum_{i=1}^N f(h_{i\to a}^{(t)})\Big\}
=\E\{f(h^{(t)})\}\, ,\label{eq:LimitBPVariables}
\end{eqnarray}
where the expectation is taken with respect to the code ensemble.
Similar expressions can be obtained for functions of several messages
(and are particularly simple when such message are asymptotically independent).
In particular\footnote{The suspicious reader will notice that this is not
exactly a particular case of the previous statement, because 
$f(x) = \ind(x<0)+\frac{1}{2}\ind(x=0)$ is not a continuous function.}, 
if we let $\bit^{(N,t)}$ be the expected (over an LDPC
ensemble) bit error rate for the decoding rule (\ref{eq:BPDecisionLLR}), 
and let $\bit^{(t)} = \lim_{N\to\infty}\bit^{(N,t)}$ be its large blocklength 
limit. Then
\begin{eqnarray}
\bit^{(t)} = \prob\big\{ B+h_1^{(t)}+\cdots+h_l^{(t)}<0\big\}+
\frac{1}{2}\, \prob\big\{ B+h_1^{(t)}+\cdots+h_l^{(t)}=0\big\}\, ,
\end{eqnarray}
where $h_1^{(t)},\dots, h_l^{(t)}$ are i.i.d. copies of $h^{(t)}$.

%
%
\subsection{The Belief Propagation Threshold}

Density evolution would not be such an useful tool if it could not
be simulated efficiently. The idea is to estimate numerically the 
distributions of the density evolution variables $\{h^{(t)},u^{(t)}\}$. 
As already discussed this gives access to a 
number of statistics on BP decoding, such as the bit error rate $\bit^{(t)}$
after $t$ iterations in the large blocklength limit.

A possible approach consists in representing the distributions
by samples of some fixed size. Within statistical physics
this is sometimes called the \emph{population dynamics algorithm}
(and made its first appearance in the study of the localization transition
on Cayley trees \cite{AbouChacra73}).
Although there exist more efficient alternatives in the 
coding context (mainly based on Fourier transform, see \cite{RiU01,RiJ06}), we shall describe
population dynamics because it is easily programmed.

Let us describe the algorithm within the setting of a general
message passing decoder, cf. Eq.~(\ref{eq:GeneralDE}). 
Given an integer $\cN\gg 1$, one represent the messages distributions
with two samples of size $\cN$: $\P^{(t)} = \{\me_1^{(t)},\dots,
\me_{\cN}^{(t)}\}$, and $\Ph^{(t)} = \{\mh_1^{(t)},\dots,
\mh_{\cN}^{(t)}\}$. Such samples are used as proxy for the corresponding 
distributions. For instance, one would approximate an expectation
as 
\begin{eqnarray}
\E f(\me^{(t)}) \approx\frac{1}{\cN}\sum_{i=1}^{\cN} \, f(\me^{(t)}_i)\, .
\label{eq:PopulationEstimate}
\end{eqnarray}
The populations are updated iteratively. For instance $\P^{(t+1)}$
is obtained from $\Ph^{(t)}$ by generating $\me_1^{(t+1)},
\dots, \me_{\cN}^{(t+1)}$ independently as follows. For each 
$i \in [\cN]$, draw indices $b_1(i),\dots, b_{l}(i)$
independently and uniformly at random from $[\cN]$, and generate
$y_i$ with distribution $Q(y|\0t)$. Then compute 
$\me^{(t+1)}_i = \Phi(\{\mh^{(t)}_{b_n(i)}\};y_i)$ and store
it in $\P^{(t+1)}$. 

An equivalent description consists in saying that we proceed as if
$\Ph^{(t)}$ exactly represents the distribution of $u^{(t)}$
(which in this case would be discrete). If this was the case,
the distribution of $h^{(t+1)}$ would be composed of $|\cA|\cdot \cN^{l-1}$
Dirac deltas. In order not to overflow memory, the algorithm samples
$\cN$ values from such a distribution.
Empirically, estimates of the form (\ref{eq:PopulationEstimate})
obtained through population dynamics have systematic errors of 
order $\cN^{-1}$ and statistical errors of order $\cN^{-1/2}$ with 
respect to the exact value.

\begin{figure}
\begin{tabular}{cc}
\hspace{-0.5cm}
\includegraphics[angle=0,width=0.5\columnwidth]{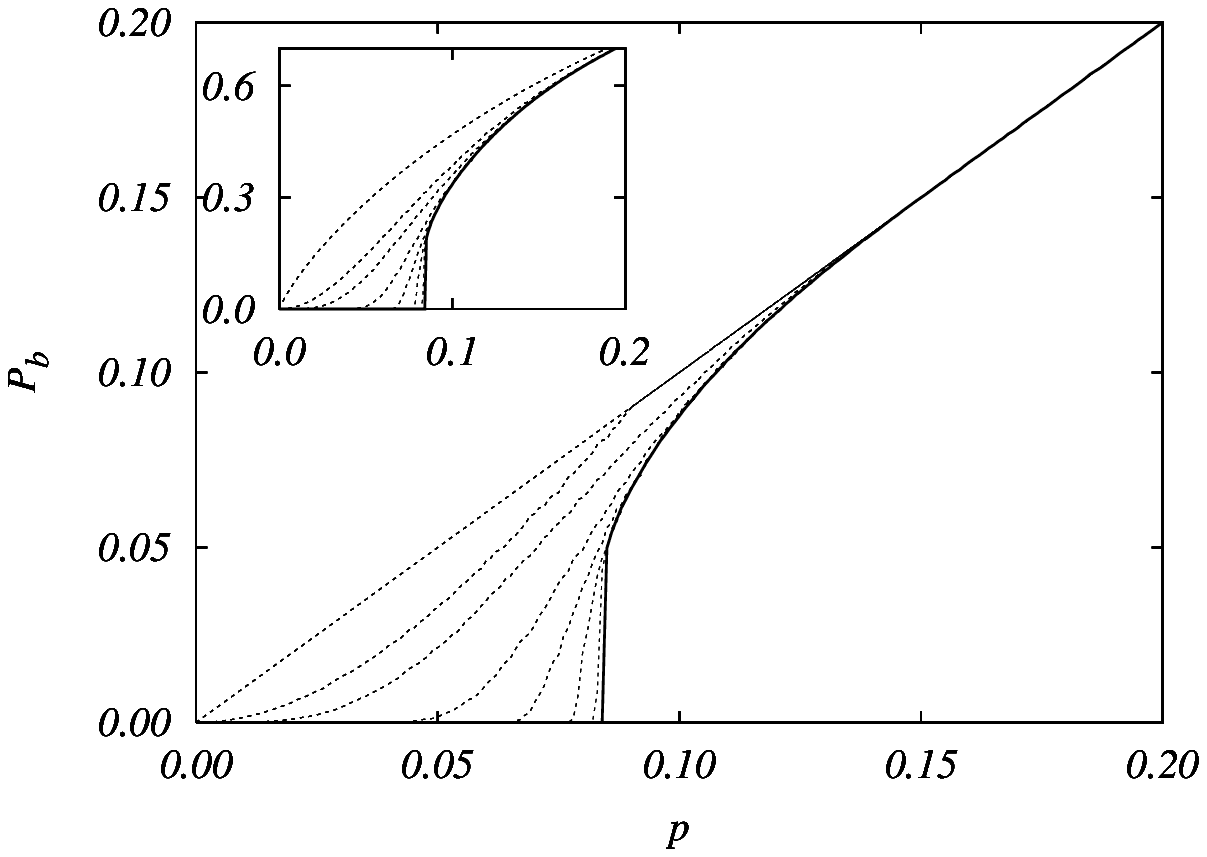}&
\includegraphics[angle=0,width=0.5\columnwidth]{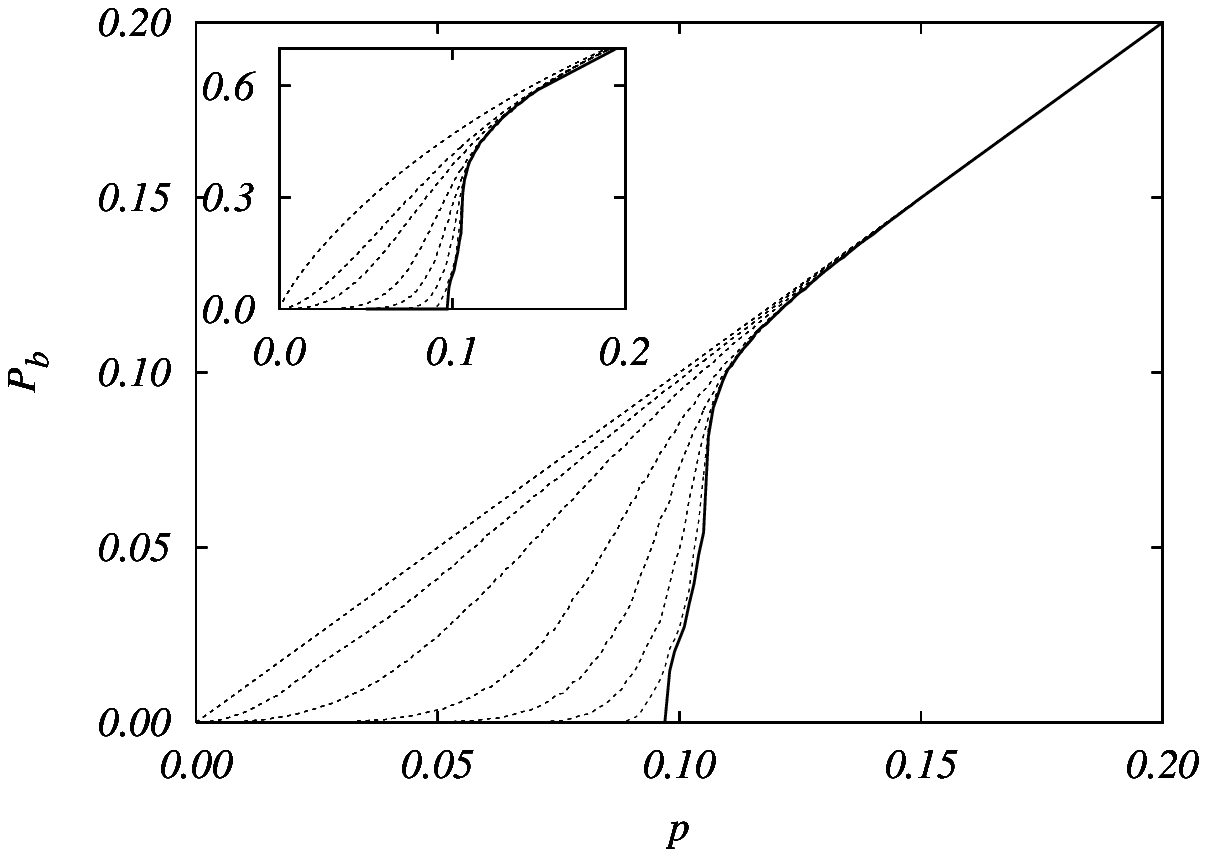}
\end{tabular}
\caption{The performances of two LDPC ensembles as predicted by
a numerical implementation of density evolution. On the left,
the $(3,6)$ regular ensemble. On the right, an optimized irregular ensemble.
 Dotted curves refer (from top to bottom) to $t=0$, 
$1$, $2$, $5$, $10$, $20$, $50$ iterations,
and bold continuous lines to the limit $t\to\infty$. In the inset we plot
the expected conditional entropy $\E\, H(X_i|\overline{\nu}_i^{(t)})$.}
\label{fig:DE}
\end{figure}
In Fig.~\ref{fig:DE} we report the results of population dynamics simulations 
for two different LDPC ensembles, with respect to the BSC.
 We consider two performance measures:
the bit error rate $\bit^{(t)}$ and the bit conditional entropy $H^{(t)}$.
The latter is defined as
\begin{eqnarray}
H^{(t)} = \lim_{N\to\infty}\frac{1}{N}\sum_{i=1}^N\E\, H(X_i|\overline{\nu}_i^{(t)})\,,
\end{eqnarray}
and encodes the uncertainty about bit $x_i$ after $t$ BP iterations. 
It is intuitively clear that, as the algorithm progresses, the
bit estimates improve and therefore $\bit^{(t)}$ and $H^{(t)}$
should be monotonically decreasing functions of the number of iterations. 
Further, they are expected to be
monotonically increasing functions of the crossover probability $p$.
Both statement can be easily checked on the above plots, and 
can be proved rigorously as well.

Since $\bit^{(t)}$ is non-negative and decreasing in $t$, it has
a finite limit
\begin{eqnarray}
\bit^{\sBP} \equiv \lim_{t\to\infty} \bit^{(t)}\, ,
\end{eqnarray}
which is itself non-decreasing in $p$. The limit curve 
$\bit^{\sBP}$ is estimated in Fig.~\ref{fig:DE} by choosing $t$ large
 enough so 
that $\bit^{(t)}$ is independent of $t$ within the numerical accuracy.

Since $\bit^{\sBP}= \bit^{\sBP}(p)$ is a non-decreasing 
function of $p$,  one can define the \emph{BP threshold}
\begin{eqnarray}
p_{\rm d} \equiv 
\sup\left\{\, p\in[0,1/2]\, : \, \bit^{\sBP}(p) = 0\, \right\}\, .
\label{eq:BPThresholdDef}
\end{eqnarray}
Analogous definitions can be provided for other channel families such 
as the BEC$(\epsilon)$.
In general, the definition (\ref{eq:BPThresholdDef}) can
be extended to any family of BMS channels BMS$(p)$  indexed by
a real parameter $p\in I$, $I\subseteq\reals$ being an interval (obviously
the $\sup$ will be then taken over $p\in I$).
The only condition is that the family is `ordered by physical degradation'.
We shall not describe this concept formally, but limit ourselves to 
say that that $p$ should be an `honest' noise parameter, in the sense
that the channel worsen as $p$ increases. 

\begin{table}
\begin{center}
\begin{tabular}{ccccc}
\hline
$l$ & $k$ & $R$ & $p_{\rm d}$ & Shannon limit\\
\hline
$3$ & $4$ & $1/4$ & $0.1669(2)$ & $0.2145018$\\
$3$ & $5$ & $2/5$ & $0.1138(2)$ & $0.1461024$\\
$3$ & $6$ & $1/2$ & $0.0840(2)$ & $0.1100279$\\
$4$ & $6$ & $1/3$ & $0.1169(2)$ & $0.1739524$\\
\hline
\end{tabular}
\label{TableBPThresholds}
\end{center}
\caption{Belief propagation thresholds for a few regular {\rm LDPC} ensembles.}
\end{table}
Analytical upper and lower bounds can be 
derived for $p_{\rm d}$. In particular it can be shown that it is strictly
larger than $0$ (and smaller than $1/2$) for all LDPC ensembles with 
minimum variable degree at least $2$. Numerical simulation of density
evolution allows to determine it numerically with good accuracy.
In Table \ref{TableBPThresholds} we report the results of a few such results.

Let us stress that the threshold $p_{\rm d}$ has an important 
practical meaning. For any $p<p_{\rm d}$ one can achieve arbitrarily 
small bit error rate with high probability by just picking
one random code from the ensemble LDPC and using BP decoding and running
it for a large enough (but independent of the blocklength)
number of iterations.
For $p>p_{\rm d}$ the bit error rate is asymptotically lower bounded 
by $\bit^{\sBP}(p)>0$ for any fixed number of iterations. 
In principle it could be that after, let's say $n^{a}$, $a>0$ iterations 
a lower bit error rate is achieved. However simulations show
quite convincingly that this is not the case.

In physics terms the algorithm undergoes a phase transition at 
$p_{\rm d}$. At first sight, such a phase transition may look
entirely dependent on the algorithm definition and not
`universal' in any sense. As we will discuss in the next 
section, this is not the case. The phase transition at
$p_{\rm d}$ is somehow intrinsic to the underlying measure $\mu(\ux)$,
and has a well studied counterpart in the theory of mean field 
disordered spin models.

Apart from the particular channel family, the BP threshold 
depends on the particular code ensemble, i.e. (for the case considered here)
on the code ensemble. 
It constitutes therefore a primary measure of the `goodness' of
such a pair. Given a certain design rate $R$,
one would like to make $p_{\rm d}$ as large as 
possible. 
This has motivated the introduction of code ensembles that generalize
the regular ones studied here (starting from `irregular' ones).
Optimized ensembles have been shown to allow for  exceptionally good
performances. In the case of the erasure channel, they 
allowed to saturate Shannon's fundamental limit \cite{LMSSS97}.
This is an important approach to the design of LDPC ensembles.

Let us finally mention that the BP threshold was defined in 
Eq.~(\ref{eq:BPThresholdDef}) in terms of the bit error rate. One may 
wonder whether a different performance parameter may yield a different 
threshold. As long as such parameter can be written in the form
$\frac{1}{N}\sum_i f(h_i^{(t)})$ this is not the case. More precisely
\begin{eqnarray}
p_{\rm d} =
\sup\left\{\, p\in I\, : \, h^{(t)} \stackrel{{\rm d}}{\to} +\infty\, 
\right\}\, ,
\end{eqnarray}
where, for the sake of generality we assumed the noise parameter to belong 
to an interval $I\subseteq \reals$. In other words, for any $p<p_{\rm d}$
the distribution of BP messages becomes a delta at plus infinity.
%
%
\subsection{Belief Propagation versus MAP Decoding}

So far we have seen that detailed predictions can be
obtained for the performance of LDPC ensembles under message 
passing decoding (at least in the large blocklength limit).
In particular the threshold noise for reliable communication
is determined in terms of a distributional recursion (density evolution). 
This recursion can in turn be efficiently approximated numerically,
leading to accurate predictions for the threshold.

It would be interesting to compare such predictions
with the performances under optimal decoding strategies. Throughout
this section we shall focus on symbol MAP decoding, which 
minimizes the bit error rate,
and consider a generic channel family $\{\BMS(p)\}$ 
ordered\footnote{Such that the channel worsen as $p$ increases. Examples
are the binary symmetric or binary erasure channels.} by the 
noise parameter $p$.

Given an LDPC ensemble, let $\bit^{(N)}$ be the expected bit error rate when 
the blocklength is $N$. 
The \emph{MAP threshold} $p_{\rm c}$
for such an ensemble can be defined as the largest 
(or, more precisely, the supremum) value of $p$ such that 
$\lim_{N\to\infty}\bit^{(N)} = 0$. 
In other words, for any $p<p_{\rm c}$ one can communicate with an 
arbitrarily small error probability, by using a random code from the
 ensemble, provided $N$ is large enough. 

By the optimality of MAP decoding, $p_{\rm d}\le p_{\rm c}$.
In coding theory some techniques have been developed to
prove upper and lower bounds on $p_{\rm c}$ \cite{Gal62,SaS06}. 
In particular it is easy 
to find ensembles for which there exist a gap between the two
thresholds (namely $p_{\rm d}<p_{\rm c}$ strictly).
Consider for instance $(k,l)$ regular ensembles 
with a fixed ratio $l/k = 1-R$. It is then possible to show that,
as $k,l\to\infty$, the BP threshold goes to $0$ while the MAP
threshold approaches the Shannon limit.

This situation is somewhat unsatisfactory. The techniques used
to estimate $p_{\rm d}$ and $p_{\rm c}$ are completely different. This is 
puzzling since the two thresholds can be extremely close and even coincide for 
some ensembles. Furthermore, we know that  $p_{\rm d}\le p_{\rm c}$
by a general argument (optimality of MAP decoding), but this inequality 
is not `built in' the corresponding derivations. Finally, it would be 
interesting to have a sharp estimate for $p_{\rm c}$.

It turns out that a sharp characterization of $p_{\rm c}$ can be 
obtained through statistical mechanics techniques \cite{MurayamaEtAl,MSK02,AM_rce}.
The statistical mechanics result has been proved to be
a rigorous upper bound for general code ensembles, and it is conjectured to be 
tight \cite{Montanari05,MMRU05}.

The starting point is to consider the conditional entropy of the 
channel input $\ux$ given the output $\uy$,
$H_N(\uX|\uY)$. As shown in Eq.~(\ref{eq:EntropyVsPartFun})
this is given by the expectation of the log partition function appearing in 
Eq.~(\ref{eq:ProbLDPC}) (apart from a trivial additive factor).

Let $\hf_N  = \E H_N(\uX|\uY)/N$ denote the entropy density averaged over
 the code ensemble.
Intuitively speaking, this quantity allows to estimate the typical 
number of inputs with non-negligible probability for a given channel output. 
If $\hf_N$ is bounded away from $0$ as $N\to\infty$, the typical
channel output  corresponds to an exponential number of 
(approximately) equally likely inputs.
If on the other hand $\hf_N\to 0$, the correct input has to be searched among
a sub-exponential number of candidates. 
This leads us to identify\footnote{A rigorous justification
of this identification can be obtained using Fano's inequality.} 
the MAP threshold as the largest noise level such that $\hf_N\to 0$ as 
$N\to\infty$.

The Bethe free energy provides a natural way to approximate 
log-partition functions on sparse graphs. It is known to be exact 
if the underlying graph is a tree and its stationary points
are in correspondence with the fixed points of BP. In statistical
physics terms, it is the correct variational formulation for the 
Bethe Peierls approximation. In random systems which are locally tree like, it
is normally thought to provide the  correct $N\to\infty$ limit unless
long range correlations set in. These are in turn described through 
`replica symmetry breaking' (see below).

As many mean field approximations, the Bethe approximation can be thought of as 
a way of writing the free energy as a function of a few correlation functions.
More specifically, one considers the single-variable marginals 
$\{b_i(x_i): \; i\in\{1,\dots,N\}\}$,
and the joint distributions of variables involved in a common check node
$\{ b_a(\ux_a):\; a\in\{1,\dots,M\}\}$.
In the present case the Bethe free energy reads
\begin{eqnarray}
\Fbe(\ub) & = & -
\sum_{i=1}^N\sum_{x_i} b_i(x_i)\log Q(y_i|x_i)+
\label{eq:BetheCode}\\
&&
+\sum_{a=1}^M\sum_{\ux_a}b_a(\ux_a)
\log b_a(\ux_a)-
\sum_{i=1}^N(|\di|-1)\sum_{x_i} b_i(x_i)\log_2  b_i(x_i)\, .\nonumber
\end{eqnarray}
The marginals $\{b_i(\,\cdot\,)\}$,  $\{b_a(\,\cdot\,)\}$ are regarded
as variables. They are constrained to be probability distributions 
(hence non-negative) and to satisfy the marginalization conditions
\begin{eqnarray}
\sum_{x_{j},\, j\in\da\backslash i} b_a(\ux_a) = b_i(x_i)
\;\;\;\;\; \forall i\in\da\, ,\;\;\;\;\;\;\;\;\;\;\;\; 
\sum_{x_i} b_i(x_i)  =  1\;\;\;\;\;\forall i\, .
\label{Marg}
\end{eqnarray}
Further, in order to fulfill the parity-check constraints $b_a(\ux_a) $ 
must be forced to vanish unless 
$x_{i_a(1)}\oplus\cdots\oplus x_{i_a(k)}=\0t$ 
(as usual we use the convention  $0\log 0=0$).
Since they do not necessarily coincide with the actual marginals of 
$\mu(\,\cdot\,)$, the $\{b_a\}$, $\{b_i\}$ are sometimes called \emph{beliefs}.

Approximating the log-partition function $-\log Z(\uy)$ requires minimizing
the Bethe free energy $\Fbe(\ub)$. 
The constraints can be resolved by introducing Lagrange multipliers,
that are in turn expressed in terms of two families of real valued messages 
$\uu \equiv \{u_{a\to i}\}$, $\hu = \{h_{i\to a}\}$. If we denote by 
$P_u(x)$ the distribution of a bit $x$ whose log likelihood ratio
is $u$ (in other words $P_u(0)= 1/(1+e^{-2u})$, $P_u(1)=
e^{-2u}/(1+e^{-2u})$), the resulting beliefs read
\begin{eqnarray}
b_a(\ux_a)  =  \frac{1}{z_a}\, \ind_a(\ux)
\prod_{j\in\da} P_{h_{j\to a}}(x_{j})\, ,\;\;\;
b_i(x_i)  =  \frac{1}{z_i}\, Q(y_i|x_i)\prod_{a\in\di}P_{u_{a\to i}}(x_i)
\, ,\label{eq:BethePar}
\end{eqnarray}
where we introduced the shorthand $\ind_a(\ux)$ to denote the indicator
function for the $a$-th parity check being satisfied.
Using the marginalization conditions (\ref{Marg}) as well as the stationarity
of the Bethe free energy with respect to variations in the beliefs,
one obtains the fixed point BP equations 
\begin{eqnarray}
h_{i\to a}  =  B_i +\sum_{b\in\di\backslash a} u_{b\to i}\, ,
\;\;\;\;\;\;
u_{a\to i} & = & \atanh \left\{\prod_{j\in\da\backslash i}
\tanh h_{j\to a}\right\}\,.\label{eq:BetheSP}
\end{eqnarray}
These in turn coincide for with the fixed point conditions for
belief propagation, cf. Eqs.~(\ref{eq:BPDec_LLR}).

The Bethe free energy can be written as a function of the messages
by plugging the expressions (\ref{eq:BethePar}) into Eq.~(\ref{eq:BetheCode}).
Using the fixed point equations, we get
\begin{eqnarray}
F_{\rm B}(\uu,\hu) & = & \sum_{(ia)\in E}\log 
\left[\sum_{x_i}P_{u_{a\to i}}(x_i)P_{h_{i\to a}}(x_i)\right]- \\
&&\!\!\!\!\!\!\!\!\!\!\!\!\!\!\!\!\!\!\!\!-
\sum_{i=1}^N\log\left[\sum_{x_i}Q(y_i|x_i)\prod_{a\in\di}
P_{u_{a\to i}}(x_i)\right]
-\sum_{a=1}^M\log\left[\sum_{\ux_a}\ind_a(\ux)\prod_{i\in\da}
P_{h_{i\to a}}(x_i)\right]\, .\nonumber
\end{eqnarray}

We are interested in the expectation of this quantity 
with respect to the code and channel realization, in the $N\to\infty$ 
limit. We assume that messages are asymptotically identically distributed 
$u_{a\to i}\ed u$, $h_{i\to a}\ed h$, and that messages incoming in 
the same node along distinct edges are asymptotically independent.
Under these hypotheses we get the limit
\begin{eqnarray}
\lim_{N\to\infty} \frac{1}{N}\E\, \Fbe(\uu,\uh) = -\phi_{u,h}+
\sum_y Q(y|0)\log_2 Q(y|0)\, ,
\end{eqnarray}
where
\begin{eqnarray}
\phi_{u,h} & \equiv & -l\,\E_{u,h}\log_2\left[ \sum_x P_u(x)
P_h(x) \right]+
\E_y\E_{\{u_i\} }\log_2\left[ \sum_x \frac{Q(y|x)}{Q(y,0)}
\prod_{i=1}^l P_{u_i}(x)\right]-
\nonumber\\
&&+\frac{l}{k}\,\E_{\{h_i\} }
\log_2\left[\sum_{x_1\dots x_k}
\ind_a(\ux)\prod_{i=1}^k P_{h_i}(x_i)\right]\,  .
\end{eqnarray}

Notice that the random variables $u$, $h$ are constrained
by Eq.~(\ref{eq:BetheSP}), which must be fulfilled
in distributional sense. In other words $u$, $h$ must form a fixed point 
of the density evolution recursion (\ref{eq:DensityEvolution}).
Given this proviso, if the above assumptions are correct and the
Bethe free energy is a good approximation for the log partition function
one expects the conditional entropy per bit to be 
$\lim_{N\to\infty} \hf_N = \phi_{u,h}$.
This guess is supported by the following rigorous result.
\begin{thm}
If $u$, $h$ are  symmetric random variables satisfying
the distributional identity
$u \ed \atanh \left\{\prod_{i=1}^{k-1}\tanh h_{i}\right\}$, then
\begin{eqnarray}
\lim_{N\to\infty}\hf_N \ge \phi_{u,h} \, .
\end{eqnarray}
\end{thm}

It is natural to conjecture that the correct limit is obtained by optimizing
the above lower bound, i.e.
\begin{eqnarray}
\lim_{N\to\infty}\hf_N = \sup_{u,h}\, \phi_{u,h} \, ,\label{eq:LDPCEntropy}
\end{eqnarray}
where, once again the $\sup$ is taken over the couples of symmetric 
random variables satisfying $u \ed \atanh \left\{\prod_{i=1}^{k-1}
\tanh h_{i}\right\}$. In fact it is easy to show that,
on the fixed point, the distributional equation
$h \ed B+\sum_{a=1}^{l-1}u_a$ must be satisfied as well. 
In other words the couple $u$, $h$ must be a density evolution fixed point.

This conjecture has indeed been proved in the 
case of communication over the binary erasure channel
for a large class of LDPC ensembles (including, for instance,
regular ones).

The expression (\ref{eq:LDPCEntropy}) is interesting because it bridges 
the analysis of BP and MAP decoding. For instance, it is immediate
to show that it implies $p_{\rm d}\le p_{\rm c}$.
\begin{exercise}{4}
This exercise aims at proving the last statement.
\begin{itemize}
\item[$(a)$] 
Recall that $u,h = +\infty$ constitute a density evolution fixed point
for any noise level. Show that $\phi_{h,u}=0$ on such a fixed point.
\item[$(b)$] Assume that, if any other 
fixed point exists, then density evolution converges to it
(this can indeed be proved in great generality).
\item[$(c)$] Deduce that $p_{\rm d}\le p_{\rm c}$.
\end{itemize}
\end{exercise}
\vspace{0.1cm}

Evaluating the expression (\ref{eq:LDPCEntropy}) implies an a priori infinite
dimensional optimization problem. In practice good approximations
can be obtained through the following procedure:
\begin{enumerate}
\item Initialize $h$, $u$ to a couple of symmetric random variables
$h^{(0)}$, $u^{(0)}$. 
\item Implement numerically the density evolution recursion 
(\ref{eq:DensityEvolution}) and iterate it until an approximate 
fixed point is attained.
\item Evaluate the functional $\phi_{u,h}$ on such a fixed point,
after enforcing  $u \ed \atanh \left\{\prod_{i=1}^{k-1}
\tanh h_{i}\right\}$ exactly.
\end{enumerate}
The above procedure can be repeated for several different 
initializations $u^{(0)}$, $h^{(0)}$. The largest of the corresponding
values of  $\phi_{u,h}$ is then picked as an estimate for 
$\lim_{N\to\infty} \hf_N$.

\begin{table}
\begin{center}
\begin{tabular}{ccccc}
\hline
$l$ & $k$ & $R$ & $p_{\rm c}$ & Shannon limit\\
\hline
$3$ & $4$ & $1/4$ & $0.2101(1)$ & $0.2145018$\\
$3$ & $5$ & $2/5$ & $0.1384(1)$ & $0.1461024$\\
$3$ & $6$ & $1/2$ & $0.1010(2)$ & $0.1100279$\\
$4$ & $6$ & $1/3$ & $0.1726(1)$ & $0.1739524$\\
\hline
\end{tabular}
\label{TableMAPThresholds}
\end{center}
\caption{MAP thresholds for a few regular LDPC ensembles and communication
over the BSC$(p)$.}
\end{table}
While his procedure is not guaranteed to exhaust all the 
possible density evolution fixed points, it allows to compute
a sequence of lower bounds to the conditional entropy density.
Further,  one expects a small finite number of density evolution 
fixed points. In particular, for regular ensembles and $p>p_{\rm d}$,  a unique 
(stable) fixed point is expected to exist apart from 
the no-error one $u,h=+\infty$. In Table \ref{TableMAPThresholds}
we present the corresponding MAP thresholds for a few regular ensembles.

For further details on these results, and complete proofs, we refer to 
\cite{Montanari05}. Here we limit ourselves to a brief discussion why the conjecture
(\ref{eq:LDPCEntropy}) is expected to hold from a statistical 
physics point of view. 

The expression (\ref{eq:LDPCEntropy}) corresponds
to the `replica symmetric ansatz' from the present problem.
This usually breaks down if some form of long-range correlation
(`replica symmetry breaking') arises in the measure $\mu(\,\cdot\,)$.
This phenomenon is however not expected to happen in the case
at hand. The technical reason is that the so-called 
Nishimori condition holds for $\mu(\, \cdot\,)$ \cite{AM_rce}. 
This condition generally holds for a large family of problems arising 
in communications and statistical inference. While Nishimori condition
does not provide an easy proof of the conjecture (\ref{eq:LDPCEntropy}),
it implies a series of structural properties of $\mu(\,\cdot\,)$ 
that are commonly regarded as incompatible with replica symmetry breaking.

Replica symmetry breaking is instead necessary to describe the structure
of `metastable states' \cite{DynamicCodes}. This can be loosely described 
as very deep local minima in the energy landscape introduced in
Section~\ref{sec:BitFlipping}. Here `very deep' means that $\Theta(N)$ 
bit flips are necessary to lower the energy (number of unsatisfied parity 
checks) when starting from such minima. As the noise level increases,
such local minima become relevant at the so called `dynamic phase transition'.
It turns out that the critical noise for this phase transition
coincides with the BP threshold $p_{\rm d}$. In other words the double
phase transition at $p_{\rm d}$ and $p_{\rm c}$ is completely analogous
to what happens in the mean field theory of structural glasses (see for
instance Parisi's lectures at this School). Furthermore, this
indicates that $p_{\rm d}$ has a `structural' rather than purely algorithmic
meaning.
%
%
\section{Belief Propagation Beyond Coding Theory}
\label{sec:beyondcoding}

The success of belief propagation as an iterative decoding 
procedure has spurred a lot of interest in its application to other
statistical inference tasks.

A simple formalization for this family of problems is provided by factor
graphs. One is given a factor graph $G=(V,F,E)$ with variable
nodes $V$, function nodes $F$, and edges $E$ and considers 
probability distributions that factorize accordingly
\begin{eqnarray}
\mu(\ux) =\frac{1}{Z}\, \prod_{a\in F} \psi_a(\ux_{\da})\, .\label{eq:GeneralFactor}
\end{eqnarray}
Here the variables $x_i$ take values in a generic finite alphabet
$\cX$, and the \emph{compatibility functions} 
$\psi_a:\cX^{\da}\to\reals_{+}$ encode dependencies among them.
The prototypical problem consists in computing marginals of
the distribution $\mu(\,\cdot\,)$, e.g.,
\begin{eqnarray}
\mu_i(x_i) \equiv \sum_{\ux_{\sim i}}\mu(\ux)\, .
\end{eqnarray}

Belief propagation can be used to accomplish this task in a fast 
and distributed (but not necessarily accurate) fashion. The general 
update rules read
\begin{eqnarray}
\me_{i\to a}^{(t+1)}(x_i)  \normeq  \prod_{b\in\di\backslash a}
\mh^{(t)}_{b\to i}(x_i)\, ,\;\;\;\;\;
\mh_{a\to i}^{(t)}(x_i)  \normeq  \sum_{\{x_j\}}
\psi_a(\ux_{\da})
\prod_{j\in\da\backslash i}\me^{(t)}_{j\to a}(x_j)\, .\label{eq:GenBP}
\end{eqnarray}
Messages are then used to estimate local marginals as follows
\begin{eqnarray}
\overline{\me}_{i}^{(t+1)}(x_i) & \normeq & \prod_{b\in\di}
\mh^{(t)}_{b\to i}(x_i)\, .
\end{eqnarray}
The basic theoretical question is of course to establish a
relation, if any between $\mu_i(\,\cdot\,)$ and $\ome_i(\, \cdot\, )$.

\begin{figure}
\begin{center}
\includegraphics[angle=0,width=0.3\columnwidth]{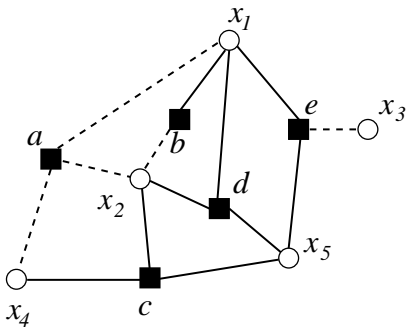}
\caption{Factor graph representation of the satisfiability formula 
(\ref{eq:SATExample}). Circle correspond to variables and squares to 
clauses.
Edges are represented as dashed line if the variable is negated in 
the corresponding clause.}
\label{fig:facgraphsatex}
\end{center}
\end{figure}
As an example, we shall consider 
\emph{satisfiability} \cite{GareyJohnson}.
Given $N$ Boolean variables $x_i$, $i\in \{1,\dots, N\}$, $x_i\in\{
{\rm True},\, {\rm False}\}$, 
a  formula is the logical expression obtained by taking the AND 
of $M$ \emph{clauses}. 
Each clause is the logical OR of a subset of the variables or their negations.
As an example, consider the formula
(here $\ox_i$ denotes the negation of $x_i$)
\begin{eqnarray}
{ ( \ox_1 \vee  \ox_{2} \vee \ox_4) }
{ \wedge }
{ ( x_{1} \vee \ox_2)}
{ \wedge }
{ ( x_2 \vee   x_{4} \vee  x_5) }
{ \wedge }
{ ( x_{1} \vee  x_2 \vee \ox_5)}
{ \wedge }
{ (  x_1 \vee \ox_{3} \vee  x_5) }\, .\label{eq:SATExample}
\end{eqnarray}
An assignment of the $N$ variables satisfies the formula if,
for each of the clause, at least one of the involved \emph{literals}
(i.e. either the variables or their negations) evaluates to True.

A satisfiability formula admits a natural factor graph representation
where each clause is associated to a factor node and each variable 
to a variable node. An example is shown in Fig.~\ref{fig:facgraphsatex}.
Admitting that the formula has at least one satisfying assignment, 
it is natural to associate to a $\cF$ the uniform measure over 
such assignments $\mu_{\cF}(\ux)$. It is easy to realize that such a distribution
takes the form (\ref{eq:GeneralFactor}) where the 
compatibility function $\psi_a(\ux_{\da})$ takes value $1$ if the assignment 
$\ux$ satisfies clause $a$ and $0$ otherwise.

Satisfiability, i.e., the problem of finding a solution to
a satisfiability formula or proving that it is usatisfiable,
is one of the prototypical NP-complete problems.
Given a satisfiability formula, computing marginals with respect to 
the associated distribution $\mu_{\cF}(\,\cdot\,)$ is relevant for tasks such
as counting the number of solutions of $\cF$ or sampling them uniformly.
These are well known \emph{$\#$-P complete} problems.\footnote{The
notation $\#$-P refers to the hardness classification for counting problems.}

The currently best algorithms for solving random instances for the $K$-SAT problem are
based on a variant of BP, which is called {\em survey propagation} \cite{BMZ03,MMW05,BMZ03}.
%
%
\subsection{Proving Correctness through Correlation Decay}

A simple trick to bound the error incurred by BP consists 
in using the correlation decay properties \cite{AldousSteele,TatikondaJordan} 
of the measure $\mu(\,\cdot\,)$.
Let $i\in \{1,\dots,N\}$ be a variable index and denote by
$\ux_{t}$ the set of variables sitting on nodes at  distance $t$ 
from $i$. 
Further, denote by $\ux_{\ge t}$ the set of variables whose distance 
from $i$ is \emph{at least} $t$.
Then the local structure of the probability distribution
(\ref{eq:GeneralFactor})
\begin{eqnarray}
\mu_i(x_i) = \sum_{\ux_{\ge t}}\mu(x_i|\ux_{\ge t})\mu(\ux_{\ge t})
=\sum_{\ux_t}\mu(x_i|\ux_t)\mu(\ux_t)\, .\label{eq:Conditional}
\end{eqnarray}
Let $\ball_i(t)$ denote the subgraph induced by nodes whose distance from 
$i$ is \emph{at most} $t$, and $\partial\ball_i(t)$ its boundary (nodes whose
distance from $i$ is exactly $t$). Further, for any $j\in\partial\ball_i(t)$
let $a(j)$ be the unique function nodes inside $\ball_i(t)$ that is 
adjacent to $j$. It is intuitively clear that belief propagation
computes the marginal at $i$ \emph{as if} the graph did not extend beyond
$\ball_i(t)$. More precisely, if the initial condition 
$\me_{i\to a}^{(0)}(x_i)$ is properly normalized, then we have the exact 
expression
\begin{eqnarray}
\ome_i^{(t)}(x_i) = \sum_{\ux_t}\mu(x_i|\ux_t)\prod_{j\in\partial
\ball_i(t)}\me^{(0)}_{j\to a(j)}(x_j)\, .\label{eq:MPTree}
\end{eqnarray}
As a consequence of Eq.~(\ref{eq:Conditional}) and (\ref{eq:MPTree})
we have 
\begin{eqnarray}
|\mu_i(x_i)-\ome_i^{(t)}(x_i)|\le \sup_{\ux_t,\ux_t'}
|\mu(x_i|\ux_t)-\mu(x_i|\ux'_t)|\, .
\end{eqnarray}
This provides an upper bound on the error incurred by BP when
computing the marginal of $x_i$ base on the local structure of the underlying graph
in terms of the influence of far away variables.
To make things fully explicit, assume that 
the graph has \emph{girth}\footnote{Recall that the girth of
a graph is the length of its shortest cycle.} $g$ and
that $ \sup_{\ux_t,\ux_t'}
|\mu(x_i|\ux_t)-\mu(x_i|\ux'_t)|\le \exp(-\kappa t)$
for some positive $\kappa$. This implies
\begin{eqnarray}
|\mu_i(x_i)-\ome_i^{(t)}(x_i)|\le e^{-\kappa g/2}\, .
\end{eqnarray}

As an example of such error estimates, we shall consider 
\emph{random $k$-satisfiability} \cite{FrancoPaull}.
This is a standard model to generate `synthetic' satisfiability 
formulae. It amounts to picking a formula uniformly at random among
all the ones including $N$ variables and $M=N\alpha$
$k$-clauses (a $k$-clause is a clause that involve \emph{exactly}
$k$ distinct variables).
We shall of course limit to $k\ge 2$, the case $k=1$ being
trivial. 

Consider a uniformly random variable node in the factor
graph associated to a random formula, and its depth-$t$
neighborhood $\ball_i(t)$.
Proceeding as in the previous section it is not hard to show that,
for any fixed $t$, $\ball_i(t)$ is with high probability 
(as $N\to\infty$) a tree.
 An appropriate model the distribution of such a tree, is given by the
tree ensemble $\T_*(t)$ described as follows.
 For $t=0$,  it is the graph containing 
a unique variable node. For any $t \ge 1$, start by a single variable node 
(the root) and add $l\ed \poisson(k\alpha)$ clauses, each one 
including the root, and $k-1$ new variables (first generation variables). 
For each one of the $l$ clauses, the corresponding literals are
non-negated or negated independently with equal probability. If $t \ge 2$, 
generate an independent copy of $\T_*(t-1)$ for each variable node
in the first generation and attach it to them. 

Assume that, for a typical random tree formula $\T(t)$, the marginal 
distribution of the variable at the root is weakly dependent on the 
values assigned at the boundary.
Following the above lines, one can use this fact to prove that BP
computes good approximations for the marginals in a random $k$-SAT 
formula. In fact it turns out that an estimate of the 
form\footnote{Here the $\sup$ is taken over assignments $\ux_t$,
$\ux'_t$ that can be extended to solutions of $\T(t)$.}
\begin{eqnarray}
\E_{\T(t)}\sup_{\ux_t,\ux_t'}
|\mu(x_i|\ux_t)-\mu(x_i|\ux'_t)|\le e^{-\kappa t}\, \label{eq:DecaySAT}
\end{eqnarray}
can be proved if the clause density $\alpha$ stays below a threshold
$\alpha_{\rm u}(k)$ that is estimated to behave as 
$\alpha_{\rm u}(k)=\frac{2\log k}{k}[1+o_k(1)]$.

While we refer to the original paper \cite{MontanariShah}
for the details of the proof
we limit ourselves to noticing that the left hand side
of Eq.~(\ref{eq:DecaySAT}) can be estimated efficiently using
a density evolution procedure. This allows to estimate
the threshold $\alpha_{\rm u}(k)$ numerically.
Consider in fact the log-likelihood (here we are identifying
$\{{\rm True}, {\rm False}\}$ with $\{+1,-1\}$)
\begin{eqnarray}
h^{(t)}(\ux_t) = \frac{1}{2}\log\, \frac{\mu(+1|\ux_t)}{\mu(-1|\ux_t)} \, .
\end{eqnarray}
This quantity depends on the assignment of the variables on the
boundary, $\ux_t$. Since we are interested on a \emph{uniform} bound over the 
boundary, let us consider the extreme cases
\begin{eqnarray}
\overline{h}^{(t)} = \max_{\ux_t}\, h^{(t)}(\ux_t) \, ,\;\;\;\;\;\;\;\;\;\;\;
\underline{h}^{(t)} = \min_{\ux_t}\, h^{(t)}(\ux_t) \, .
\end{eqnarray}
It is then possible to show that the couple $(\overline{h}^{(t)},\underline{h}^{(t)})$ obeys a recursive distributional equation that, as mentioned,
can be efficiently implemented numerically.
%
%

\section{Belief Propagation Beyond the Binary Symmetric Channel}
\label{sec:beyondbms}
So far we have considered mainly the case of transmission over 
BMS channels, our reference example being the BSC.
There are many other channel models that are important and 
are encountered in practical situations. Fortunately, it is relatively straightforward
to extend the previous techniques and statements to a much larger class, and we review
a few such instances in this section. 

%
%
\subsection{Binary Memoryless Symmetric Channels}

In order to keep the notation simple, we assumed channel output to 
belong to a finite alphabet $\cA$.
In our main example, the BSC, we had $\cA = \{\0t,\1t\}$.
But in fact all results are valid for a wider class 
of {\em binary memoryless symmetric (BMS) channels}.
One can prove that there is no loss of generality
in assuming the output alphabet to be the real line $\reals$ 
(eventually completed with $\overline{\reals}=\reals\cup\{\pm \infty\}$).

Let $\uy = (y_1,\dots,y_N)$ be the  vector of channel outputs
on input $\ux = (x_1,\dots,x_N)$. For a BMS the input is binary, i.e.
$\ux\in\{\0t,\1t\}^N$. Further the channel is \emph{memoryless},
i.e. the probability density
of getting $\uy\in\overline{\reals}^N$ at output when the input is
$\ux$, is 
\begin{align*}
Q(\uy | \ux) & = \prod_{t=1}^N Q(y_t | x_t).
\end{align*}
Finally, the \emph{symmetry} property can be written without loss
of generality, as $Q(y_t | x_t=\1t)=Q(-y_t | 
x_t = \0t)$.

One of the most important elements in this class is the
{\em additive white Gaussian noise} (AWGN) channel, defined by
\begin{align*}
y_t & = x_t + z_t\, ,\;\;\;\;\;\;  t\in\{1,\dots,N\}\, ,
\end{align*}
where the sequence $\{z_t\}$ is i.i.d. consisting of Gaussian random
variables with mean zero and variance $\sigma^2$. It is common in this setting to
let $x_i$ take values in $\{+ 1,-1\}$ instead of $\{\0t, \1t\}$ as we have assumed so far. 
The AWGNC transition probability density function is therefore
\begin{eqnarray*}
Q(y_t|x_t) = \frac{1}{\sqrt{2\pi\sigma^2}}\; e^{-\frac{(y-x)^2}{2\sigma^2}}\, .
\end{eqnarray*}
The AWGNC is the basic model of transmission of an electrical signal over a cable (here the
noise is due to thermal noise in the receiver) and
it is also a good model of a wireless channel in free space (e.g., transmission from a satellite).

Although the class of BMS channels is already fairly large, it is important in practice to
go beyond it. The extension to the non-binary case is quite straightforward and so we will
not discuss it in detail. The extension to channels with memory or the asymmetric case
are more interesting and so we present them in the subsequent two sections.
%
%
\subsection{Channels With Memory}
\label{sec:channelwithmemory}
Loosely speaking, in a memoryless channel the channel acts on each transmitted bit independently.
In a channel with memory, on the other hand, the channel acts generally on the whole block
of input bits together. An important special case of a channel with memory is if the
channel can be modeled as a Markov chain, taking on a sequence of ``channel states.''
Many physical channels posses this property
and under this condition the message-passing approach can still be applied.
For channels with memory there are two problems. First, we need to determine the
capacity of the channel. Second, we need to devise efficient coding schemes
that achieve rates close to this capacity. It turns out that both problems can 
be addressed in a fairly similar framework. Rather than discussing the general case
we will look at a simple but typical example.   

Let us start by computing the information rate/capacity of channels with memory, assuming
that the channel has a Markov structure. As we discussed in Section~\ref{sec:shannontheorem}
in the setting of BMS channels, the channel 
capacity can be expressed as the difference of two entropies, namely 
as $H(X)-H(X| Y)$. Here, $X$ denotes the binary input and $Y$ denotes
the observation at the output of the channel whose input is $X$. 
Given two random variables $X$, $Y$, this entropy difference is
called the {\em mutual information} and is typically denoted by 
$I(X; Y) = H(X)-H(X;Y)$. 

A general channel, is defined by a channel transition probability
$Q(y_1^N|x_1^N)$ (here and below $x_1^N$ denotes the vector
$(x_1,\dots,x_N)$). In order to define a joint distribution of the 
input and output vectors, we have to prescribe a distribution on the 
channel input, call it $p(x_1^N)$. The channel capacity 
is obtained by maximizing the mutual information over 
all possible distributions of the input, and eventually taking the $N\to\infty$
limit. In formulae
\begin{eqnarray*}
\capac(Q) = \lim_{N\to\infty}\sup_{p(\,\cdot\,)}
\, I(X_{1}^N;Y_1^N)/N\, .
\end{eqnarray*}
For BMS channels
it is possible to show that the maximum occurs for the uniform prior:
$p(x_1^N) = 1/2^N$. Under this distribution, $I(X_{1}^N;Y_1^N)/N$
is easily seen not to depend on $N$ and we recover the expression in 
Sec.~\ref{sec:shannontheorem}.

For channels with memory we have to maximize the mutual information over
all possible distributions over $\{\0t,\1t\}^N$ (a space whose dimension is
exponential in $N$), and take the limit $N\to\infty$.
An easier task is to choose a convenient input distribution $p(\,\cdot\,)$
and then compute the corresponding mutual information in the 
$N\to\infty$ limit:
\begin{align}
I & =   \lim_{N \rightarrow \infty} I(X_1^N; Y_1^N)/N\, .\label{eq:LimitMutualInfo}
\end{align}
Remarkably, this quantity has an important operational meaning.
It is the largest rate at which we can transmit reliably across the channel 
using a coding scheme such that the resulting input distribution 
matches $p(\,\cdot\,)$.

\begin{figure}[htp]
\centering
\setlength{\unitlength}{1bp}%
\begin{picture}(0,0)
\includegraphics[scale=1.0]{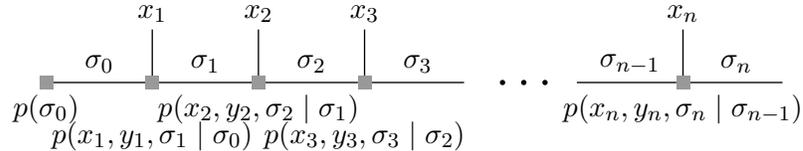}
\end{picture}%
\begin{picture}(300,50)
\put(40, 27){\makebox(0,0)[c]{$\sigma_0$}}
\put(80, 27){\makebox(0,0)[c]{$\sigma_1$}}
\put(120, 27){\makebox(0,0)[c]{$\sigma_2$}}
\put(160, 27){\makebox(0,0)[c]{$\sigma_3$}}
\put(240, 27){\makebox(0,0)[c]{$\sigma_{n-1}$}}
\put(280, 27){\makebox(0,0)[c]{$\sigma_n$}}

\put(60, 46){\makebox(0,0)[c]{$x_1$}}
\put(100, 46){\makebox(0,0)[c]{$x_2$}}
\put(140, 46){\makebox(0,0)[c]{$x_3$}}
\put(260, 46){\makebox(0,0)[c]{$x_n$}}

\put(20, 10){\makebox(0,0)[c]{$p(\sigma_0)$}}
\put(60, 0){\makebox(0,0)[c]{$p(x_1, y_1, \sigma_{1} \mid \sigma_0)$}}
\put(100, 10){\makebox(0,0)[c]{$p(x_2, y_2, \sigma_{2} \mid \sigma_1)$}}
\put(140, 0){\makebox(0,0)[c]{$p(x_3, y_3, \sigma_{3} \mid \sigma_2)$}}
\put(260, 10){\makebox(0,0)[c]{$p(x_n, y_n, \sigma_{n} \mid \sigma_{n-1})$}}
\end{picture}
\caption{\label{fig:channelswithmemoryfsfg}
The FSFG corresponding to (\ref{equ:channelcapfactorization}).}
\end{figure}

To be definite, assume that the channel is defined by a state sequence 
$\{\sigma_{t}\}_{t \geq 0}$, taking values in a finite alphabet, such 
that the joint probability distribution
factors in the form
\begin{equation} \label{equ:channelcapfactorization}
p(x_1^n, y_1^n, \sigma_0^n) =
p(\sigma_0) \prod_{i=1}^{n} 
p(x_i, y_i, \sigma_i \mid \sigma_{i-1}).
\end{equation}
We will further assume that the transmitted bits $(x_1,\cdot,x_N)$
are iid uniform in $\{\0t,\1t\}$.
The factor graph corresponding to (\ref{equ:channelcapfactorization}) is shown in Fig.~\ref{fig:channelswithmemoryfsfg}.
It is drawn in a somewhat different way compared to the factor graphs we have seen so far.
Note that in the standard factor graph corresponding to this factorization 
all variable nodes have degree two. In such a case it is convenient not to draw the factor
graph as a bipartite graph but as a standard graph in which the nodes correspond to the factor nodes
and the edges correspond to the variable nodes (which have degree two and therefore connect exactly two factors).
Such a graphical representation is also known as {\em normal} graph or as Forney-style factor graph (FSFG), in honor
of Dave Forney who introduced them \cite{For01}.
Let us now look at a concrete example.
\begin{example}{5}[Gilbert-Elliott Channel]
\label{exa:isichannel}
The Gilbert-Elliot channel is a model for a {\em fading} channel, i.e., a channel where 
the quality of the channel is varying over time. In this model we assume that
the channel quality is evolving according to a Markov chain.
In the simplest
case there are exactly two states, and this is the original Gilbert-Elliott channel (GEC) model.
More precisely, consider the two-state Markov chain depicted in Fig.~\ref{fig:gemodel}.
Assume that $\{X_{t}\}_{t \geq 1}$
is i.i.d., taking values in $\{ \pm 1\}$ with uniform probability.
\begin{figure}[htp]
\centering
\setlength{\unitlength}{1.0bp}%
\begin{picture}(180,50)
\put(0,0){\includegraphics[scale=1.0]{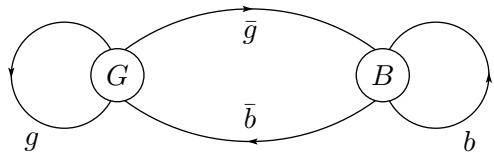}}
{
\put(40,25){\makebox(0,0){$G$}}
\put(140,25){\makebox(0,0){$B$}}
\put(5,0){\makebox(0,0)[l]{$g$}}
\put(90,45){\makebox(0,0)[t]{$\bar{g}$}}
\put(175,0){\makebox(0,0)[r]{$b$}}
\put(90,5){\makebox(0,0)[b]{$\bar{b}$}}
}
\end{picture}
\caption{\label{fig:gemodel} The Gilbert-Elliott channel with two states.}
\end{figure}

The channel is either in a {\em good} state,
denote it by $G$, or in a {\em bad} state, call it $B$. In either state the channel is a BSC.
Let the crossover probability in the good state be $\epsilon_{\text{G}}$
and in the bad state be $\epsilon_{\text{B}}$, with
$0 \leq \epsilon_{\text{G}} < 
\epsilon_{\text{B}} \leq 1/2$. Let $P$ be the $2 \times 2$ matrix
\begin{align*}
P
= 
\left(
\begin{array}{cc}
g & \bar{b} \\
\bar{g}   & b 
\end{array}
\right)
\end{align*}
which encodes the transition probabilities between the states (the columns
indicate the present state and the rows the next state). Define the {\em steady
state} probability vector
$p=(p_{\text{G}}, p_{\text{B}})$, i.e., the vector which
fulfills $P p^T= p^T$. This means that in steady state the system spends
a fraction $p_{\text{G}}$
of the time in state $G$
and a fraction $p_{\text{B}}$ of the time in state $B$.
If we consider e.g. the state $G$ then the detailed balance condition reads
$p_{\text{G}} \bar{g}=p_{\text{B}} \bar{b}$.
From this we get
$p=(\bar{b}/(\bar{g}+\bar{b}), \bar{g}/(\bar{g}+\bar{b}))$.
More generally, let us assume that we have $s$ states, $s \in \naturals$,
and that the channel in state $i$, $i \in [s]$, is the BSC$(\epsilon_i)$.
Let $P$ be the $s \times s$ matrix encoding the transition probabilities
between these states. Let $p$ denote the steady-state probability distribution
vector. If $(I-P^T+E)$ is invertible then a direct check shows that
$p=e(I-P^T+E)^{-1}$, where $e$ is the all-one vector of length $s$,
$I$ is the $s \times s$ identity matrix and $E$ is the $s \times s$ all-one matrix.

Note that the state sequence is ergodic as long as the Markov
chain is irreducible  
(i.e. there is a path of strictly positive probability
from any state to any other state) and aperiodic
(i.e. there there exists such a path for any number of steps large enough).  
In the original Gilbert-Elliot model this is true
as long as $0 < g, b < 1$.
\end{example}
Consider the computation of the maximal rate at which we can transmit reliably.
We have
\begin{align*}
I(X_1^N; Y_1^N) & =  H(Y_1^N) - H(Y_1^N | X_1^N).
\end{align*}
Let us see how we can compute $\lim_{N \rightarrow \infty} H(Y_1^N)/N$.
Because of the ergodicity assumption on the state sequence, $-\frac{1}{N}\log p(y_1^N)$ converges 
with probability one to $\lim_{N \rightarrow \infty} H(Y_1^N)/N$. It follows that if we
can compute $-\frac{1}{N}\log p(y_1^N)$ for a very large sequence, then with
high probability the value will be close to the desired entropy rate. Instead of computing
$p(y_1^N)$, let us compute $p(\sigma_N, y_1^N)$. From this we trivially
get our desired quantity by summing,
\[
p(y_1^N) = \sum_{\sigma_N} p(\sigma_N, y_1^N).
\]
Note that
\begin{align}
p(\sigma_N, y_1^N) & =  \sum_{x_N, \sigma_{N-1}} 
p(x_N, \sigma_{N-1}, \sigma_N, y_1^N) \nonumber \\
& =  \sum_{x_N, \sigma_{N-1}} 
\underbrace{p(x_N, \sigma_N, y_N \mid \sigma_{N-1})}_{\text{kernel}}
\underbrace{p(\sigma_{N-1}, y_1^{N-1})}_{\text{message}}.
\label{equ:recursion}
\end{align}
From this we see that $p(\sigma_N, y_1^N)$ can be computed recursively. In fact this recursion
corresponds to running the BP message-passing rules on the factor graph depicted in Fig.~\ref{fig:channelswithmemoryfsfg} (which is a tree):
denote the message which is passed 
along the edge labeled by $\sigma_N$ by  $\me_N(\sigma_N)$. Then according to the BP
message-passing rules we have
\[
\me_N(\sigma_N) 
= \sum_{x_N, \sigma_{N-1}}
p(x_N, \sigma_N, y_N \mid \sigma_{N-1})\,
\me_{N-1}(\sigma_{N-1}).
\]
If we compare this to the recursion stated in (\ref{equ:recursion}) we see that these two
recursions are identical.
In other words,
$\me_N(\sigma_N)=
p(\sigma_N, y_1^N)$, so that
\begin{align}
\label{equ:capwithmemory}
\lim_{N \rightarrow \infty} H(Y_1^N)/N = 
-\lim_{N \rightarrow \infty} \log \Bigl(\sum_{\sigma_N} \me_N(\sigma_N)\Bigr)/N.
\end{align}
From a practical perspective it is typically more convenient to pass
{\em normalized} messages $\tilde{\me}_N(\sigma_N)$
so that $\sum_{\sigma} \tilde{\me}_N(\sigma_N)=1$.
The first message $\me_0(\sigma_0) = p(\sigma_0)$
is already a probability distribution and, hence, normalized,
$\tilde{\me}_0(\sigma_0)=\me_0(\sigma_0)$.
Compute $\me_1(\sigma_1)$ and let
$\lambda_1 = \sum_{\sigma_1} \me_1(\sigma_1)$.
Define $\tilde{\me}_1(\sigma_1)=\me_1(\sigma_1)/\lambda_1$.
Now note that by definition of the message-passing rules all subsequent
messages in the case of rescaling differ from the messages which
are sent in the unscaled case only by this scale factor.
Therefore, if $\lambda_i$ denotes the normalization constant by which we have
to divide at step $i$ so as to normalize the message then
$\tilde{\me}_N(\sigma_N)=
\me_N(\sigma_N)/(\prod_{i=1}^{N} \lambda_i)$.
It follows that
\begin{align*}
\lim_{N \rightarrow \infty} H(Y_1^N)/N & =
-\lim_{N \rightarrow \infty} 
\log\Bigl(\sum_{\sigma_N} 
\alpha_{N}(\sigma_N)\Bigr)/N
\\
& = -\lim_{N \rightarrow \infty} 
\log\Bigl(\bigl(\prod_{i=1}^{N} \lambda_i \bigr) \sum_{\sigma_N} \tilde{\alpha}_{N}(\sigma_N)\Bigr)/N
=\lim_{N \rightarrow \infty} \Bigl(\sum_{i=1}^N \log(\lambda_i) \Bigr)/N.
\end{align*}

It remains to compute $H(Y_1^N \mid X_1^N)$.
We write $H(Y_1^N \mid X_1^N)/N=H(Y_1^N, X_1^N)/N-H(X_1^N)/N$. 
The second part is trivial since
the inputs are i.i.d. by assumption so that $H(X_1^N)/N=1$. For the term
$H(Y_1^N, X_1^N)/N$ we use the same technique as for the computation of $H(Y_1^N)/N$.
Because of the ergodicity assumption on the state sequence, $-\frac{1}{N}\log p(y_1^N, x_1^N)$ converges
with probability one to $\lim_{N \rightarrow \infty} H(Y_1^N, X_1^N)/N$. We write
$p(y_1^N, x_1^N)=\sum_{\sigma_N} p(\sigma_N, y_1^N, x_1^N)$
and use the factorization
\begin{align*}
p(\sigma_N, y_1^N, x_1^N) = &  \sum_{\sigma_{N-1}}
p(\sigma_{N-1}, \sigma_N, y_1^N, x_1^N) \\
= &   \sum_{\sigma_{N-1}}
\underbrace{p(x_N, \sigma_N, y_N \mid \sigma_{N-1})}_{\text{kernel}} \cdot 
 \phantom{\sum_{\sigma_{N-1}}} \underbrace{p(\sigma_{N-1}, y_1^{N-1}, x_1^{N-1})}_{\text{message}}.
\end{align*}
In words, we generate a random instance $X_1^N$ and $Y_1^N$
and run the BP algorithm on the FSFG shown in Fig.~\ref{fig:channelswithmemoryfsfg}
assuming that {\em both} $Y_1^N$ and $X_1^N$ are `quenched.' Taking the logarithm, multiplying by minus one
and normalizing by $1/N$ gives us an estimate of the desired entropy.

Now that we can compute the maximal rate at which we can transmit reliably, let us consider coding.
The symbol MAP decoder is
\begin{align*}
\xh_i(\uy)
& =  \text{argmax}_{x_i}
p(x_i \mid y_1^N) \\
& =  \text{argmax}_{x_i}
\sum_{\{x_j,\,j\neq i\}} p(x_1^N, y_1^N, \sigma_0^N) \\
& =  \text{argmax}_{x_i} 
\sum_{\{x_j,\,j\neq i\}} p(\sigma_0) \prod_{j=1}^{N} 
p(x_j, y_j, \sigma_j \mid \sigma_{j-1}) 
\ind_{\Code}(x_1^N).
\end{align*}
In words, the FSFG in Fig.~\ref{fig:channelswithmemoryfsfg} describes also the
factorization for the message-passing decoder if we add to it the factor nodes describing
the definition of the code.
As always, this factor graph together with
the initial messages stemming from the channel completely specify the message-passing
rules, except for the message-passing schedule. Let us agree that
we alternate one round of decoding with one round of channel estimation. No claim as
to the optimality of this scheduling rule is made.

Notice that the correlations induced by the markovian structure of the channel
are in general short ranged in time. This is analogous to what happens with a 
one-dimensional spin model, whose correlation length is always finite
(at non-zero temperature). A good approximation to the above message passing
schedule is therefore obtained by a `windowed' decodes. 
This means that the state at time $t$ is estimated only of the basis of observations between time $t-R$ and $t+R$, for some finite $R$.
 
Assuming windowed decoding for channel estimation, it is not hard to 
show that after a fixed number of iterations, the decoding neighborhood
is again asymptotically tree-like.
In the case of the GEC the channel symmetry can be used to reduce
to the all-zero codeword. 
Therefore,  we
can employ the technique of density evolution to determine thresholds
and to optimize the ensembles.
\begin{example}{6}[GEC: State Estimation]
\label{exa:gec}
\begin{figure}[ht]
\centering
\setlength{\unitlength}{1.0bp}%
\begin{picture}(253,318)(-13,-8)
\put(0,0)
{
\graphtextsize
\put(0,0){\includegraphics[scale=1.0]{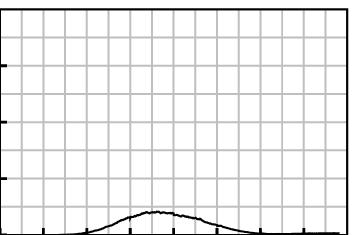}}
\multiputlist(-12,16.25)(0,16.25)[l]{$0.1$,$0.2$,$0.3$}
\multiputlist(0,-8)(12.5,0)[b]{$\text{-}10$,$0$,$10$,$20$,$30$,$40$,$50$,$60$,$70$}
}
\put(140,0)
{
\graphtextsize
\put(0,0){\includegraphics[scale=1.0]{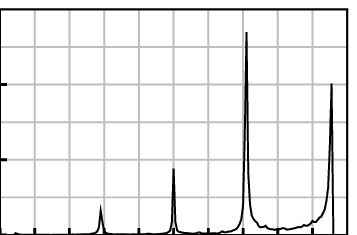}}
\multiputlist(-6,21.6667)(0,21.6667)[l]{$1$,$2$}
\multiputlist(0,-8)(10,0)[b]{$\text{-}5$,$\text{-}4$,$\text{-}3$,$\text{-}2$,$\text{-}1$,$0$,$1$,$2$,$3$,$4$,$5$}
}
\put(0,80)
{
\graphtextsize
\put(0,0){\includegraphics[scale=1.0]{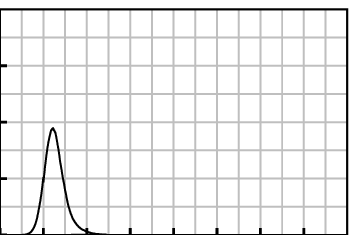}}
\multiputlist(-12,16.25)(0,16.25)[l]{$0.1$,$0.2$,$0.3$}
\multiputlist(0,-8)(12.5,0)[b]{$\text{-}10$,$0$,$10$,$20$,$30$,$40$,$50$,$60$,$70$}
}
\put(140,80)
{
\graphtextsize
\put(0,0){\includegraphics[scale=1.0]{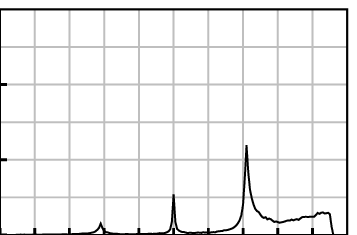}}
\multiputlist(-6,21.6667)(0,21.6667)[l]{$1$,$2$}
\multiputlist(0,-8)(10,0)[b]{$\text{-}5$,$\text{-}4$,$\text{-}3$,$\text{-}2$,$\text{-}1$,$0$,$1$,$2$,$3$,$4$,$5$}
}
\put(0,160)
{
\graphtextsize
\put(0,0){\includegraphics[scale=1.0]{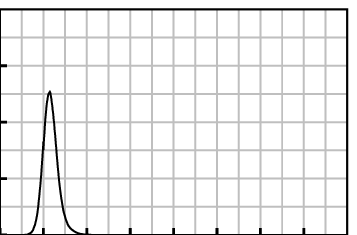}}
\multiputlist(-12,16.25)(0,16.25)[l]{$0.1$,$0.2$,$0.3$}
\multiputlist(0,-8)(12.5,0)[b]{$\text{-}10$,$0$,$10$,$20$,$30$,$40$,$50$,$60$,$70$}
}
\put(140,160)
{
\graphtextsize
\put(0,0){\includegraphics[scale=1.0]{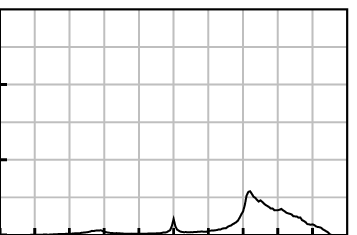}}
\multiputlist(-6,21.6667)(0,21.6667)[l]{$1$,$2$}
\multiputlist(0,-8)(10,0)[b]{$\text{-}5$,$\text{-}4$,$\text{-}3$,$\text{-}2$,$\text{-}1$,$0$,$1$,$2$,$3$,$4$,$5$}
}
\put(0,240)
{
\graphtextsize
\put(0,0){\includegraphics[scale=1.0]{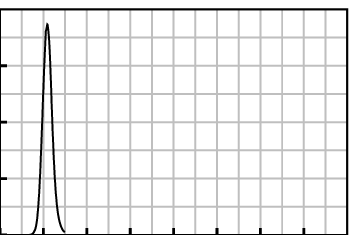}}
\multiputlist(-12,16.25)(0,16.25)[l]{$0.1$,$0.2$,$0.3$}
\multiputlist(0,-8)(12.5,0)[b]{$\text{-}10$,$0$,$10$,$20$,$30$,$40$,$50$,$60$,$70$}
}
\put(140,240)
{
\graphtextsize
\put(0,0){\includegraphics[scale=1.0]{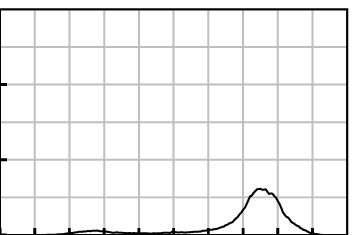}}
\multiputlist(-6,21.6667)(0,21.6667)[l]{$1$,$2$}
\multiputlist(0,-8)(10,0)[b]{$\text{-}5$,$\text{-}4$,$\text{-}3$,$\text{-}2$,$\text{-}1$,$0$,$1$,$2$,$3$,$4$,$5$}
}
\end{picture} 
\caption{Density evolution for the GEC at iteration 1, 2, 4, and 10.
The left pictures show the densities of the messages which are passed
from the code towards the part of the FSFG which estimates the channel state.
The right hand side shows the density of the messages which are the estimates of
the channel state and which are passed to the FSFG corresponding to
the code. \label{fig:deforgec}}
\end{figure}
For the case of transmission over the GEC the iterative decoder
implicitly also estimates the state of the channel.
Let us demonstrate this by means of the following example.
We pick a GEC with three states. Let
\begin{align*}
P
=
\left(
\begin{array}{ccc}
0.99 & 0.005 & 0.02 \\
0.005 & 0.99 & 0.02 \\
0.005 & 0.005 & 0.96
\end{array}
\right),
\end{align*}
which has a steady state probability vector of $p \approx (0.4444, 0.4444, 0.1112)$.
Finally, let the channel parameters of the BSC in these three states be
$(\epsilon_1, \epsilon_2, \epsilon_3) \approx
(0.01, 0.11, 0.5)$. This corresponds to an {\em average} error probability of
$\epsilon_{\text{avg}} = \sum_{i=1}^{3} p_i \epsilon_i \approx 0.108889$.
Using the methods described above, the capacity of this channel (assuming
uniform inputs) can be computed
to be equal to $C \approx 0.583$ bits per channel use. This is markedly higher than
$1-h(\epsilon_{\text{avg}}) \approx 0.503444$, which is the capacity
of the BSC$(\epsilon_{\text{avg}})$. The last channel is
the channel which we experience if we ignore the Markov structure.

Fig.~\ref{fig:deforgec} shows the evolution of the densities
for an optimized ensemble of rate $r \approx 0.5498$. 
The pictures on the right correspond to the messages which are passed from 
the part of the factor graph which estimates the state towards the part of the
factor graph which describes the code. These messages therefore can be interpreted
as the current estimate of the state the channel is in at a given point in time.
Note that after 10 iterations 5 clear peaks emerge.
These peaks are at $\pm \log(0.99/0.01) \approx \pm 4.595$,
$\pm \log(0.9/0.1) \approx \pm 2.197$, $\pm \log(0.5/0.5)=0$.
They correspond to the received likelihoods in the three possible channel states.
In other words, the emergence of the peaks shows that at this stage the
system has identified the channel states with high reliability.
This is quite pleasing. Although the channel state is not known to the receiver and can
not be observed directly, in the region where the iterative decoder works reliably it
also automatically estimates the channel state with high confidence. 
\end{example}

Although we only looked a very particular example it is quite typical of the general
situation: as long as the channel memory can be described by a Markov chain the factor
graph approach applies and we can use message-passing schemes to construct efficient
coding schemes \cite{DPJDCG,BKH97,GaV97,GFV98a,KiE03,Neu04}.
%
%
\subsection{Asymmetric Channels - The {Z} Channel}
\label{sec:zc}
Let us now consider the second generalization, namely the case of {\em non-symmetric} channels.

Consider the channel depicted on the right of Fig.~\ref{fig:Channels}. For obvious
reasons it is called the Z channel (ZC).
This channel has binary input and it is memoryless but it is {\em not} symmetric.
Nevertheless,
essentially the same type of analysis which we performed in
Section~\ref{sec:decodingproblem} can be applied to this case as well. Symmetry is
therefore a {\em nice} property to have but it is {\em not essential}.

\begin{figure}[htp]
\centering
\setlength{\unitlength}{1bp}%
\begin{picture}(153,83)(-13,-8)
\put(0,0){\includegraphics[scale=1.0]{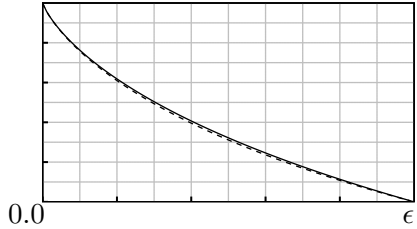}}
{
\put(-13,-8){\makebox(0,0)[lb]{$0.0$}}
\put(140,-8){\makebox(0,0)[rb]{$\epsilon$}}
}
\end{picture}
\caption{\label{fig:zccap}
Comparison of $C_{\text{ZC}(p)}$ (solid curve)
with $I_{\alpha=\frac{1}{2}}(X; Y)$ (dashed curve), both measured in bits.}
\end{figure}

Consider the capacity of this channel. Since the channel is not symmetric the capacity
is not necessarily given by the mutual information between channel input and channel output for
a uniform input distribution of the input. We must instead maximize the
mutual information over the input distribution.
Since the channel is memoryless, it can be assumed that the input is given by a sequence of i.i.d. bernoulli variables.
Assuming that $p(x_i=\0t)=\alpha$, the output distribution is
\begin{align*}
(p(y_i=\0t), p(y_i=\1t)) & =  (\alpha \bar{p}, 1-\alpha \bar{p}),
\end{align*}
so that the mutual information $I_{\alpha}(X; Y)$ for a fixed $\alpha$ is equal to
\begin{equation} \label{equ:zcinformationrate}
I_{\alpha}(X; Y) = H(Y)-H(Y \mid X) = h(\alpha \bar{p})-\alpha h(p).
\end{equation}
Some calculus reveals that the optimal choice of $\alpha$ is
\begin{equation} \label{equ:zcoptimalalpha}
\alpha(p) = \frac{p^{p/\bar{p}}}{1+\bar{p} p^{p/\bar{p}}},
\end{equation}
so that
\[
C_{\text{ZC}(p)} = h(\alpha(p) \bar{p})-\alpha(p) h(p).
\]
Fig.~\ref{fig:zccap} compares $C_{\text{ZC}(p)}$ with
$I_{\alpha=\frac{1}{2}}(X; Y)$, i.e., it compares the capacity with
the transmission rate which is achievable with {\em uniform} input distribution.
This is important and surprising --
only little is lost by insisting on an uniform input distribution:
the rate which is achievable by using a uniform input distribution is at least a fraction
$\frac{1}{2} e \ln(2) \approx 0.924$ of
capacity over the entire range of $p$ (with equality when $p$ approaches one).
Even more fortunate, from this
perspective the Z channel is the extremal case \cite{MaR91,ShF04}: the information rate of
any binary-input memoryless channel when the input distribution is the uniform one
is at least a fraction $\frac{1}{2} e \ln(2)$ of its capacity.
From the above discussion we conclude that, when
dealing with asymmetric channels, not much is lost if we use a binary linear coding
scheme (inducing a uniform input distribution).

Consider the density evolution analysis.
Because of the lack of symmetry
we can no longer make the all-one codeword assumption.
Therefore, it seems at first that we have to analyze the behavior of the decoder
with respect to each codeword.
Fortunately this is not necessary.
First note that, since we consider an ensemble average,
only the \emph{type} of the codeword matters. More precisely, let us say that
a codeword has type $\tau$ if the fraction of zeros
is $\tau$.
For $\ux \in \Code$, let $\tau(\ux)$ be its type.
Let us assume that we use an LDPC ensemble
whose dominant type is one-half. This means that ``most'' codewords contain
roughly as many zeros as one. Although it is possible
to construct degree-distributions which violate this constraint,
``most'' degree distributions do fulfill it.
Under this assumption there exists
some strictly positive constant $\gamma$  such that
\begin{equation}
\label{equ:typicaltype}
\prob \left\{\tau(\ux) \not\in 
[ 1/2-\delta/\sqrt{n}, 1/2+ \delta/\sqrt{n} ] \right\} 
\le e^{-\delta^2 \gamma}\, , 
\end{equation}
where the probability is with respect to a uniformly random codeword $\ux$.
We can therefore analyze the performance of
such a system in the following way:
determine the error probability assuming that the
type of the transmitted codeword is ``close'' to the typical one.
Since sublinear changes in the type do not figure in the density analysis,
this task can be accomplished by a straightforward density evolution analysis.
Now add to this the probability
that the type of a random codeword deviates significantly from the typical one.
The second term can be made arbitrarily small (see right hand side of (\ref{equ:typicaltype}))
by choosing $\delta$ sufficiently large.

We summarize: if we encounter a non-symmetric channel and we are willing
to sacrifice a small fraction of capacity then we can still use standard LDPC ensembles 
(which impose a uniform input distribution) to transmit at low complexity. 
If it is crucial that we approach capacity even closer, a more sophisticated approach is
required. We can combine LDPC ensembles with non-linear mappers which map the uniform
input distribution imposed by linear codes into a non-uniform input distribution
at the channel input in order to bring the mutual information closer to capacity.
For a detailed discussion on coding for the $Z$-channel we refer the reader to 
\cite{McE01,WKP03,BeB03}.

%
\section{Open Problems}
\label{sec:openproblems}
Let us close by reviewing some of the most important open challenges
in the channel coding problem.

\subsection{Order of Limits}

Density evolution computes the limit
\begin{align*}
\lim_{\iteration \rightarrow \infty} \lim_{N \rightarrow \infty} \expectation [\bit^{(N, \iteration)}].
\end{align*}
In words we determined the limiting
performance of an ensemble under a {\em fixed} number of iterations
as the blocklength tends to infinity and then let the number of iterations
tend to infinity. As we have seen, this limit is relatively easy to compute.
What happens if the order of limits is exchanged, i.e., how does the limit
\begin{align*}
\lim_{N \rightarrow \infty} \lim_{t \rightarrow \infty} \expectation [\bit^{(N, \iteration)}]
\end{align*}
behave? This limit is closer in spirit 
to the typical operation in practice: for each fixed length the BP
decoder continues until no further progress is achieved. We are interested
in the limiting performance as the blocklength tends to infinity.

For the BEC it is known that the two limits coincide. If we combine this with
the fact that for the BEC the performance is a monotone function in the number of
iterations (any further iteration can only make the result better) then we get the
important observation that regardless of how we take the limit (jointly or sequentially),
as long as both the blocklength as well as the number of iterations tend to infinity 
we get the same result. From a practical perspective this is comforting to know: it shows
that we can expect a certain robustness of the performance with respect to the 
details of the operation.

It is conjectured that the same statement holds for general BMS channels. 
Unfortunately, no proof is known.

\subsection{Finite-Length Performance}
\begin{figure}[htp]
\centering
\setlength{\unitlength}{1.5bp}%
\begin{picture}(120,108)(-12,-8)
\put(0,0)
{
\put(0,0){\includegraphics[scale=1.5]{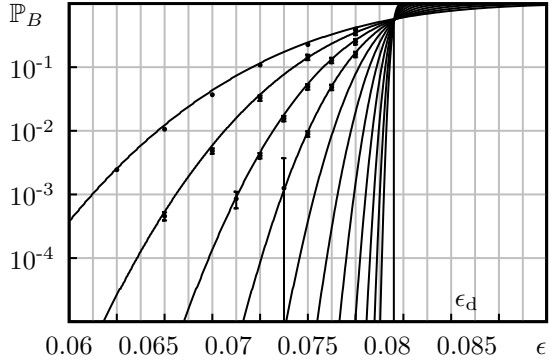}}
\graphtextsize
\multiputlist(0,-8)(20,0)[b]{$0.06$,$0.065$,$0.07$,$0.075$,$0.08$,$0.085$}
\multiputlist(-15,16)(0,16)[l]{$10^{\text{-4}}$,$10^{\text{-3}}$,$10^{\text{-2}}$,$10^{\text{-1}}$}
\put(120,-8){\makebox(0,0)[rb]{$\epsilon$}}
\put(-15,80){\makebox(0,0)[lt]{$\prob_{B}$}}
\put(97,3){\makebox(0,0)[lb]{$\epsilon_{\rm d}$}}
}
\end{picture} 
\caption{\label{fig:scaling}
Performance of the BP decoder for the $(3, 6)$-regular ensemble
when transmission takes place over the BSC.  The block lengths are $n=2^i$, $i=10, \cdots, 20$.
The dots correspond to simulations.
For most simulation points the $95\%$ confidence intervals are smaller than the dot size.
The lines correspond to the analytic approximation based on scaling laws.
}
\end{figure}

The threshold gives an indication of the asymptotic performance: for
channel parameters which are better than the threshold sufficiently long codes 
allow transmission at an arbitrarily low probability of bit error.
If, on the other hand, we transmit over a channel which has a parameter that
is worse than the threshold then we can not hope to achieve a low probability of error.
This is an important insight but from a practical perspective we would like to know 
how fast the finite-length performance approaches this asymptotic limit.
There can be many different ensembles that all have the same asymptotic performance
but that might have a substantially different finite-length behavior. 
Can we predict which one we should choose a priori without having to resort
to simulations? The typical convergence of the performance to the asymptotic limit
is shown in Fig.~\ref{fig:scaling}.
The points correspond to simulation results whereas the solid curves correspond to
a general scaling conjecture \cite{AMRU03}. Roughly speaking, this scaling conjecture 
states that around the threshold
the error probability behaves as follows: 
Let the channel be parameterized by $\epsilon$ with increasing $\epsilon$ indicating a worsening
of the channel.
Let $\epsilon_{\rm d}$ be the BP threshold, and define $z = \sqrt{n} (\epsilon-\epsilon_{\rm d})$. Then for $z$ fixed 
and $n$ tending to infinity we have
\begin{align*}
\block(n, \epsilon)
& = \Phi\left(z/\alpha \right)\bigl(1+o(1) \bigr), \\
\end{align*}
where $\Phi(\,\cdot\,)$ is the error function (i.e. $\Phi(x)$ 
is the probability that a standard normal random variable is smaller than 
$x$), and $\alpha$ is a constant which depends on the channel as well as on the
channel.

For the BEC this scaling law has been shown to be correct \cite{AMRU03}. In fact, even a refined version is
known \cite{DeM07}:  define $z = \sqrt{n} (\epsilon-\epsilon_{\rm d}+\beta n^{-\frac{2}{3}})$
where $\beta$ is a constant depending on the ensemble. Then for $z$ fixed
and $n$ tending to infinity we have
\begin{align*}
\block(n,  \epsilon)
& = Q\left(z/\alpha \right)\bigl(1+O \bigl(n^{-1/3} \bigr) \bigr).
\end{align*}

For general channels on the other hand the problem is largely open. 
If proved to be correct, finite length scaling laws could be used as a tool for
an efficient finite-length optimization.

\subsection{Capacity-Achieving Codes}
For the most part we have taken the point of view that we are given
an ensemble of codes and a family of channels and  we would like to
determine the performance of this combination. For instance, the most
fundamental question is to determine the threshold noise
for such a code/decoder  combination.
Hopefully this threshold is close to the best possible as determined by
Shannon's capacity formula.

But we can take a  more active point of view. Given a family of channels
how should we choose the ensemble in order for the threshold noise
to be as high as possible. In other words, can we approach the capacity of the channel?

For the BSC this question has been answered by Luby, Mitzenmacher, Shokrollahi, Spielman,
and Steman in \cite{LMSSS97}. These authors showed that by a suitable choice of the degree distribution
one can approach the capacity arbitrarily closely. More precisely, in order to 
approach capacity up to a fraction $\delta$ the average degree has to grow like $\log(1/\delta)$
and this is the best possible. For general channels it is not known whether capacity can be achieved.
Although the resolution of this problem will most likely only have a small practical implication
it is without doubt the most important open theoretical problem in this area.

\appendix

\section{A Generating Function Calculation}
\label{app:GenFun}

We want to compute the number of ways of
picking $n$ distinct objects from $M$ groups each containing $k$ objects, in such
a way  that the number of selected elements in each group is even.
Let us denote this number by $C_n$.
The result of this computation is used in Section~\ref{sec:we}, where we claimed the result to be
$C_n = \coeff[q_k(z)^M,z^n]$ (in that case $n=lw$).

First notice that given $m_1, \dots, m_M$ all even, the number of ways of 
choosing $m_1$ objects from the first group, $m_2$ from the second, etc
is
\begin{eqnarray}
\binom{k}{m_1}\binom{k}{m_2}\cdots\binom{k}{m_M}\, .
\end{eqnarray}
The desired number $C_n$ is the sum of this quantity over 
the even numbers $m_1,\cdots, m_n$ such that 
$m_1,\dots, m_M\in\{0,\dots k\}$ and
$m_1+\dots +m_M = n$. The corresponding generating
function is
\begin{eqnarray}
C(z) \equiv\sum_{n=0}^{kM} C_n z^n = 
\sum^{\mbox{\tiny even}}_{m_1\dots m_n} \binom{k}{m_1}\binom{k}{m_2}\cdots\binom{k}{m_M}
z^{m_1}\cdots z^{m_N}\, ,
\end{eqnarray}
the sum being restricted over even integers $m_1,\dots, m_M\in\{0,\dots k\}$.
We now notice that the sum factorizes yielding
\begin{eqnarray}
C(z) = \left\{\sum_{m\mbox{\tiny even}}\binom{k}{m}z^m\right\}^M
= q_k(z)^{M}\, ,
\end{eqnarray}
which proves the claim.

\section*{Acknowledgments}
We would like to thank Jean-Philippe Bouchaud and Marc M{\'e}zard for organizing the
2006 Summer School on Complex Systems in Les Houches.
We would also like to thank Cyril M{\'e}asson and Christine Neuberg with their help 
in creating Fig.~\ref{fig:deforgec}.

\bibliographystyle{siam}

\begin{thebibliography}{10}

\bibitem{AldousSteele}
{\sc D.~Aldous and J.~M. Steele}, {\em The {O}bjective {M}ethod:
  {P}robabilistic {C}ombinatorial {O}ptimization and {L}ocal {W}eak
  {C}onvergence}, in Probability on discrete structures, H.~Kesten, ed.,
  Springer, New York, 2004.

\bibitem{AMRU03}
{\sc A.~Amraoui, A.~Montanari, T.~Richardson, and R.~Urbanke}, {\em
  Finite-length scaling for iteratively decoded {LDPC} ensembles}, in Proc.
  41th Annual Allerton Conference on Communication, Control and Computing,
  Monticello, IL, 2003.

\bibitem{BaB05}
{\sc O.~Barak and D.~Burshtein}, {\em Lower bounds on the spectrum and error
  rate of {LDPC} code ensembles}, in Proc. of the IEEE Int. Symposium on
  Inform. Theory, Adelaide, Australia, September 4--9 2005, pp.~42--46.

\bibitem{BKH97}
{\sc G.~Bauch, H.~Khorram, and J.~Hagenauer}, {\em Iterative equalization and
  decoding in mobile communications systems}, in Proceedings of GLOBECOM, 1997.

\bibitem{BeB03}
{\sc A.~Bennatan and D.~Burshtein}, {\em Iterative decoding of {LDPC} codes
  over arbitrary discrete-memoryless channels}, in Proceedings of the 41-st
  Allerton Conference on Communication, Control, and Computing, Monticello, IL,
  Oct. 2003, pp.~1416--1425.

\bibitem{Ber84}
{\sc E.~R. Berlekamp}, {\em Algebraic Coding Theory}, Aegean Park Press, 1984.

\bibitem{BGT93}
{\sc C.~Berrou, A.~Glavieux, and P.~Thitimajshima}, {\em Near {S}hannon limit
  error-correcting coding and decoding}, in Proceedings of ICC'93, Geneve,
  Switzerland, May 1993, pp.~1064--1070.

\bibitem{Bethe}
{\sc H.~A. Bethe}, {\em Statistical theory of superlattices}, Proc. Roy. Soc.
  London A, 150 (1935), pp.~552--558.

\bibitem{Bla84}
{\sc R.~E. Blahut}, {\em Theory and Practice of Error Control Codes},
  Addison-Wesley, 1983.

\bibitem{Boltzmann}
{\sc L.~E. Boltzmann}, {\em Vorlesungen {\"u}ber Gastheorie}, J.A. Barth,
  Leipzig, 1896.

\bibitem{BMZ03}
{\sc A.~Braunstein, M.~M{\'e}zard, and R.~Zecchina}, {\em Survey propagation:
  algorithm for satisfiability}.
\newblock arXiv:cond-math/cond-mat/0212002.

\bibitem{CoT91}
{\sc T.~M. Cover and J.~A. Thomas}, {\em Elements of Information Theory},
  Wiley, New York, 1991.

\bibitem{DeM07}
{\sc A.~Dembo and A.~Montanari}, {\em Finite size scaling for the core of large
  random hypergraphs}.
\newblock Xarch:math.PR/0702007, 2007.

\bibitem{DPJDCG}
{\sc C.~Douillard, A.~Picart, M.~J{\'e}z{\'e}quel, P.~Didier, C.~Berrou, and
  A.~Glavieux}, {\em Iterative correction of intersymbol interference:
  Turbo-equalization}, European Trans. on Commun., 6 (1995), pp.~507--511.

\bibitem{For01}
{\sc G.~D. Forney, Jr.}, {\em Codes on graphs: Normal realizations}, IEEE
  Trans. Inform. Theory, 47 (2001), pp.~520--548.

\bibitem{FrancoPaull}
{\sc J.~Franco and M.~Paull}, {\em Probabilistic analysis of the {Davis-Putnam}
  procedure for solving satisfiability}, Discrete Appl.~Math, 5 (1983),
  pp.~77--87.

\bibitem{DynamicCodes}
{\sc S.~Franz, M.~Leone, A.~Montanari, and F.~Ricci-Tersenghi}, {\em Dynamic
  phase transition for decoding algorithms}, Phys.~Rev.~E, 22 (2002),
  p.~046120.

\bibitem{FraPar_pot}
{\sc S.~Franz and G.~Parisi}, {\em Recipes for metastable states in spin
  glasses}, J. Physique I, 5 (1995), p.~1401.

\bibitem{Gal62}
{\sc R.~G. Gallager}, {\em Low-density parity-check codes}, IRE Transactions on
  Information Theory, 8 (1962), pp.~21--28.

\bibitem{Gal68}
\leavevmode\vrule height 2pt depth -1.6pt width 23pt, {\em Information theory
  and reliable communication}, Wiley, 1968.

\bibitem{GaV97}
{\sc J.~Garcia-Frias and J.~D. Villasenor}, {\em Combining hidden {M}arkov
  source models and parallel concatenated codes}, IEEE Communications Letters,
  1 (1997), pp.~111--113.

\bibitem{GFV98a}
\leavevmode\vrule height 2pt depth -1.6pt width 23pt, {\em Turbo decoders for
  {M}arkov channels}, IEEE Commun. Lett., 2 (1998), pp.~257--259.

\bibitem{GareyJohnson}
{\sc M.~R. Garey and D.~S. Johnson}, {\em {C}omputers and {I}ntractability: {A}
  {G}uide to the {T}heory of {NP-C}ompleteness}, W. H. Freeman \& Co., New
  York, 1979.

\bibitem{Kikuchi}
{\sc R.~Kikuchi}, {\em A theory of cooperative phenomena}, Phys. Rev., 81
  (1951), pp.~988--1003.

\bibitem{KiE03}
{\sc F.~R. Kschischang and A.~W. Eckford}, {\em Low-density parity-check codes
  for the gilbert-elliot channel}, in Proc. 41th Annual Allerton Conference on
  Communication, Control and Computing, Monticello, IL, 2003.

\bibitem{LiC04}
{\sc S.~Lin and D.~J. Costello, Jr.}, {\em Error Control Coding},
  Prentice-Hall, 2nd~ed., 2004.

\bibitem{LMSSS97}
{\sc M.~Luby, M.~Mitzenmacher, A.~Shokrollahi, D.~A. Spielman, and V.~Stemann},
  {\em Practical loss-resilient codes}, in Proceedings of the $29$th annual ACM
  Symposium on Theory of Computing, 1997, pp.~150--159.

\bibitem{Mac97}
{\sc D.~J.~C. MacKay}, {\em Good error correcting codes based on very sparse
  matrices}, IEEE Trans. Inform. Theory, 45 (1999), pp.~399--431.

\bibitem{MacKay}
{\sc D.~J.~C. MacKay}, {\em Information Theory, Inference \& Learning
  Algorithms}, Cambridge University Press, Cambridge, 2002.

\bibitem{MaN95}
{\sc D.~J.~C. MacKay and R.~M. Neal}, {\em Good codes based on very sparse
  matrices}, in Cryptography and Coding. 5th {IMA} Conference, C.~Boyd, ed.,
  no.~1025 in Lecture Notes in Computer Science, Springer, Berlin, 1995,
  pp.~100--111.

\bibitem{Mac05}
{\sc N.~Macris}, {\em Correlation inequalities: a useful tool in the theory of
  {LDPC} codes}, in Proc. of the IEEE Int. Symposium on Inform. Theory,
  Adelaide, Australia, Sept. 2005, pp.~2369 -- 2373.

\bibitem{MaR91}
{\sc E.~E. Majani and H.~Rumsey, Jr.}, {\em Two results on binary-input
  discrete memoryless channels}, in Proc. of the IEEE Int. Symposium on Inform.
  Theory, June 1991, p.~104.

\bibitem{MMW05}
{\sc E.~Maneva, E.~Mossel, and M.~J. Wainwright}, {\em A new look at survey
  propagation and its generalizations}, in SODA, Vancouver, Canada, 2005.

\bibitem{McE01}
{\sc R.~J. McEliece}, {\em Are turbo-like codes effective on nonstandard
  channels?}, IEEE inform. Theory Soc. Newslett., 51 (2001), pp.~1--8.

\bibitem{MMRU05}
{\sc C.~M{\'e}asson, A.~Montanari, T.~Richardson, and R.~Urbanke}, {\em {The
  Generalized Area Theorem and Some of its Consequences}}.
\newblock submitted to IEEE IT, 2005.

\bibitem{MeM07}
{\sc M.~Mezard and A.~Montanari}, {\em Information, Physics and Computation},
  Clarendon Press - Oxford, 2007.
\newblock to be published.

\bibitem{MezardParisiBethe}
{\sc M.~M\'ezard and G.~Parisi}, {\em The bethe lattice spin glass revisited},
  Eur. Phys. J. B, 20 (2001), p.~217.

\bibitem{AM_rce}
{\sc A.~Montanari}, {\em The glassy phase of {G}allager codes},
  Eur.~Phys.~J.~B, 23 (2001), pp.~121--136.

\bibitem{Montanari05}
\leavevmode\vrule height 2pt depth -1.6pt width 23pt, {\em {Tight bounds for
  LDPC and LDGM codes under MAP decoding}}, IEEE Trans. Inform. Theory, 51
  (2005), pp.~3221--3246.

\bibitem{MontanariShah}
{\sc A.~Montanari and D.~Shah}, {\em Counting good truth assignments of random
  k-sat formulae}, in SODA, New Orleans, USA, Jan. 2007, pp.~1255--1264.

\bibitem{MoR95}
{\sc R.~Motwani and P.~Raghavan}, {\em Randomized Algorithms}, Cambridge
  University Press, Cambridge, 1995.

\bibitem{MurayamaEtAl}
{\sc T.~Murayama, Y.~Kabashima, D.~Saad, and R.~Vicente}, {\em Statistical
  physics of regular low-density parity-check error-correcting codes}, Phys.
  Rev. E, 62 (2000), p.~1577.

\bibitem{Neu04}
{\sc C.~Neuberg}, {\em {G}ilbert-{E}lliott channel and iterative decoding}.
\newblock EPFL, Semester Project (Supervisor: Cyril M{\'e}ason), 2004.

\bibitem{NishimoriBook}
{\sc H.~Nishimori}, {\em Statistical Physics of Spin Glasses and Information
  Processing}, Oxford University Press, Oxford, 2001.

\bibitem{Per86}
{\sc J.~Pearl}, {\em Fusion, propagation, and structuring in belief networks},
  Artificial Intelligence, 29 (1998), pp.~241--288.

\bibitem{AbouChacra73}
{\sc P.~W.~A. R.~Abou-Chacra and D.~J. Thouless}, {\em A selfconsistent theory
  of localization}, J. Phys C, 6 (1973), p.~1734.

\bibitem{Rat06}
{\sc V.~Rathi}, {\em On the asymptotic weight and stopping set distribution of
  regular {LDPC} ensembles}, IEEE Trans. Inform. Theory, 52 (2006),
  pp.~4212--4218.

\bibitem{RiJ06}
{\sc T.~Richardson and H.~Jin}, {\em A new fast density evolution}, in Proc. of
  the IEEE Inform. Theory Workshop, Monte Video, Uruguay, Feb. 2006.
\newblock pp. 183--187.

\bibitem{RiU01}
{\sc T.~Richardson and R.~Urbanke}, {\em The capacity of low-density parity
  check codes under message-passing decoding}, IEEE Trans. Inform. Theory, 47
  (2001), pp.~599--618.

\bibitem{RiU07}
\leavevmode\vrule height 2pt depth -1.6pt width 23pt, {\em Modern Coding
  Theory}, Cambridge University Press, 2007.
\newblock to be published.

\bibitem{Rujan}
{\sc P.~Rujan}, {\em Finite temperature error-correcting codes},
  Phys.~Rev.~Lett., 70 (1993), pp.~2968--2971.

\bibitem{SaS06}
{\sc I.~Sason and S.~Shamai}, {\em Performance Analysis of Linear Codes under
  Maximum-Likelihood Decoding: A Tutorial}, vol.~3 of Foundations and Trends in
  Communications and Information Theory, NOW, Delft, the Netherlands, July
  2006.

\bibitem{Sha48}
{\sc C.~E. Shannon}, {\em A mathematical theory of communication}, Bell System
  Tech. J., 27 (1948), pp.~379--423, 623--656.

\bibitem{ShF04}
{\sc N.~Shulman and M.~Feder}, {\em The uniform distribution as a universal
  prior}, IEEE Trans. Inform. Theory, 50 (2004), pp.~1356--1362.

\bibitem{SiS96}
{\sc M.~Sipser and D.~A. Spielman}, {\em Expander codes}, IEEE Trans. Inform.
  Theory, 42 (1996), pp.~1710--1722.

\bibitem{Sou89}
{\sc N.~Sourlas}, {\em Spin-glass models as error-correcting codes}, Nature,
  339 (1989), pp.~693--695.

\bibitem{TatikondaJordan}
{\sc S.~Tatikonda and M.~Jordan}, {\em Loopy belief propagation and {Gibbs}
  measures}, in Proc. Uncertainty in Artificial Intell., Alberta, Canada, Aug.
  2002, pp.~493--500.

\bibitem{MSK02}
{\sc J.~van Mourik, D.~Saad, and Y.~Kabashima}, {\em Critical noise levels for
  {LDPC} decoding}, Physical Review E, 66 (2002).

\bibitem{WKP03}
{\sc C.-C. Wang, S.~R. Kulkarni, and H.~V. Poor}, {\em Density evolution for
  asymmetric memoryless channels}, IEEE Trans. Inform. Theory, 51 (2005),
  pp.~4216--4236.

\bibitem{WiS05}
{\sc G.~Wiechman and I.~Sason}, {\em On the parity-check density and achievable
  rates of {LDPC} codes for memoryless binary-input output-symmetric channels},
  in Proc. of the Allerton Conf. on Commun., Control and Computing, Monticello,
  IL, USA, September 28--30 2005, pp.~1747--1758.

\bibitem{YedidiaFreemanWeiss}
{\sc J.~S. Yedidia, W.~T. Freeman, and Y.~Weiss}, {\em Constructing free energy
  approximations and generalized belief propagation algorithms}, IEEE Trans.
  Info.Theory, 51 (2005), pp.~2282--2313.

\end{thebibliography}

\newcommand{\SortNoop}[1]{}

\end{document}